\documentclass[pre,twocolumn,preprintnumbers,showpacs,floatfix,superscriptaddress]{revtex4-1}
\usepackage{epsf}
\usepackage{graphicx}
\usepackage{dcolumn}
\usepackage{bm}
\usepackage{subfigure}
\usepackage{color}
\usepackage{amsthm}
\usepackage{amssymb}
\usepackage{amsmath}
\usepackage{pb-diagram}

\usepackage{hyperref}

\newcommand{\pd}{{\mathsf{PD}}}

\newcommand{\setof}[1]{\left\{ {#1}\right\}}

\newcommand{\R}{{\mathbb{R}}}

\def\setof#1{\left\{{#1}\right\}}

\definecolor{gray85}{gray}{0.85} 
\definecolor{gray8}{gray}{0.8} 
\definecolor{gray7}{gray}{0.7} 
\definecolor{gray6}{gray}{0.6} 
\definecolor{gray5}{gray}{0.5} 
\definecolor{gray4}{gray}{0.4} 
\definecolor{gray35}{gray}{0.35} 

\begin{document}

\preprint{}
\title[]{Structure of force networks in tapped particulate systems of disks and pentagons (Part 2): Persistence analysis}

\author{L. Kondic}
\affiliation{Department of Mathematical Sciences, New Jersey Institute of Technology, Newark, New Jersey 07102, USA}

\author{M. Kram\'ar}
\affiliation{Department of Mathematics, Rutgers University, Piscataway, New Jersey 08854-8019, USA}

\author{Luis A. Pugnaloni}
\affiliation{Dpto.~de Ingenier\'ia Mec\'anica, Facultad Regional La Plata, Universidad Tecnol\'ogica Nacional, Av. 60 Esq. 124, 1900 La Plata, Argentina}
\affiliation{Consejo Nacional de Investigaciones Cient\'ificas y T\'ecnicas, Argentina}

\author{C. Manuel Carlevaro}
\affiliation{Instituto de F\'isica de L\'iquidos y Sistemas Biol\'ogicos (CONICET La Plata,
UNLP), Calle 59 Nro 789, 1900 La Plata, Argentina}
\affiliation{Universidad Tecnol\'ógica Nacional-FRBA, UDB F\'isica, Mozart 2300, C1407IVT Buenos Aires, Argentina}

\author{K. Mischaikow}
\affiliation{Department of Mathematics, Rutgers University, Piscataway, New Jersey 08854-8019, USA}


\begin{abstract}
In the companion paper~\cite{paper1}, we use classical measures based on 
force probability density functions (PDFs), as well as Betti numbers (quantifying the number of components, related to force chains, and
loops), to describe the force networks in tapped systems of disks and pentagons. 
In the present work, we focus on the use of persistence analysis, that allows to describe these
networks in much more detail.   This approach allows not only to describe, but also to quantify the differences
between the force networks in different realizations of a system, in different parts of the considered domain, or 
in different systems.   We show that persistence analysis clearly distinguishes the systems that are very difficult
or impossible to differentiate using other means.   One important finding is that the differences in force networks 
between disks and pentagons are most apparent when loops are considered: the quantities describing properties 
of the loops may differ significantly even if other measures (properties of components, Betti numbers, or force
PDFs) do not distinguish clearly the investigated systems.
\end{abstract}

\volumeyear{}
\volumenumber{}
\issuenumber{}
\eid{}
\date{\today}
\startpage{1}
\endpage{}
\maketitle

\section{Introduction}

In the companion paper~\cite{paper1} we compare the force networks in tapped systems
by using relatively simple measures: probability density functions (PDFs) for normal and tangential forces between the particles, 
correlation functions describing positional order of the considered particles, as well as possible correlations of the 
emerging force networks.  These classical measures are supplemented by analysis of cluster sizes and distributions at different force levels (i.e., by considering the part of the force network that only includes contacts involving forces exceeding a threshold).  
These results have uncovered some differences between the force networks in the considered systems. For example, 
we have found that the number of clusters as a function of the force level is heterogeneous in the tapped systems under gravity, with different 
distributions deeper in the samples compared to the ones measured closer to the surface.   However, some of the
differences remain unclear.  For example,  tapped disks exposed to different tap intensities that lead to the 
same (average) packing fraction, are found to have similar PDFs and similar cluster size distributions, although it is known~\cite{arevalo_pre13} that there are some differences in the geometrical properties of the contact networks in these systems. 

In the present paper, we focus on a different approach, based on persistence analysis.
This approach has been successfully used to explain and quantify the properties of force 
networks in the systems exposed to compression~\cite{pre13,pre14,physicaD14}.   In essence, persistence analysis allows to quantify the 
force network `landscapes' in a manner that is global in character, but it still includes detailed information about the geometry at all force levels.   
The global approach to the analysis of force networks makes it complementary to other works that have considered in detail
the local structure of force networks~\cite{tordesillas_pre10}, and attention to geometry distinguishes this approach from network type of
analysis~\cite{daniels_pre12, herrera_pre11,walker_pre12,bassett_soft_mat15}. We will use persistence analysis to compare the force networks between the systems of disks exposed to different tapping intensities,  as well as to discuss similarities and differences between the systems of 
disks and pentagons.  As we will see, some differences 
between the considered networks that could not be clearly observed (and even less quantified) using classical
measures become obvious when persistence analysis is used.     Furthermore, persistence analysis allows for formulating
measures that can be used to quantify, in a precise manner, differences in force networks between realizations of a nominally same system.   
We note  that persistence has been used to quantify the features in other physical systems such as isotropically compressed granular 
media \cite{pre13,pre14,physicaD14}. It was also used to study dynamics of the 
Kolmogorov flow and Rayleigh-B\'enard convection~\cite{PDSubmited}.

This paper is organized as follows.  In Sec.~\ref{sec:methods}, we discuss briefly the persistence approach and 
also provide some examples to illustrate its use in the present context.    In Sec.~\ref{sec:results}, we discuss 
the outcome of persistence approach and quantify the differences between the considered systems. Section~\ref{sec:conclusions} is devoted to the  conclusions and suggestions for the future work. 

\section{Methods}
\label{sec:methods}

\subsection{Simulations}

The simulations utilized in this paper are  described in detail in~\cite{paper1}; here we provide a brief overview.
We consider  tapped systems of disks and pentagons in a gravitational field. The particles are confined in two-dimensional (2D) rectangular box 
with solid (frictionless) side walls,. Initially $500$ particles are placed at random (without overlaps) into the box,
and the  particles are allowed to settle to create the initial packing. Then,  $600$ vertical taps are applied to each system considered; we discard the
initial $100$ taps and analyze the remaining $500$.  After each tap, 
we wait for the particles to dissipate their kinetic energy and achieve a mechanical equilibrium.  We record the particles positions  and the forces   
acting between them; the interactions between the particles and the walls are not included.   For more direct comparison, 
the forces are normalized by the average contact force. 

In addition to discussing the influence of particle shape, we consider
two different tapping intensities, $\Gamma$ (called `high' and `low' tap in what follows) that lead to the same packing fractions for disks
($\Gamma = 3.83\sqrt{dg}$ (low) and $\Gamma = 12.14\sqrt{dg}$  (high), where $d$ is the disk radius and $g$ the acceleration of gravity).
We also discuss the influence of gravitational compaction, and for this 
purpose we consider `slices' of the systems, $10$ particle diameters thick: bottom slice positioned deep inside the domain, 
and the top slice close  to the surface.   See~\cite{paper1} for more details.

\subsection{Persistent homology}

We are interested in understanding the geometry exhibited by force networks.
Their complete numerical representation contains far too much information.
With this in mind, we make use of the tools from algebraic topology, in particular homology, to reduce this information by 
counting simple geometric structures.  In the two dimensional setting of interest in the present context, 
fixing a magnitude, $F$, of the force and considering the particles which interact with a force at or above $F$ yields a 
2D topological space, $X(F)$.  Two simple geometric properties of $X(F)$  are the number of components (clusters), 
$\beta_0(X(F))$, and the number of loops (holes), $\beta_1(X(F))$.

In~\cite{paper1} it is shown that even though we are counting very simple geometric objects, by varying the 
threshold $F$, the set of  Betti numbers  $\beta_0(X(F))$ and $\beta_1(X(F))$  provides novel 
distinctions between the behavior of the above mentioned  systems. 
However, there is an obvious limitation to just using the Betti numbers to describe a system.  
Consider two different thresholds and assume that the values of the Betti numbers are the same. 
Does this mean that the geometric structures, e.g.\ components and loops, are the same at these two thresholds, or have 
some components or loops disappeared and been replaced by an equal number of  different components or loops?
This distinction cannot be determined from the Betti number count alone.

\begin{figure}
\centering
\subfigure[ $\pd_0$, disks.]{\includegraphics[width=1.6in]{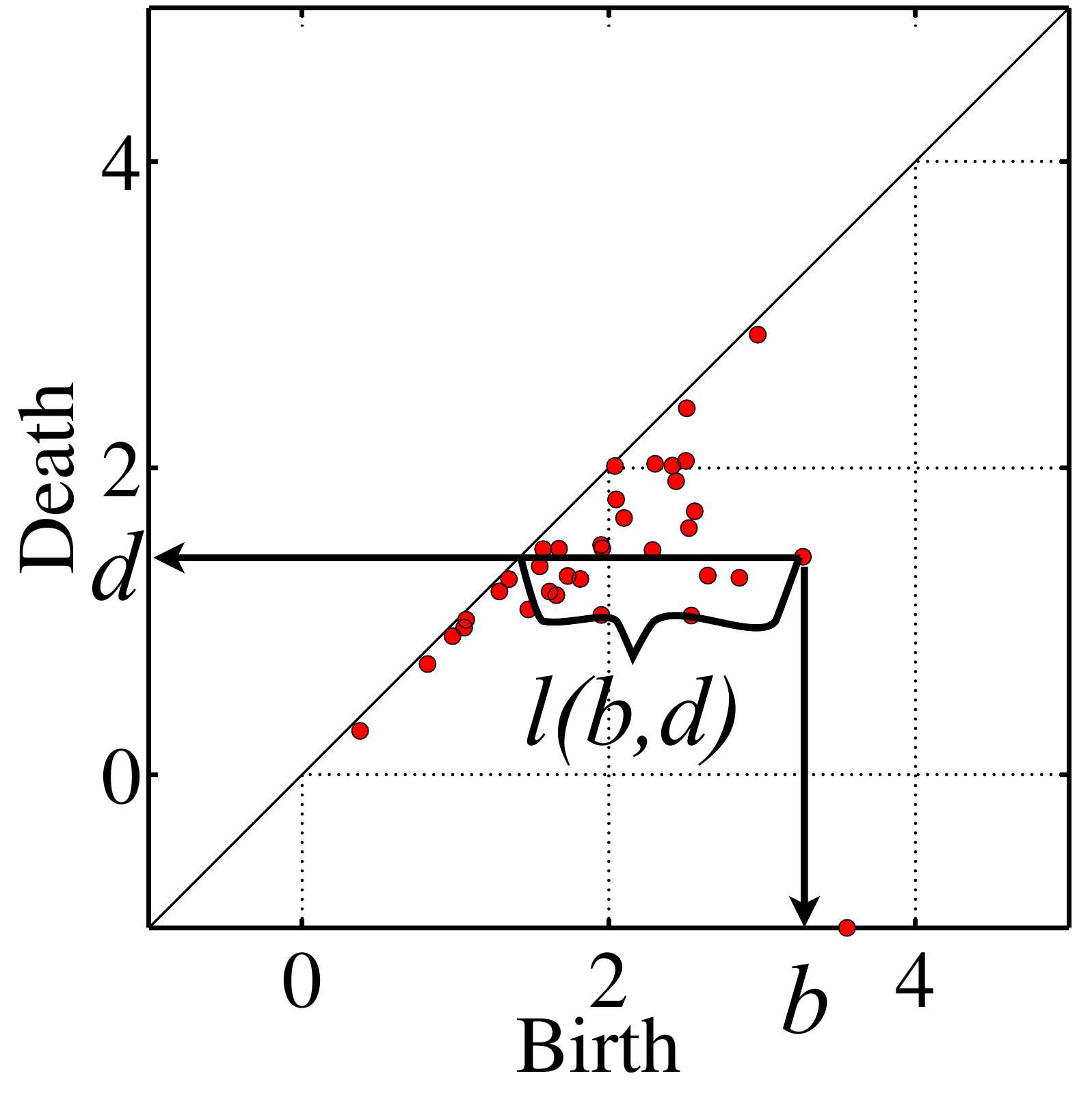}} 
\subfigure[ $\pd_1$ disks.]{\includegraphics[width = 1.69in]{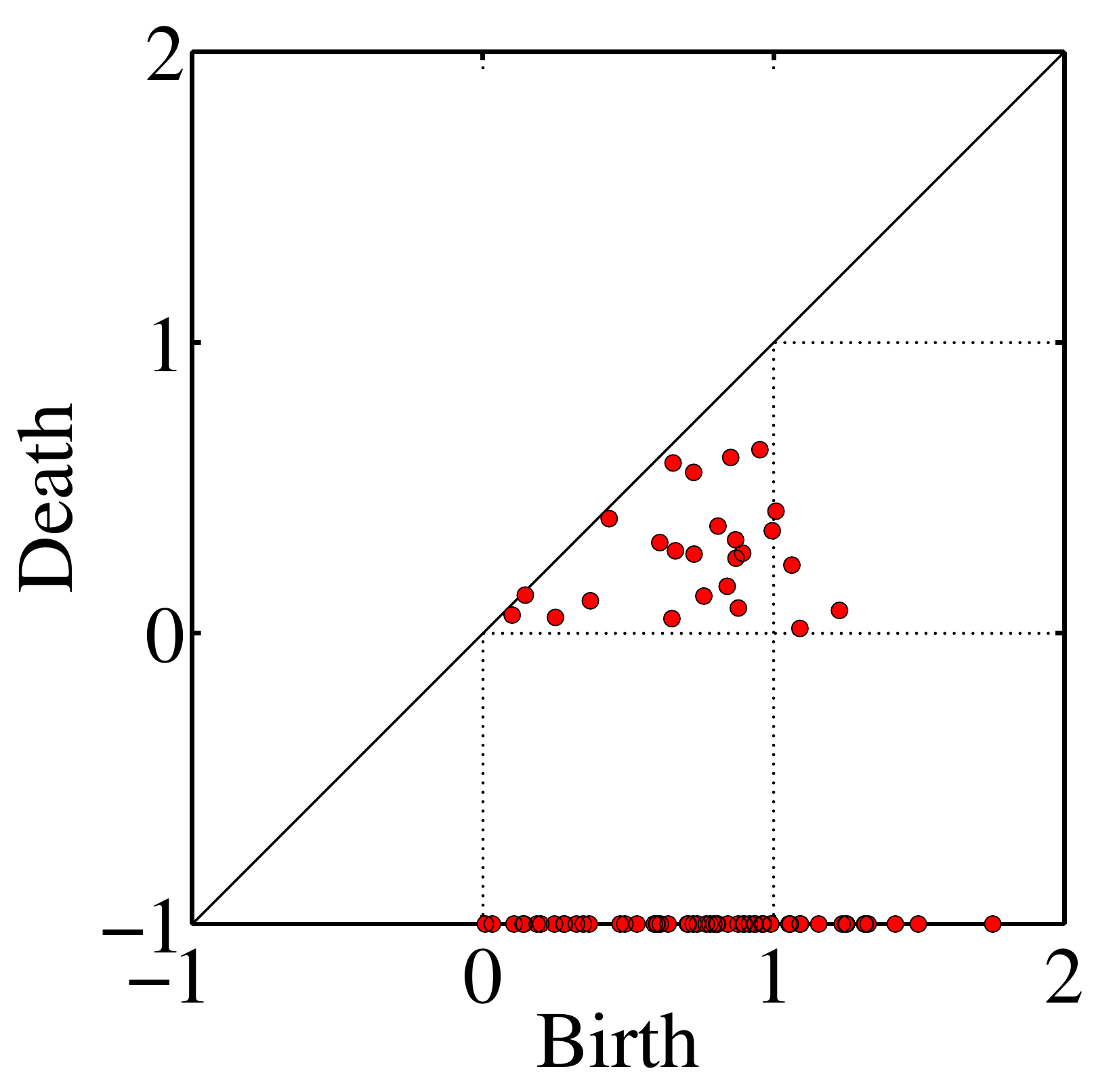}}\\
\subfigure[ $\pd_0$ pentagons.]{\includegraphics[width=1.6in]{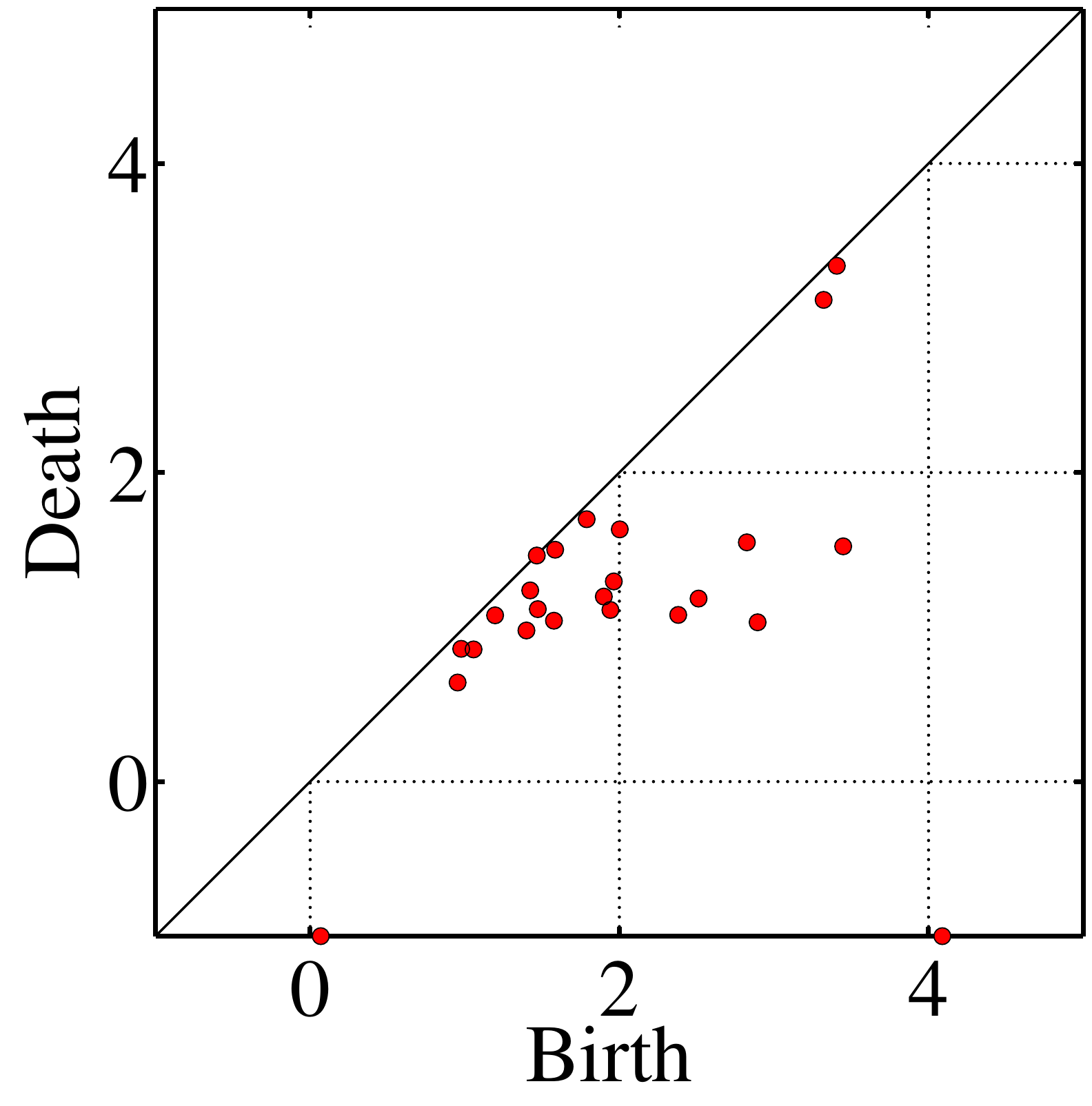}} 
\subfigure[ $\pd_1$ pentagons.]{\includegraphics[width = 1.69in]{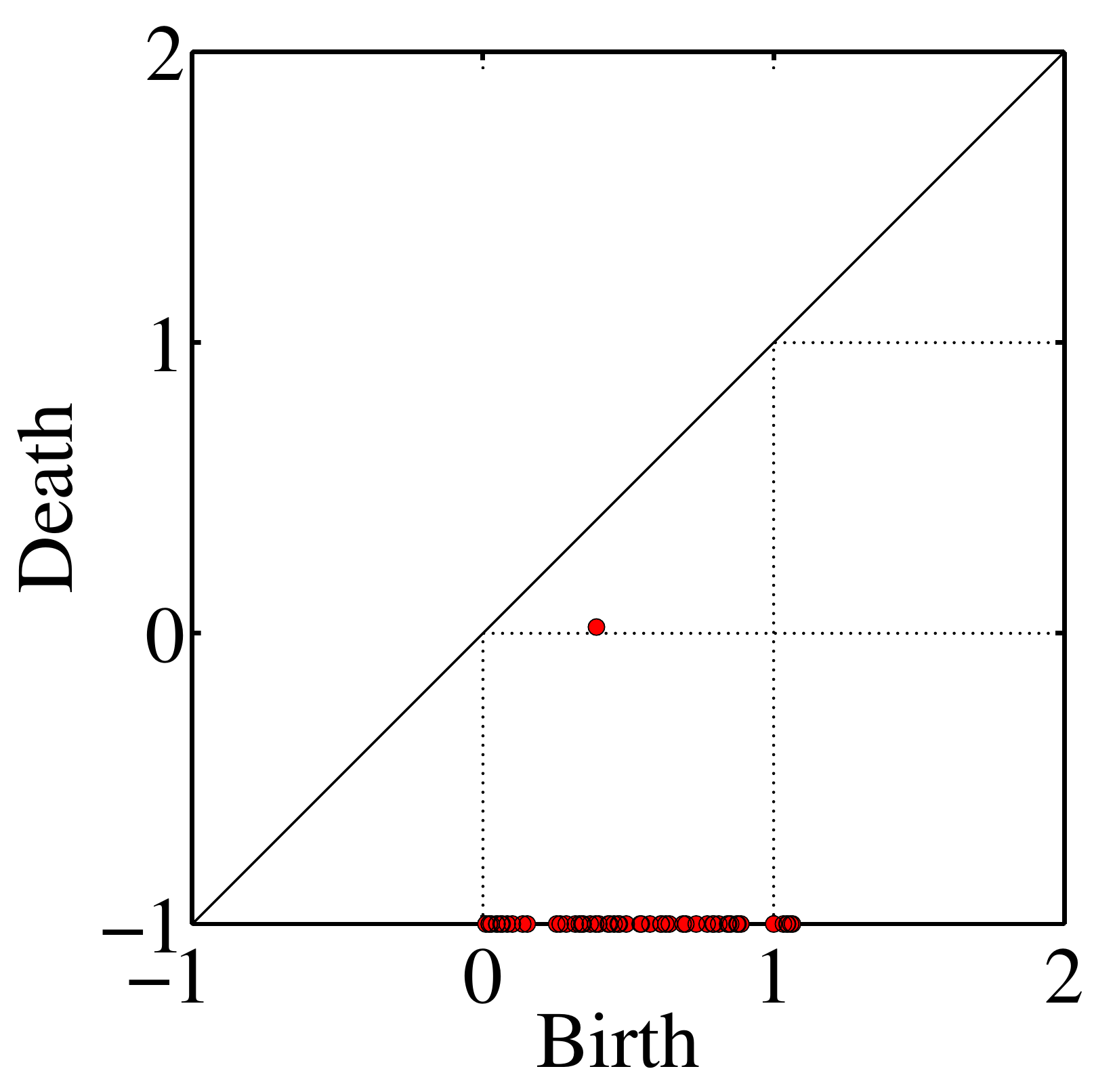}}
\caption{Examples of $\pd$s corresponding to the normal force network of one realization of systems of disks and pentagons (low tapping,
bottom slice).
In the part (a)  we also illustrate some of the concepts that will be used later in the paper: $b$ (birth time), $d$ (death time) and $l(b,d)$ (lifespan). 
}
\label{fig:diagrams}
\end{figure}

To provide more complete description, we make use of a relatively new algebraic topological tool called {\em persistent homology}.
In the context of the 2D systems that we are considering here, it is sufficient to remark that to each force network landscape persistent 
homology assigns two {\em persistence diagrams}, $\pd_0$ and $\pd_1$, such as those shown in Fig.~\ref{fig:diagrams}.
Each persistence diagram consists of a collection of pairs of points $(b,d)\in\R^2$ where $b$, the birth, indicates the threshold 
value at which a geometric structure (a component/cluster for $\pd_0$ or a loop for $\pd_1$) first appears and $d$, the death, indicates the 
threshold value at which the geometric structure disappears.  In this paper we measure the geometry of the part of the contact 
network with force interactions greater than a given threshold, and thus $b \geq d$.  The value $b-d$ is called the {\em lifespan}.
Note that  the component  represented by the point $(b,d)$  `dies' when it merges with some other component 
with the birth coordinate larger or equal to $b$. In particular, the single generator in $\pd_0$ with death coordinate 
$-1$, see Fig.~\ref{fig:diagrams}(a), represents the component that contains the strongest 
force `chain' in the system: the one that formed at the highest force level.  Note that it has both the highest birth value and the longest lifespan. 
More detailed interpretation of $\pd_0$ in 1D can be found in~\cite{pre14}, while a rigorous presentation for 2D is given in~\cite{physicaD14}.

The loop structure of a force network is described by $\pd_1$. 
A loop in the network is a closed path of the edges connecting centers of the particles.    
Similarly to  $\pd_0$, the point $(b,d) \in \pd_1$ indicates that a loop appears in the part of the network exceeding the force threshold $b$.  
This loop is present for all the values of the threshold in $(d,b]$.  
At the value $d$, this loop is filled in, that is, the interior of the loop is filled in with particles that form a crystalline structure, and the forces between the interacting particles inside of the loop are larger or equal to $d$. This fill-in process can be also seen as filling the loop by `trivial' loops formed by exactly three particles with  forces stronger or equal to $d$. 
To visually distinguish the loops that  do not get filled in at any force level (including the threshold $0$), we set their death time to $d =-1$. 
However, in the distance computations (discussed below) we follow the convention presented  in~\cite{physicaD14} and replace them by $0$.   
Note that in the example shown in Fig.~\ref{fig:diagrams}, for pentagons there is a single loop that gets filled in (close to threshold level $0$), 
while for disks there is a number of loops that gets filled in at a variety of thresholds.   

A particularly simple descriptor based on the lifespans is the total persistence, 
$
TP(\pd) = \sum_{(b,d) \in \pd}(b - d),
$
i.e.\ the sum of all lifespans of the points in $\pd$. 
In the context of Fig.~\ref{fig:diagrams}, the total persistence of (a) and (c) is roughly the same, while the total persistence of (b) is roughly twice that of (d).
Thus, this gives a simple measure that can, at least in some settings, distinguish persistence diagrams arising from different systems. 
To further distinguish the diagrams, we will also consider the distribution of lifespans
(representing the number of times a value of lifespan in a 
specified range is found).

\begin{figure}
\centering
\subfigure[ $\pd_0$ disks.]{\includegraphics[width=1.6in]{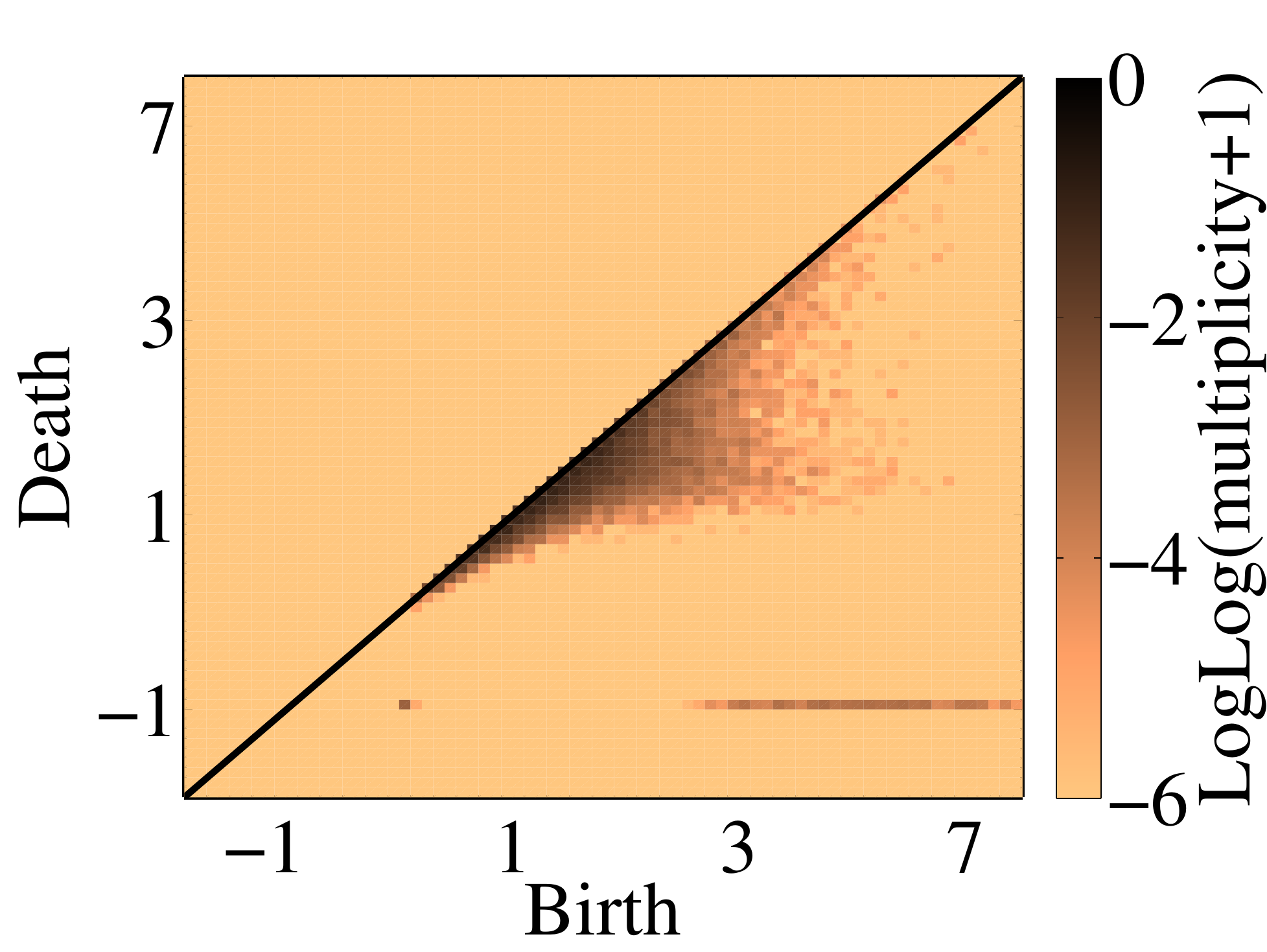}} 
\subfigure[ $\pd_1$ disks.]{\includegraphics[width=1.6in]{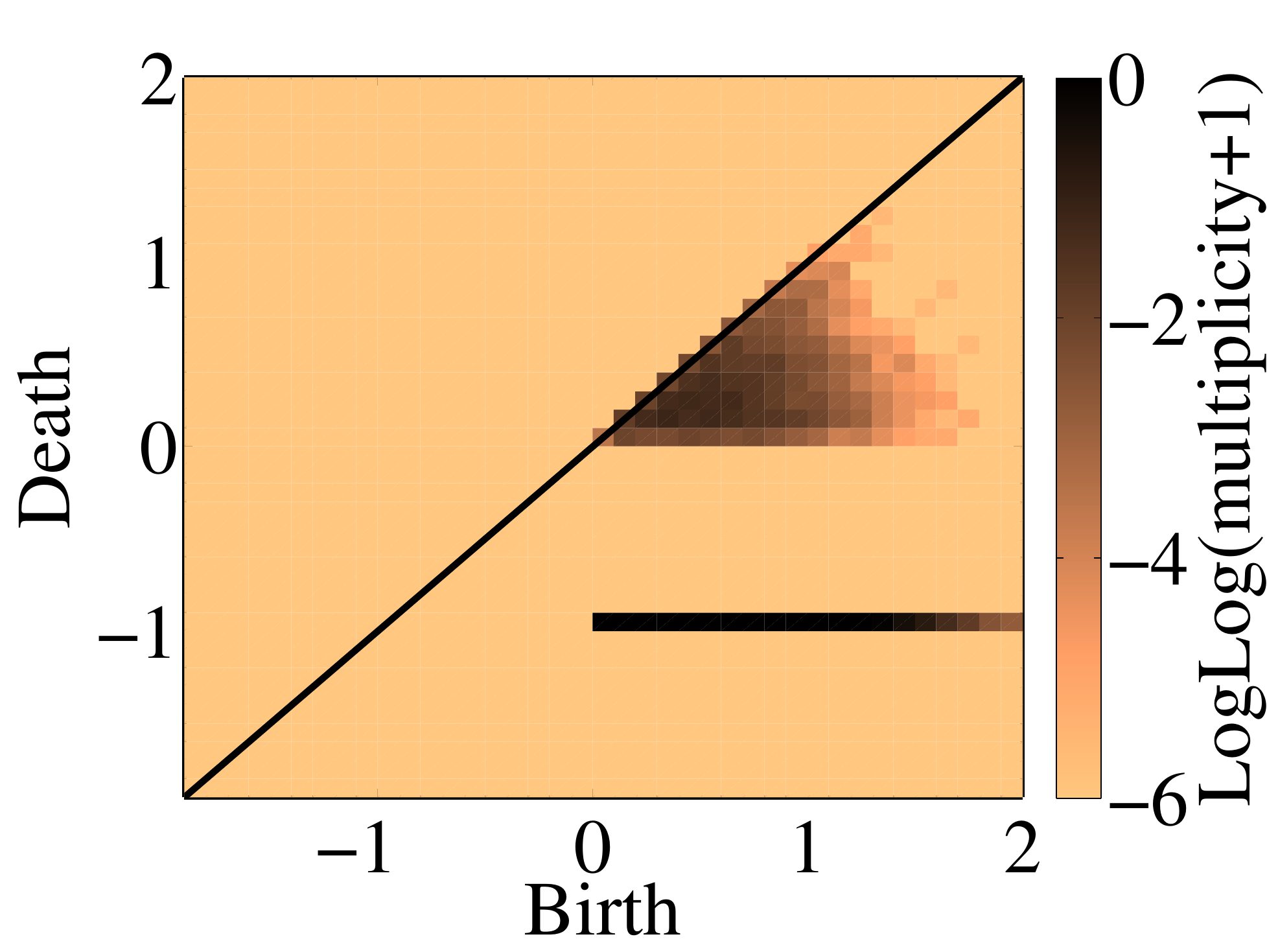}} 
\subfigure[ $\pd_0$ pentagons.]{\includegraphics[width=1.6in]{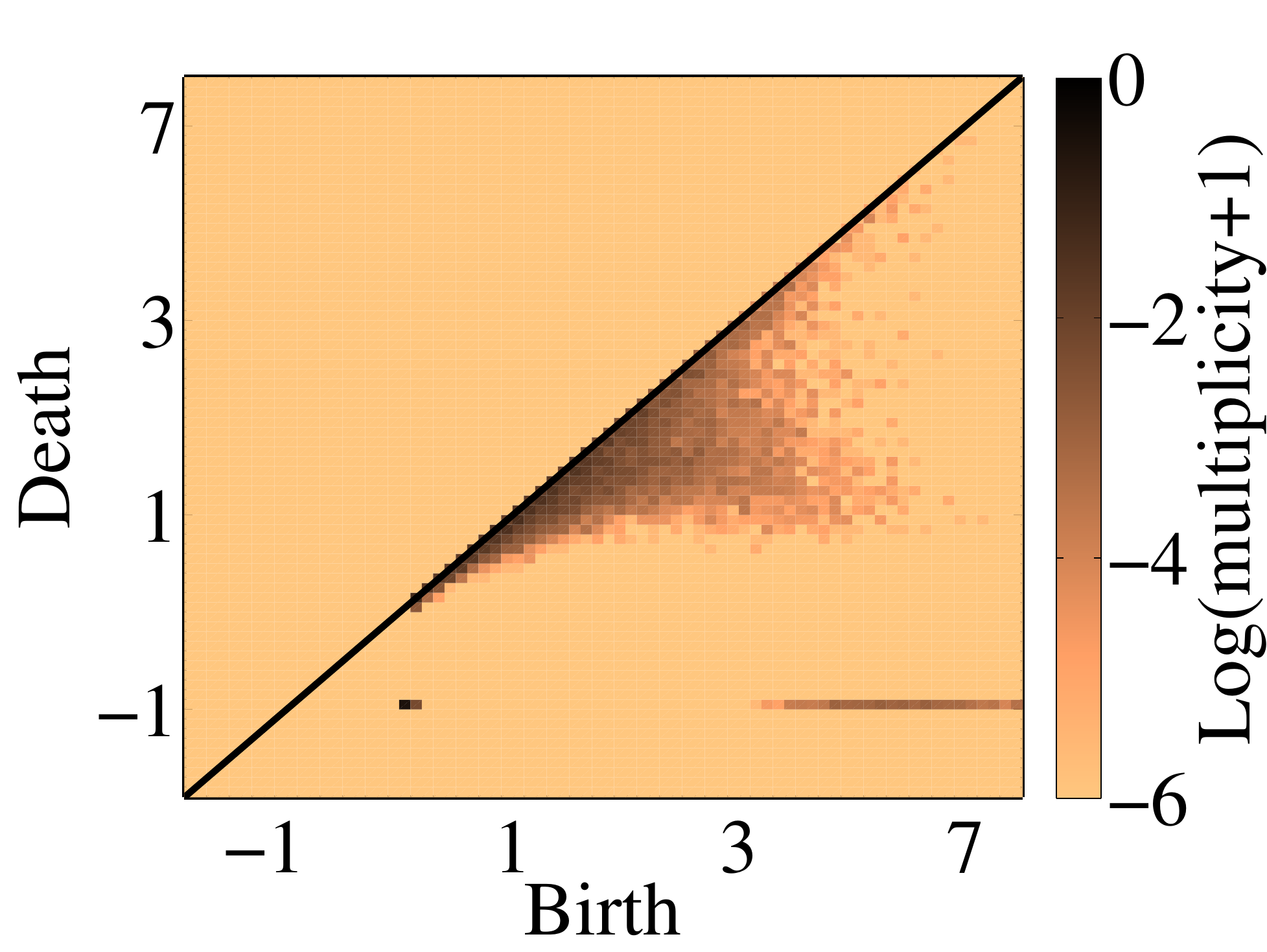}} 
\subfigure[ $\pd_1$ pentagons.]{\includegraphics[width=1.6in]{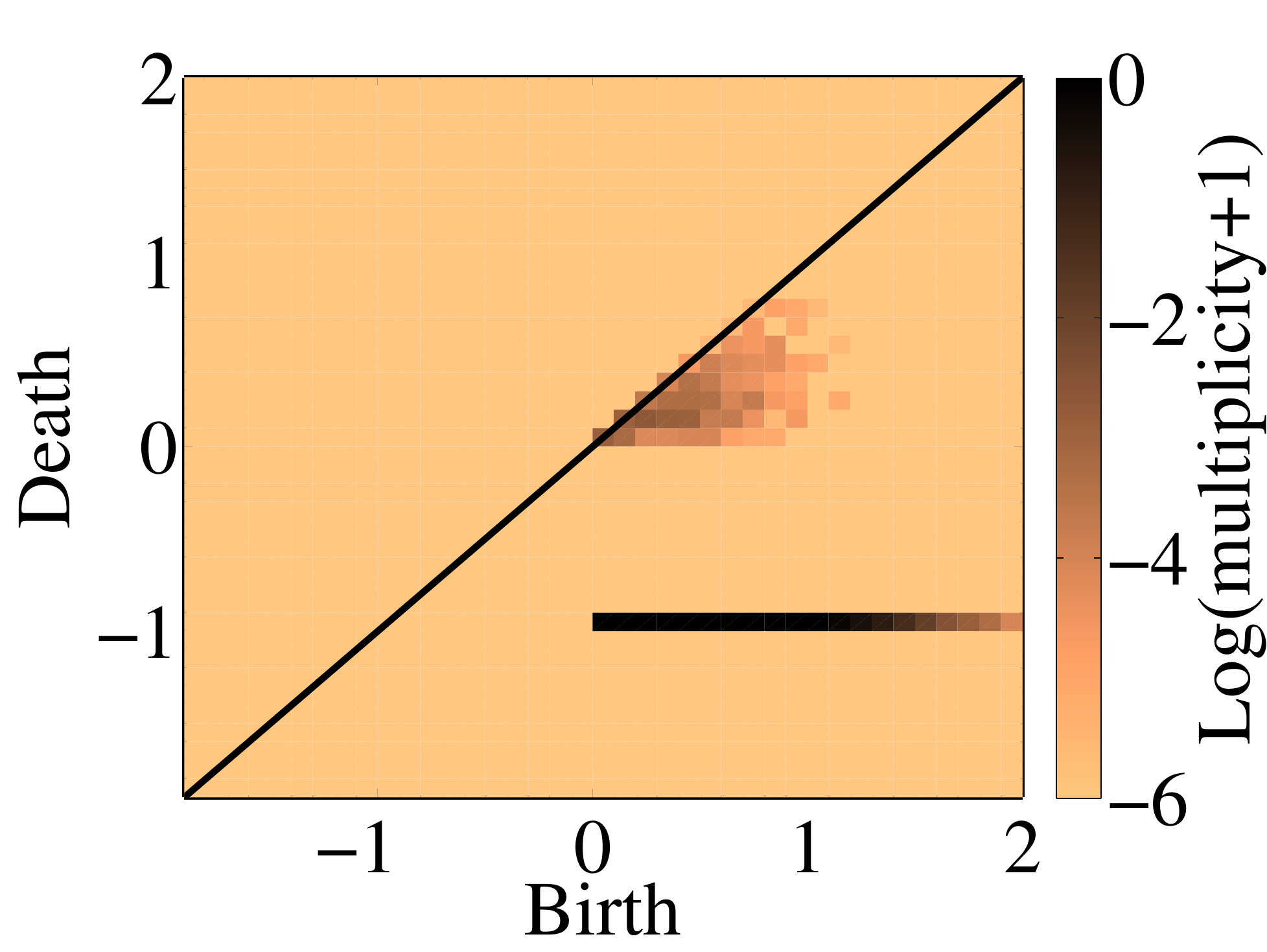}} 
\caption{Superimposed ${\pd}$s for the normal force networks of disks and pentagons including superimposed points from  $500$ realizations
(bottom slice, low tapping). }
\label{fig:diagrams_all}
\end{figure}

As is made clear in Section~\ref{sec:results}, even within a single system there can be considerable variability in the 
force networks from tap to tap.   Unfortunately, the concept of an average persistence diagram is not yet 
well defined~\cite{0266-5611-27-12-124007, munch2015, JMLR:v16:bubenik15a}. 
However, using the above mentioned measures we avoid this difficulty by applying them to an aggregate 
persistence diagram obtained by considering, as is done in Fig.~\ref{fig:diagrams_all}, all persistence points from the 
$500$ simulated taps on a single diagram.  The distributions of birth times presented in Section~\ref{sec:results} are based on these diagrams.
Note that for the remainder of the paper the distributions of birth times and lifespans are {\it normalized by the number of particles} in the 
domain used to define the force network under consideration.

We note that the $\pd$s provide a compressed and simplified description of the underlying force network landscape.
Thus, some information, such as size, shape and position of the components or loops, is discarded while passing from a force network to the corresponding $\pd$.  
Therefore, different force networks, may produce the same $\pd$s.
However, as is discussed in Sec.~\ref{sec:distance}, there are metrics that can be imposed on the space of all persistence diagrams such that if 
two force network landscapes are similar, then their associated persistence diagrams are similar. On the other hand, if the diagrams differ considerably, then so do the corresponding force networks.
Hence, in summary, persistent homology provides a continuous reduction of information that captures geometric information.

It is worth mentioning that persistence computations extend to higher dimensions.  
In particular, for $3$D, the diagram $\pd_0$ describes the structure of the connected components, as in $2$D. 
The features in $\pd_1$ are interpreted as tunnels rather than loops. 
Finally, there is an additional diagram $\pd_2$ that describes the structure of cavities. 

The computational codes used to construct the force networks and persistence diagrams are available at~\cite{miro,perseus}, respectively.
There are also  other publicly available  packages for computing persistent homology \cite{bauer2014phat}.

\subsection{Distance between persistence diagrams}
\label{sec:distance}

In the previous section, we introduced specific descriptors of persistence diagrams based on a single feature of the points, e.g. lifespan or birth value.
In this section we discuss metrics on the space of persistence diagrams that are based on the entire diagram, i.e.\ 
we compare two diagrams by comparing all points in each diagram.   Note that this comparison does not involve force thresholding: this measure
compares the force networks at all force levels.

Consider two persistence points $p_0=(b_0,d_0)$ and $p_1=(b_1,d_1)$. 
The distance between $p_0$ and $p_1$ is defined by
\[
\|(b_0,d_0)-(b_1,d_1)\|_\infty := \max\setof{|b_0-b_1| , |d_0-d_1|}.
\]
Now, given two persistence diagrams $\pd$ and $\pd'$  let $\gamma\colon \pd \to\pd'$ be a bijection between points in the two persistence diagrams where we are allowed to match points of one diagram with points on the diagonal of the other diagram. 
The {\em degree-$q$ Wasserstein distance}, {$d_{W^q}(\pd,\pd')$, is obtained by considering for each bijection, $\gamma$, the quantity
\[
\left(\sum_{p\in\pd} \| p -\gamma(p)\|^q_\infty \right)^{1/q}
\]
and defining the distance between $\pd$ and $\pd'$ to be the minimum value of this quantity over all possible bijections.
Stated formally
\[
d_{W^q}(\pd,\pd') = \inf_{\gamma\colon \pd \to \pd'} \left(\sum_{p\in\pd} \| p -\gamma(p)\|^q_\infty \right)^{1/q}. 
\] 
The {\em bottleneck distance} $d_B(\pd,\pd')$ is given by
 \[
d_B(\pd,\pd') = \inf_{\gamma\colon \pd \to \pd'} \sup_{p\in\pd} \| p -\gamma(p)\|_\infty.
\]

The cost of `moving the points'  (i.e., selecting a given bijection) varies for different distances. The bottleneck distance captures only the largest move  corresponding to the largest difference between the diagrams. On the other hand, the  Wasserstein distance $d_{W^1}$  sums up all the differences with equal weight. If all the points in one diagram are close to the points in the other diagram (or close to the  diagonal), then  the bottleneck distance is small. However, $d_{W^1}$ tends to be large since it is a sum over a large number of small differences. In the case that $d_{W^1}$ is close to $d_B$  the diagrams tend to differ in a small number of points.  There is the following relation between the distances: $d_B \leq d_{W^p} \leq d_{W^1}$ for $p> 1$ and $d_{W^p}$ converges to $d_B$ as $p$ goes to infinity. Hence, using the $d_{W^2}$ distance keeps track of all the changes but the small differences contribute less. Comparing the $d_B, d_{W^1} $ and the $d_{W^2}$ distances allows to better understand the difference between the diagrams. For example if $d_B$ and $d_{W^2}$ are similar but $d_{W^1}$ is large, then there is only one dominant difference between the diagrams and a large number of small differences. 

To compare a large number of $\pd$s representing the steady states of the tapped systems, we use the distance matrix (heat map) $D$. The entry $D(i,j)$ is the distance between the diagrams corresponding to the taps $i$ and $j$. Clearly, $D(i,i) = 0$ and if  the states are similar, then the value $D(i,j)$ is small. The distance matrix provides a detailed information about the differences between all the states. Sometimes a more condensed representation of the differences between the states  is desirable; 
for this purpose we will use  distribution of the values of $D$.

To illustrate that the distance matrices are sensitive to the structure of the force networks, and furthermore that they allow for a simple visual 
inspection of the similarities and differences both between the considered systems and between different realizations/taps for the same system, we provide
here an example.  We consider a system of disks exposed to high intensity tapping, record all the force information, and then randomly mix
up the forces.  This randomization is done after each tap by performing $1000$ force swaps (i.e., picking any two contacts at random and swapping the forces between them).  This procedure leaves the
PDFs of the force networks unchanged, however the force network may undergo a dramatic change, since one   
expects the force chains to be broken into a large number of short chains and therefore the differences between the states should be larger. 

Figure~\ref{fig:mixed} shows the $d_{W^1}$ distance matrix for the original and randomized system.   Considering the distances between $\pd_0$ 
diagrams first, we see that the distances between `real' realizations are small compared to the ones between randomized systems, and 
also that the distances between the original and randomized states are larger than the ones between the randomized states.  On the contrary, 
the variation in $\pd_1$ diagrams is smaller for the randomized systems. This is due to the fact that by randomly reassigning the forces, 
the loops with strong force interactions are destroyed and the points in $\pd_1$ tend to be close to the diagonal. 

We conclude that the distance matrices clearly identify  the differences between original simulation results and the randomized ones.   As we will
see in what follows, they can be also used to identify the differences between various simulation results considered in the present work.  

\begin{figure}
\centering
\subfigure[$\pd_0$ normal forces.]{\includegraphics[width=1.6in]{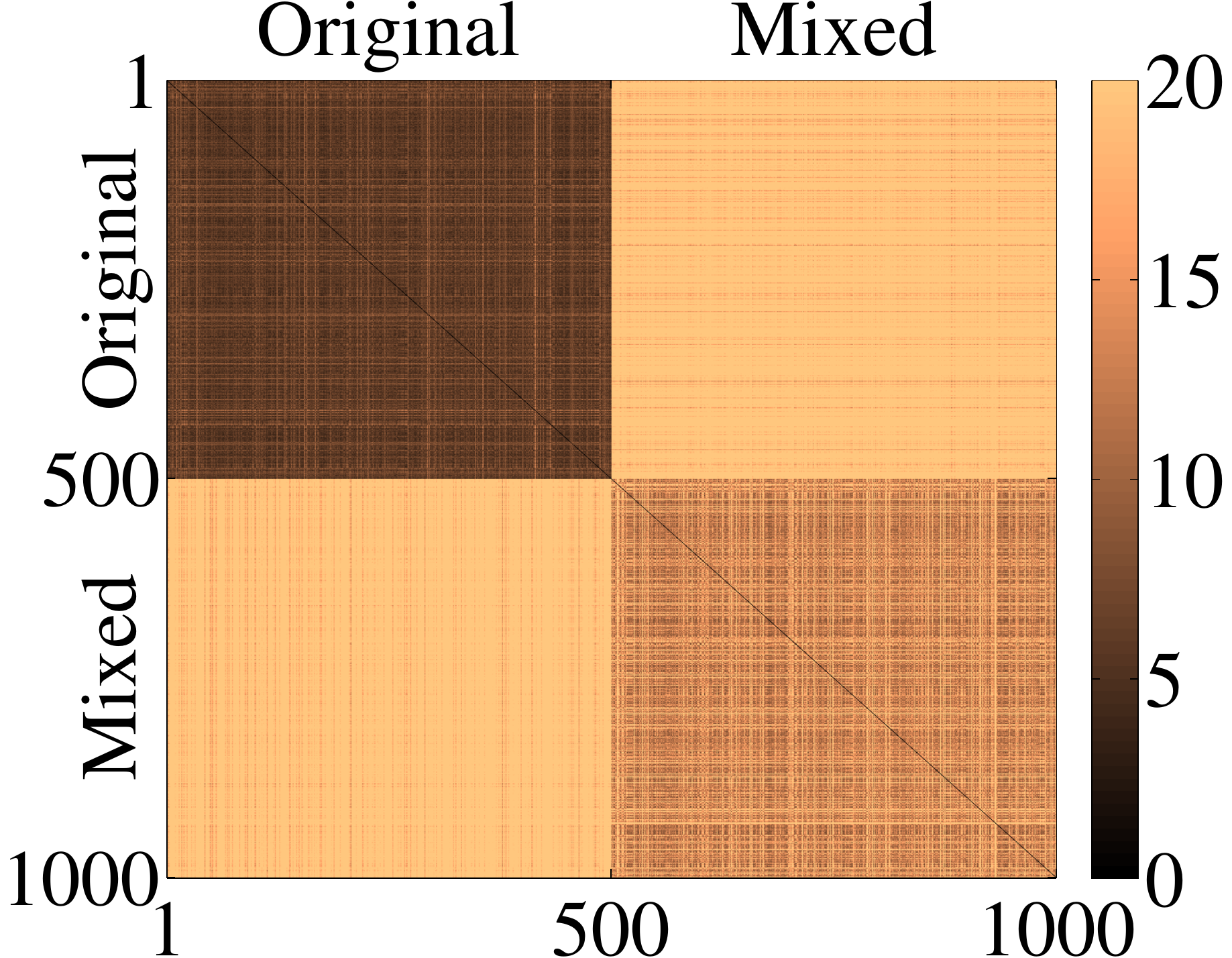}} 
\subfigure[$\pd_1$ normal forces.]{\includegraphics[width=1.6in]{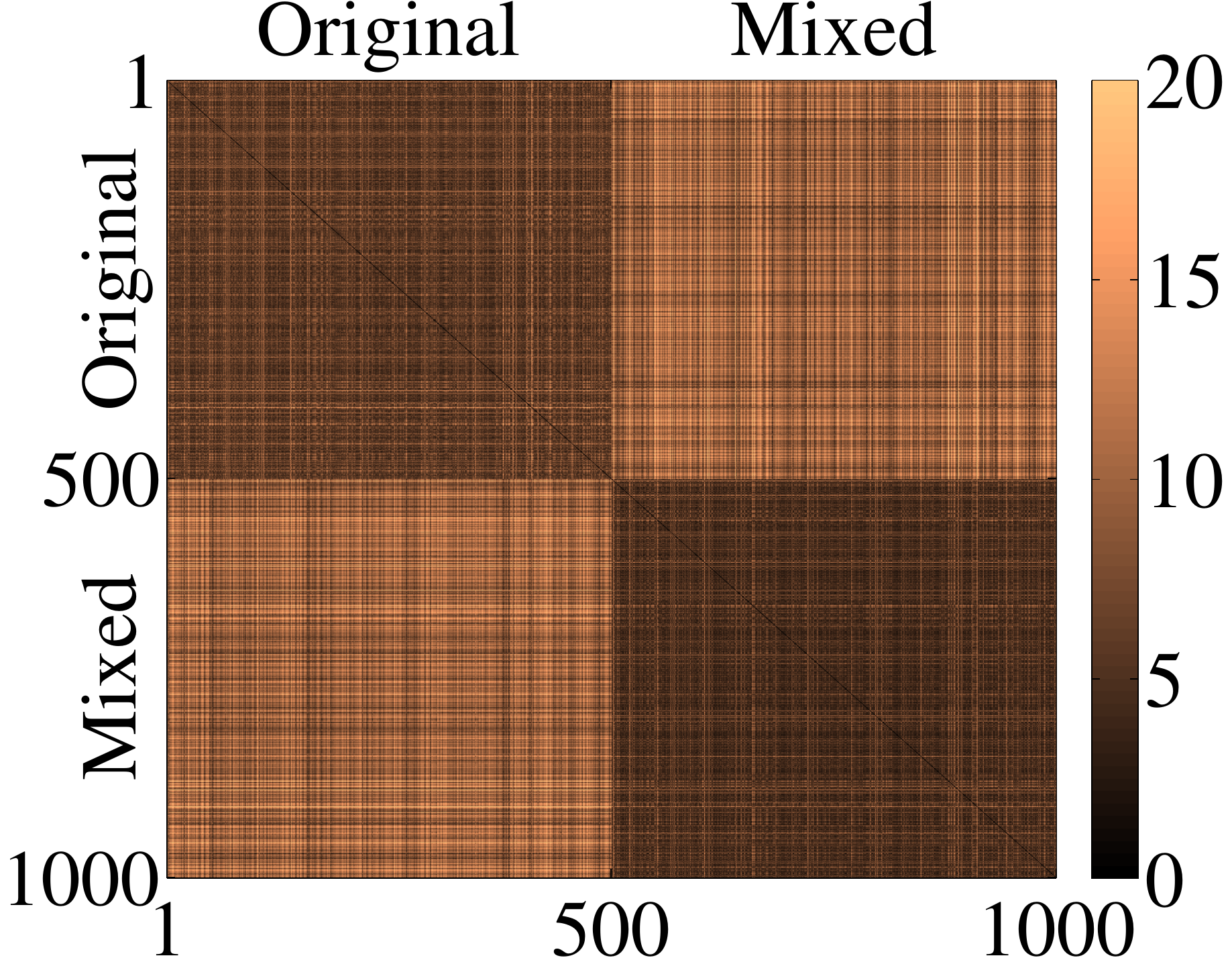}} 
\caption{Distance matrices showing $d_{W1}$ for randomly mixed-up forces between the particles (disks, bottom slice, high tapping).
The axes show the number of tap/realization, and the colors illustrate the value of the distance.
}
\label{fig:mixed}
\end{figure}

\section{Results}
\label{sec:results}

\subsection{Influence of gravitational compaction on force network properties }

Gravitational compaction and its influence  on force networks is  discussed in~\cite{paper1}, where it is shown that PDFs of the normalized 
forces do not depend on the depth in the sample. 
However, using $\beta_0$ to count  components shows differences: for both normal and tangential forces the number of 
components for the bottom slices  is considerably larger than  for the top ones.   
In this paper we compare top and bottom slices using the corresponding distance matrices and the descriptors 
discussed in the previous section.

\begin{figure}[thb]
\centering
\subfigure[$\pd_0$ normal forces.]{\includegraphics[width=1.6in]{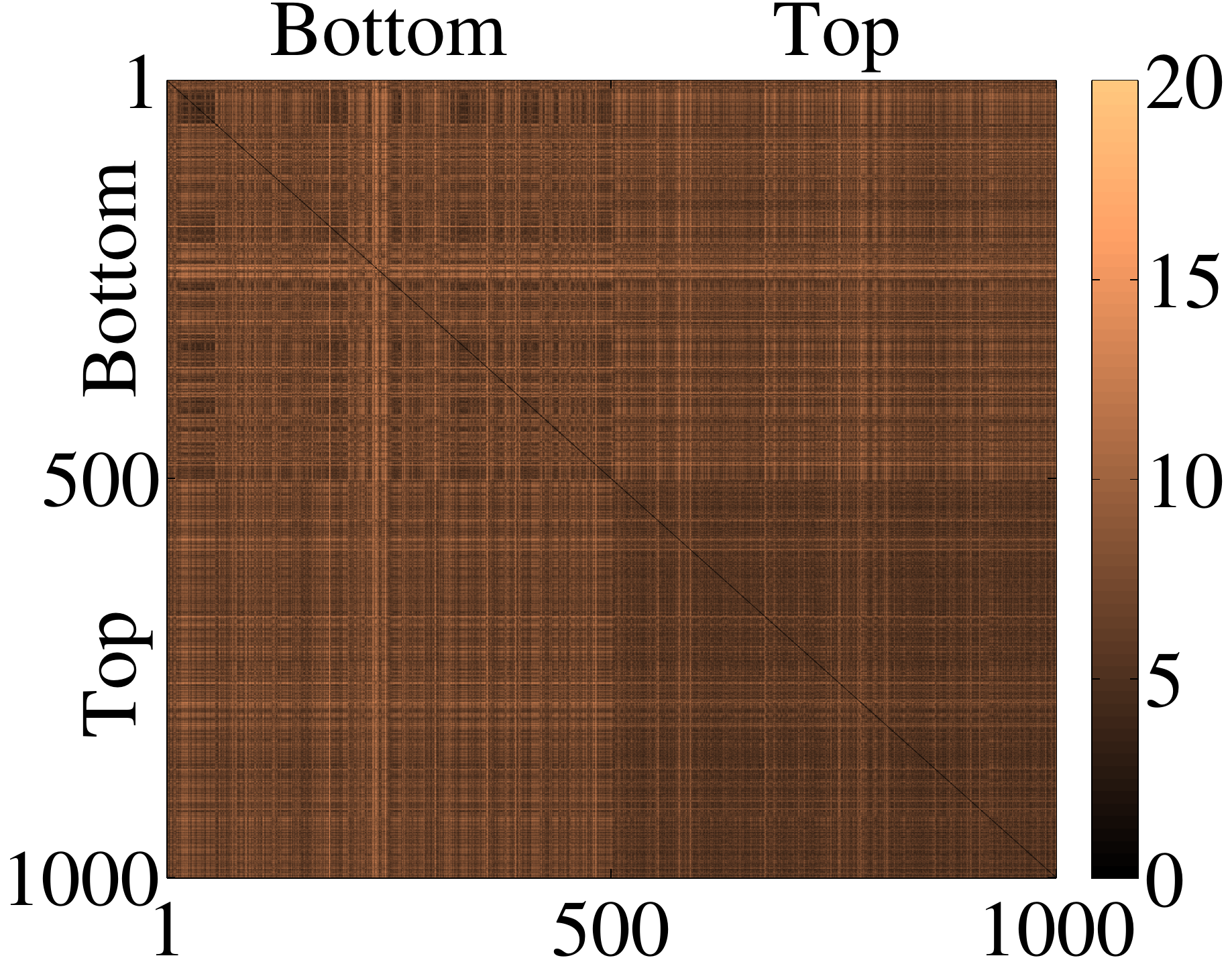}} 
\subfigure[$\pd_0$ tangential forces.]{\includegraphics[width = 1.6in]{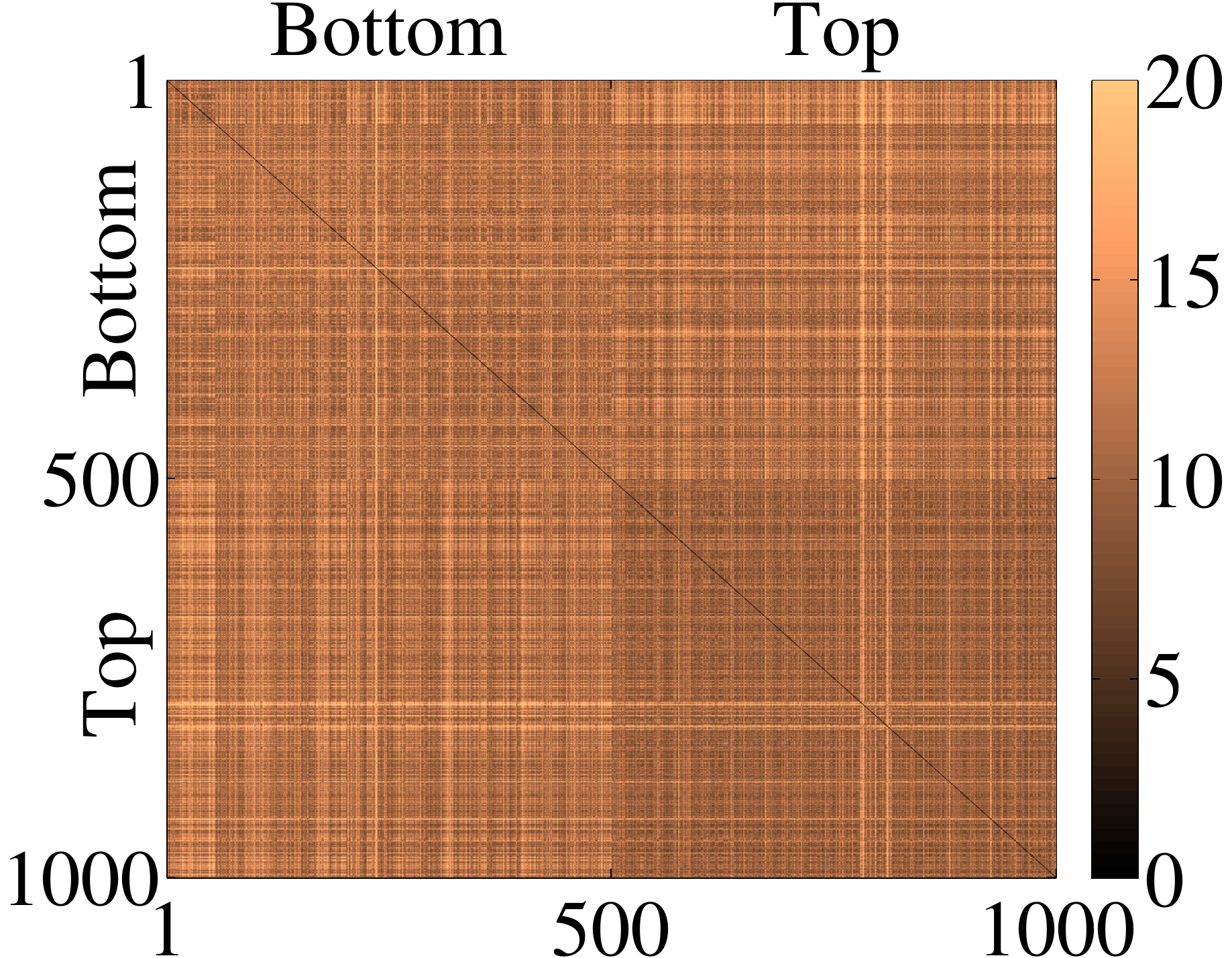}}\\
\subfigure[$\pd_1$ normal forces.]{\includegraphics[width=1.6in]{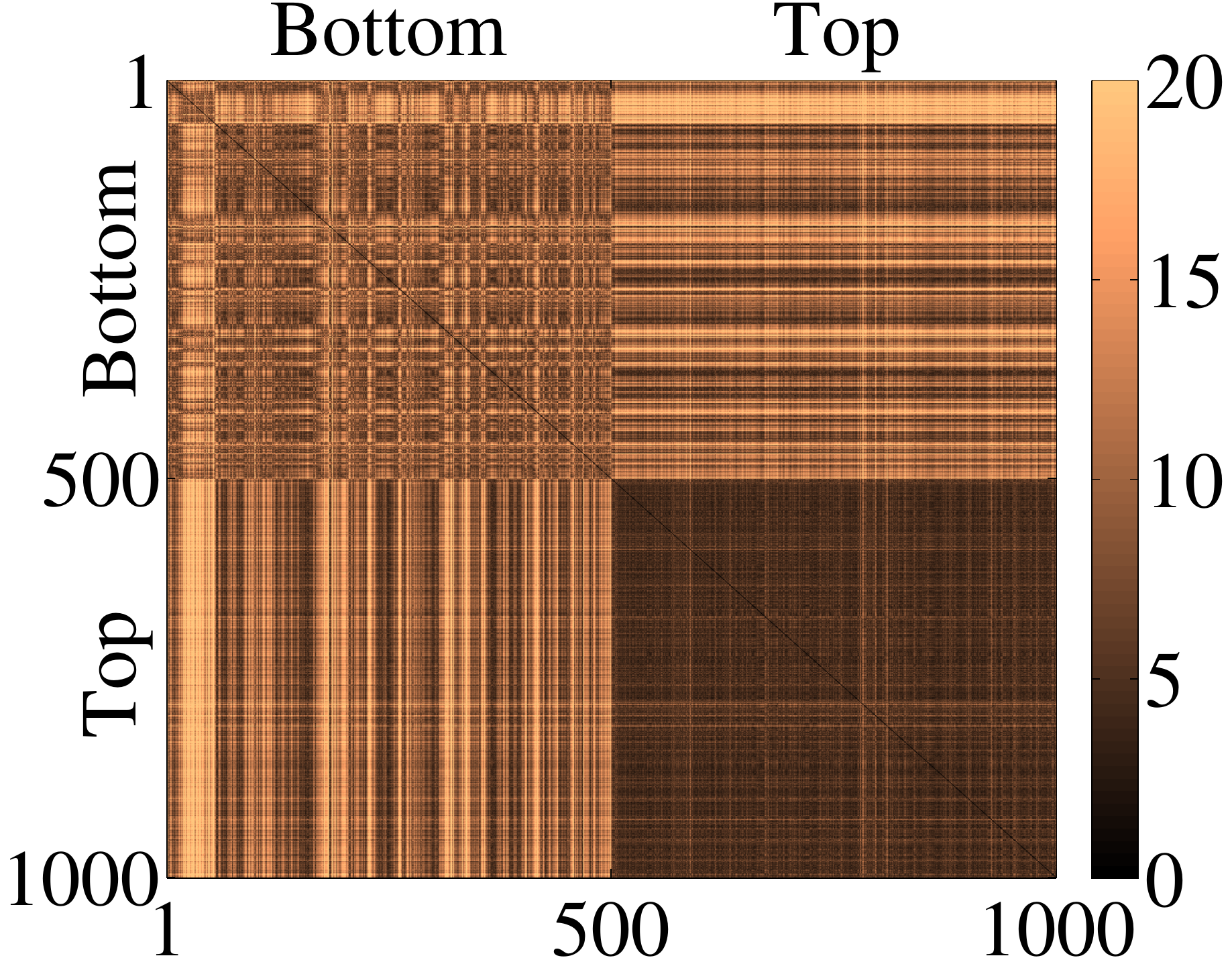}} 
\subfigure[$\pd_1$ tangential forces.]{\includegraphics[width = 1.6in]{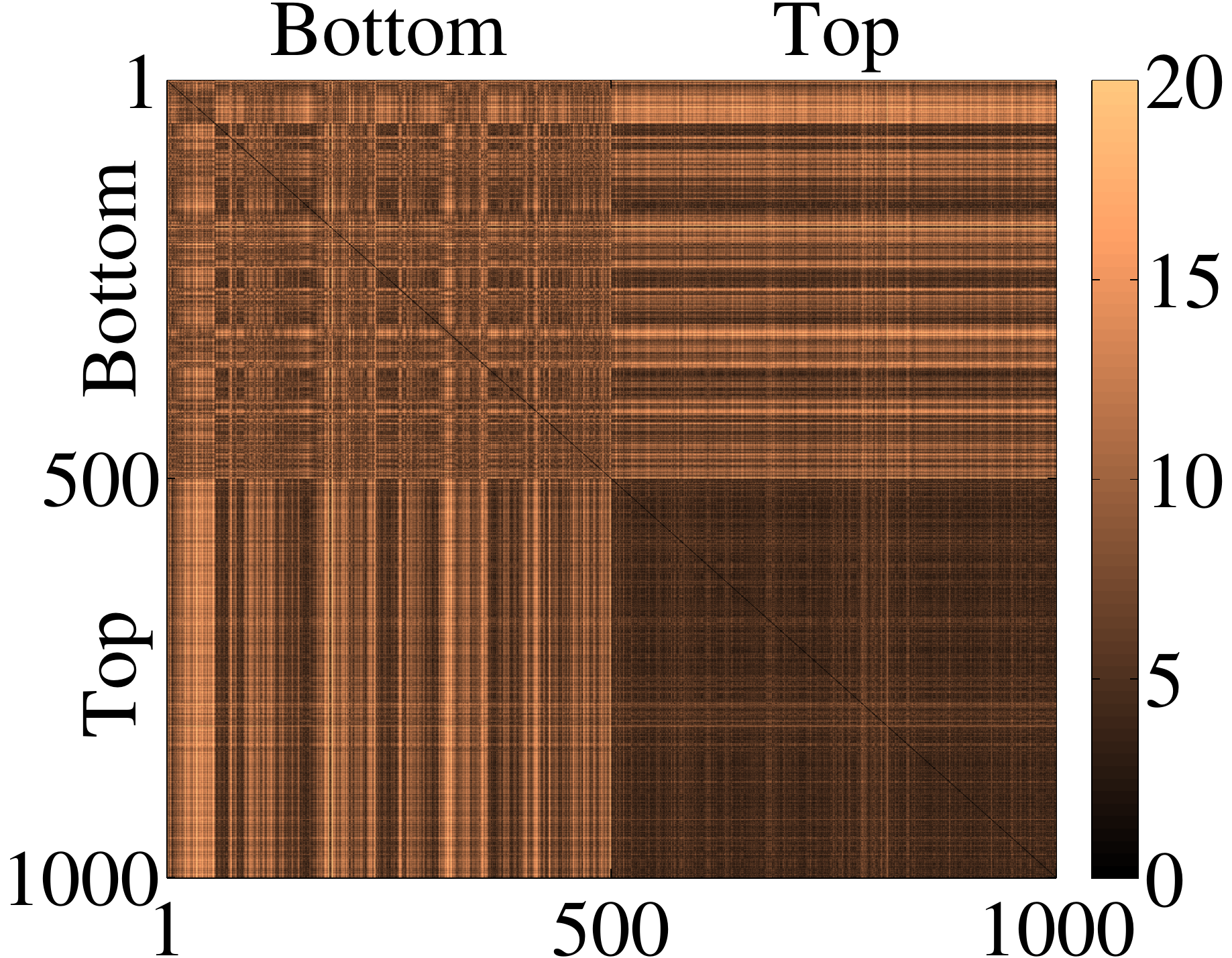}}
\caption{Distance matrices showing $d_{W1}$ (disks, low tapping). 
}
\label{fig:disks_low_heat}
\end{figure}

\begin{figure}[thb]
\centering
\subfigure[$\pd_0$ normal forces.]{\includegraphics[width=1.6in]{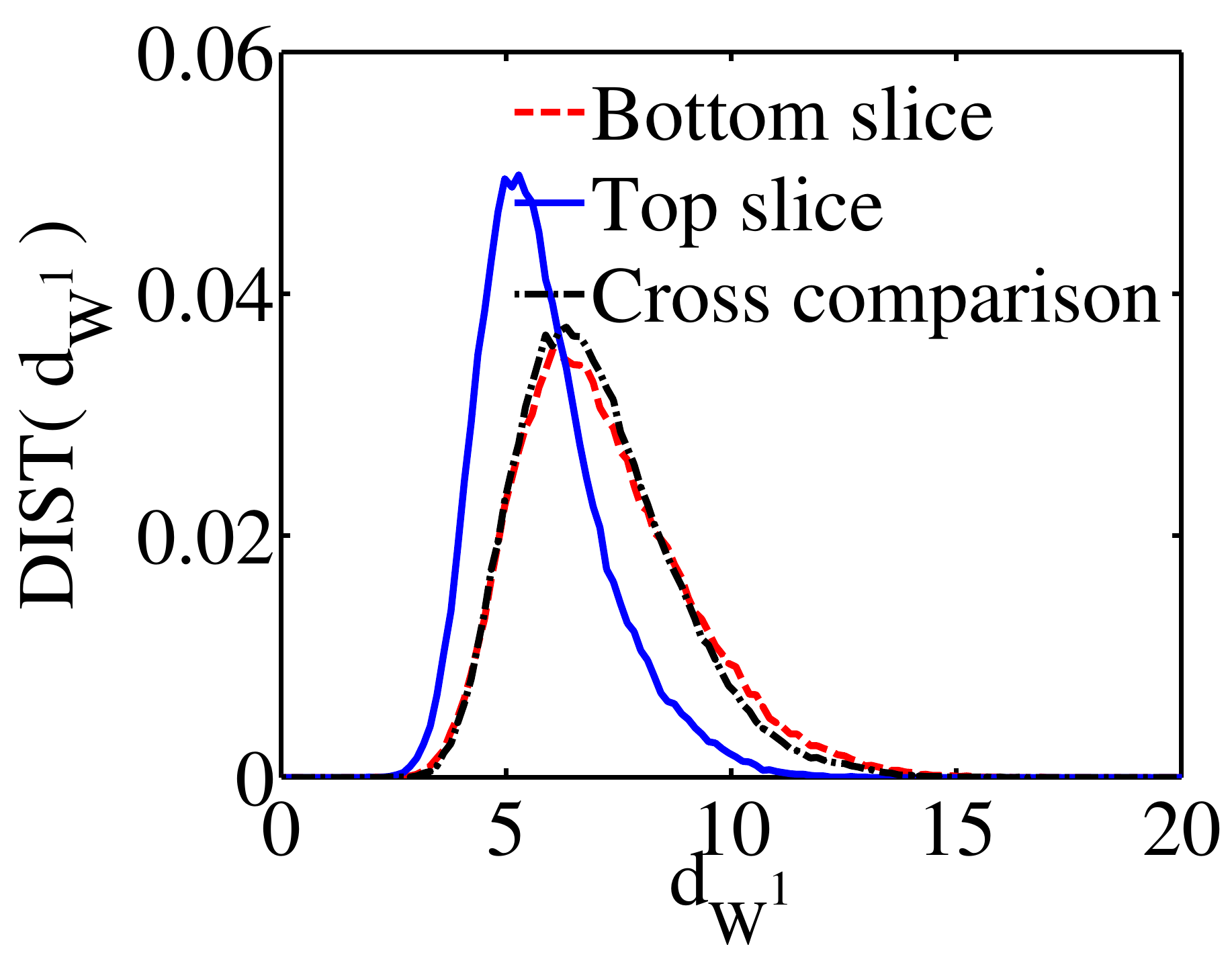}} 
\subfigure[$\pd_0$ tangential forces.]{\includegraphics[width = 1.6in]{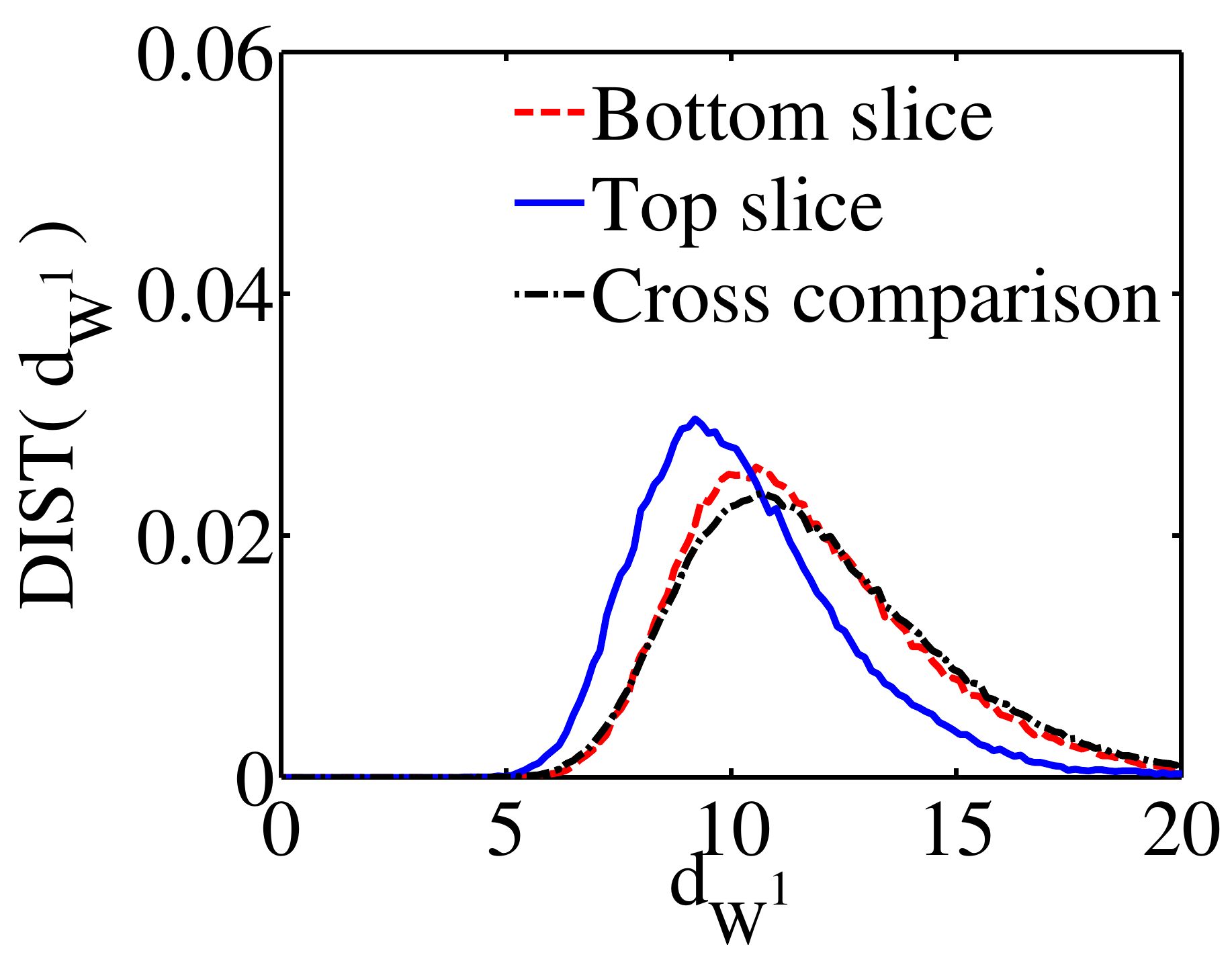}}\\
\subfigure[$\pd_1$ normal forces.]{\includegraphics[width=1.6in]{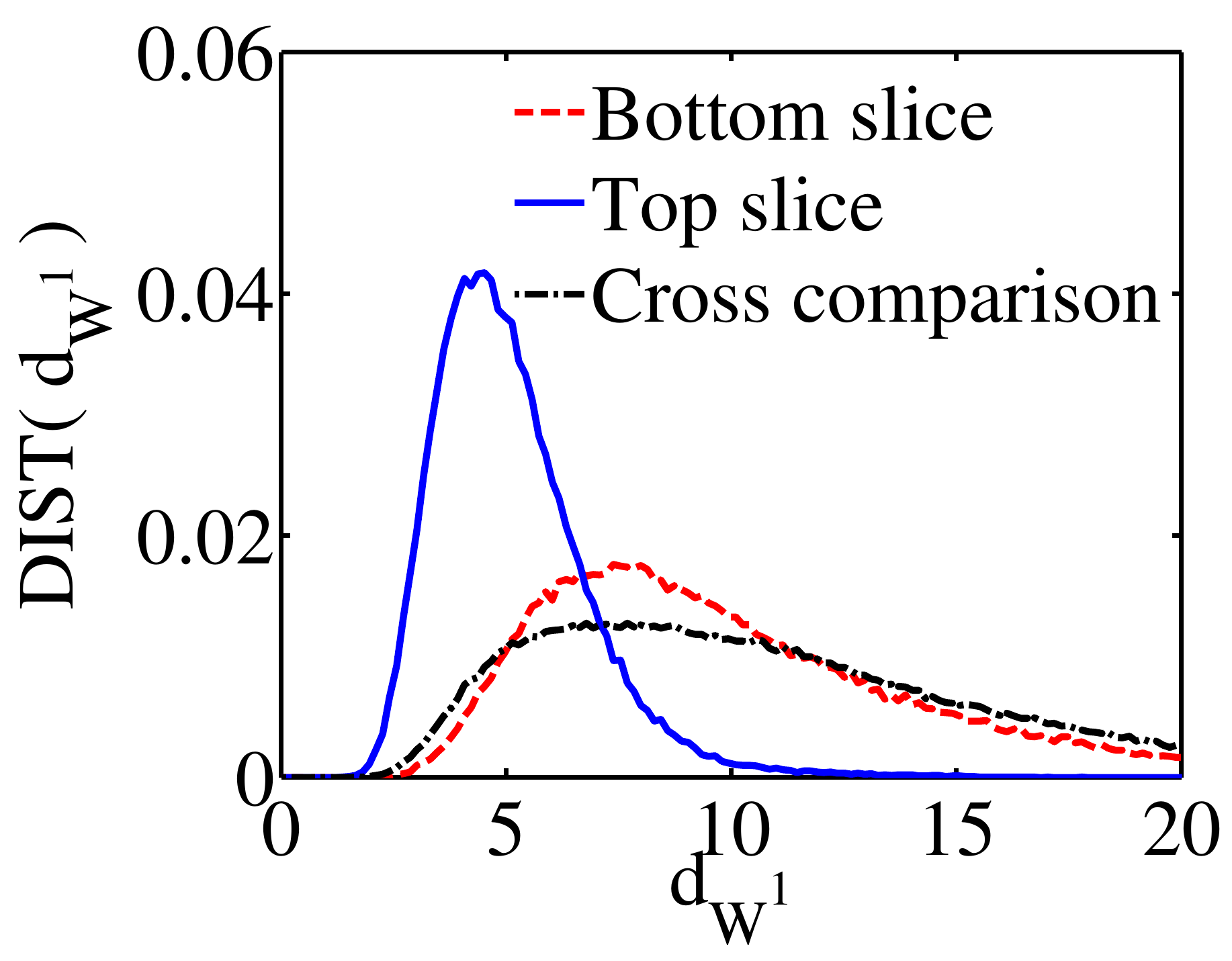}} 
\subfigure[$\pd_1$ tangential forces.]{\includegraphics[width = 1.6in]{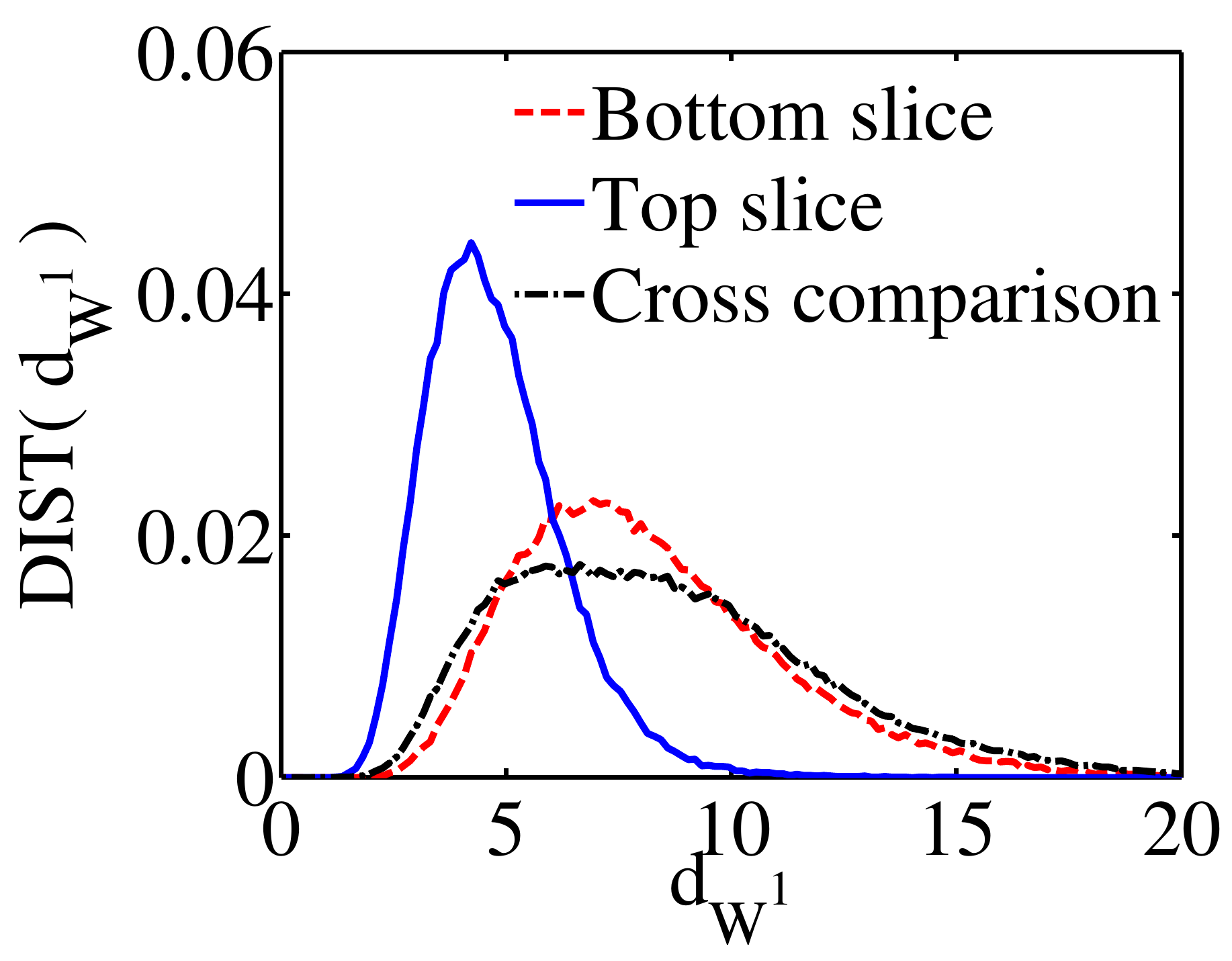}}
\caption{Distributions of $d_{W1}$ distance (disks, low tapping).  
}
\label{fig:disks_low_hist}
\end{figure}

\begin{figure}[thb]
\centering
\subfigure[$\pd_0$ normal forces.]{\includegraphics[width=1.6in]{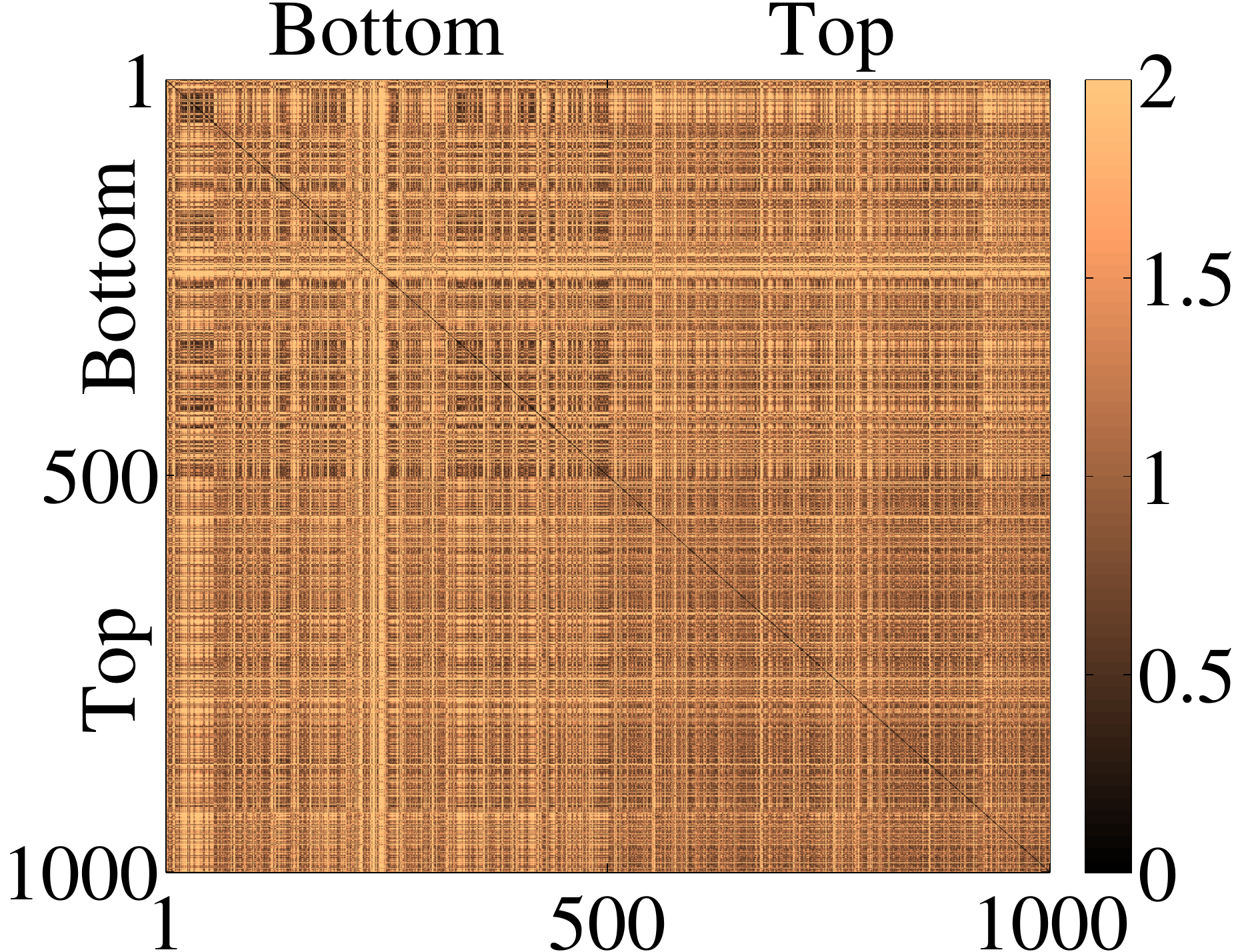}} 
\subfigure[$\pd_0 $ tangential forces.]{\includegraphics[width = 1.6in]{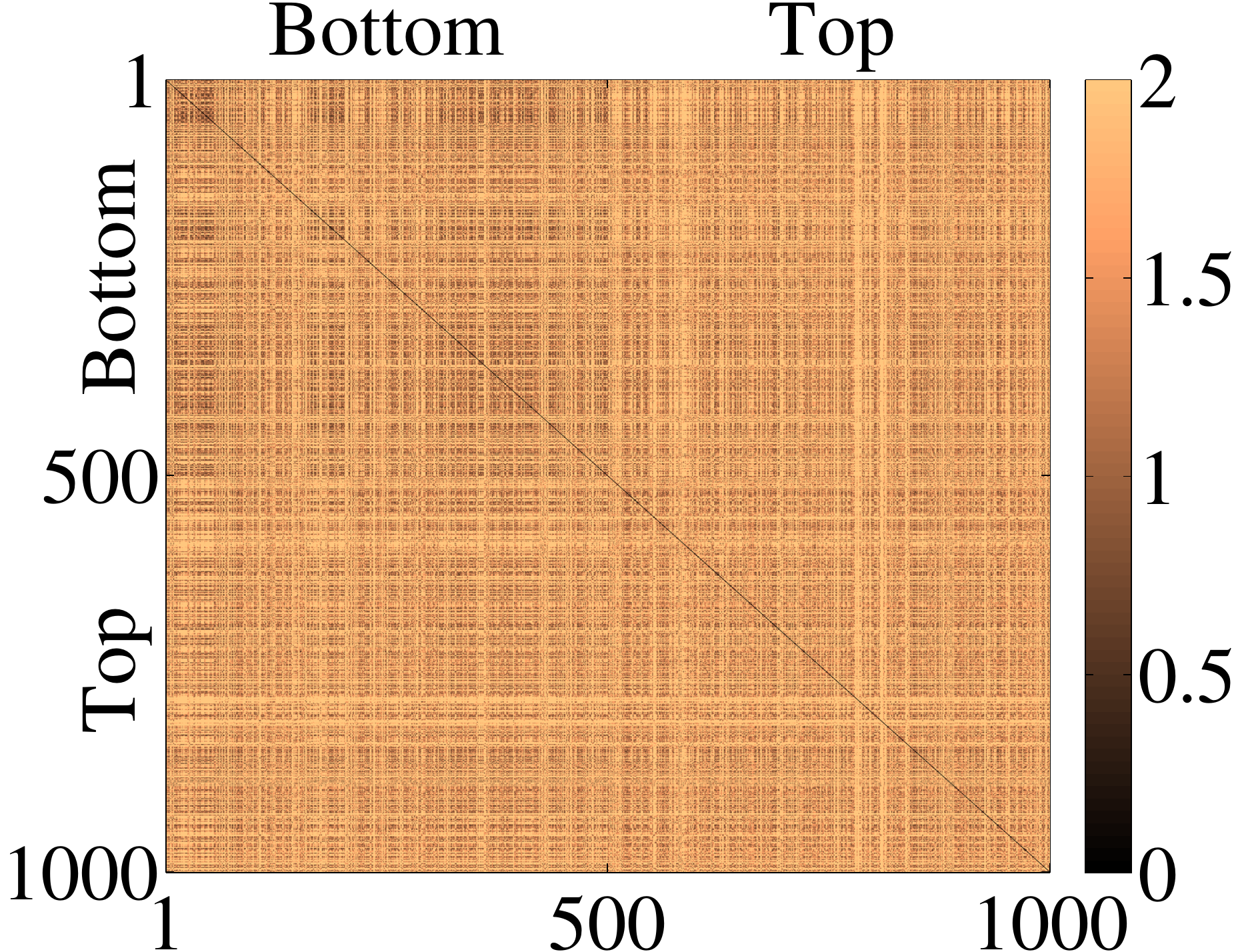}}\\
\subfigure[$\pd_1$ normal forces.]{\includegraphics[width=1.6in]{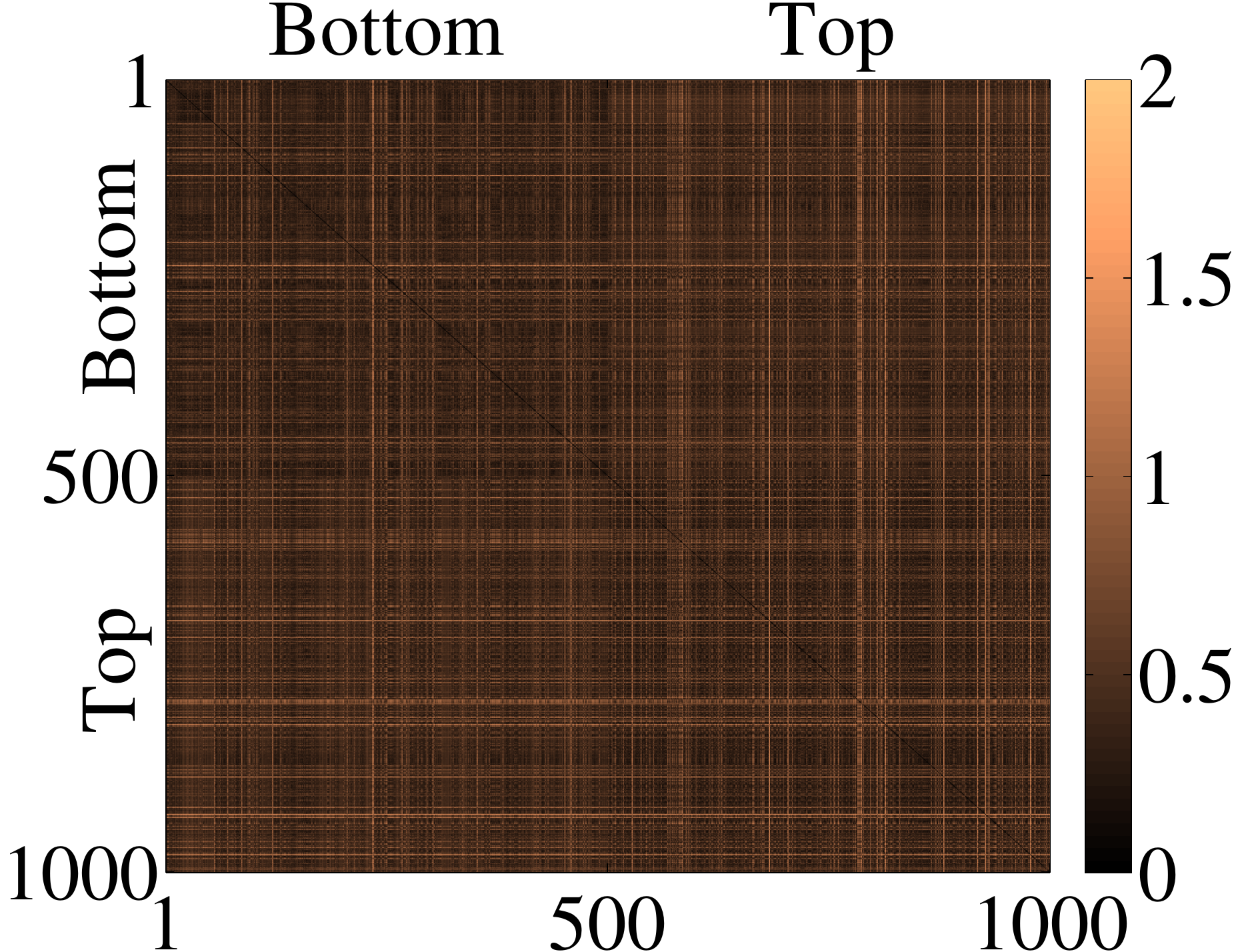}} 
\subfigure[$\pd_1$ tangential forces.]{\includegraphics[width = 1.6in]{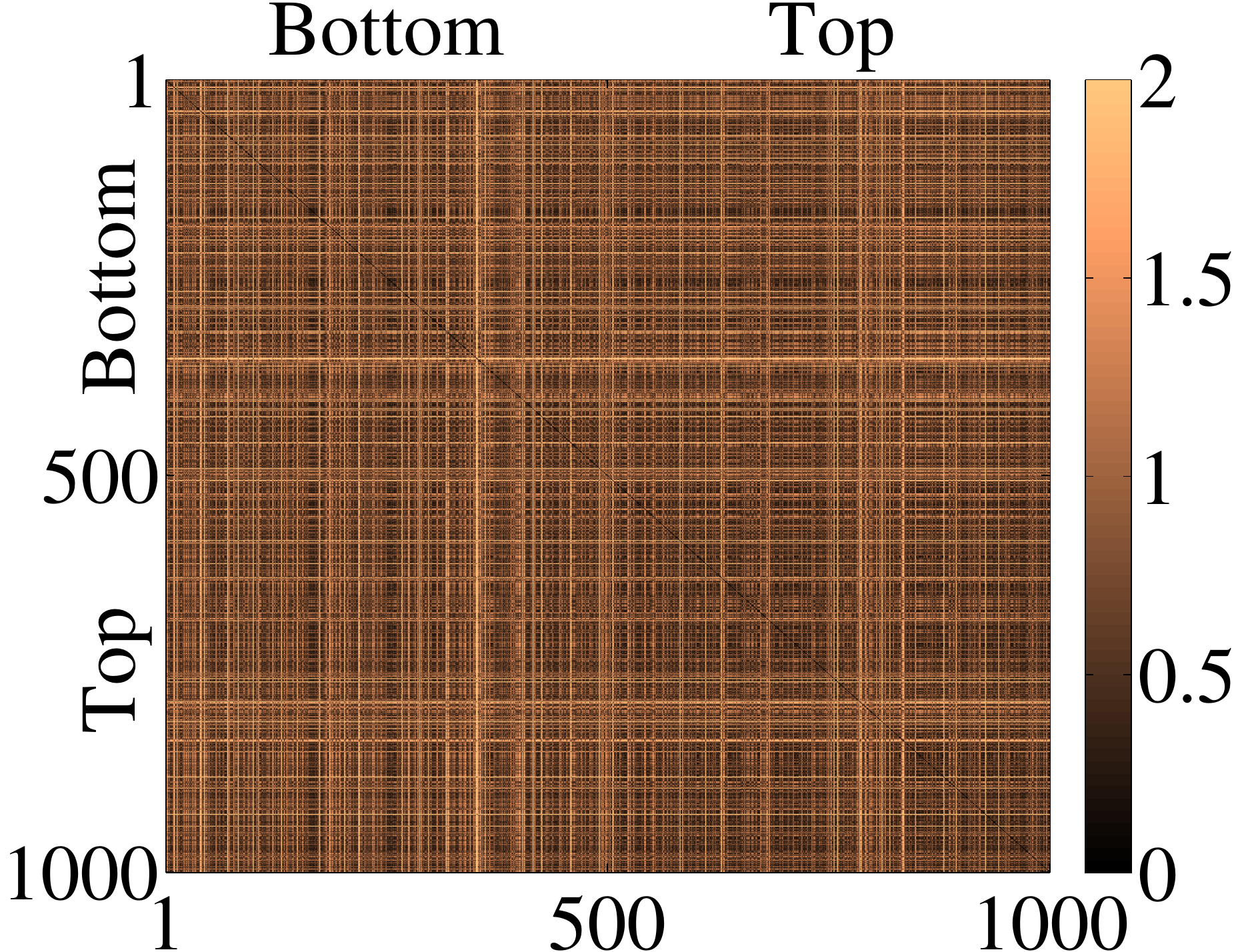}}
\caption{Distance matrices showing $d_{B}$ (disks, low tapping). 
}
\label{fig:disks_low_heat_B}
\end{figure}

Figures~\ref{fig:disks_low_heat} shows the $d_{W^1}$ distance matrices for the persistence diagrams of 
top and bottom slices of disks packings, respectively.    While distances between the components do not appear to 
vary significantly, three observations can be made regarding $\pd_1$ distances, shown in the parts (c) and (d):
(1) the distance between persistence diagrams of the top slice associated with different taps is relatively small (see lower right corners);  
(2) persistence diagrams of the bottom slice exhibit more variability (see upper left corners), and (3) the distances between 
persistence diagrams for different taps from the top slice and bottom slice range from small to large (see upper right corners).  
To reinforce these observations, we include Fig.~\ref{fig:disks_low_hist} that contains plots of the distribution of the distances. 
The distributions are computed only from the upper triangle of the distance matrix. Hence the diagonal (zero) entries are not included.
One immediate conclusion from  the observation (3) is that the geometry as measured by persistence diagrams for the force network 
observed after each tap for the top slice are not far away from those for the bottom slice, but lay on a small 'subdomain' of this 
second more scattered set. Hence, the cross comparison distances are dominated by the scattered distances of the bottom slice networks.

These distinctions are much less pronounced in Fig.~\ref{fig:disks_low_heat_B}, that shows $d_B$ distance.
This suggests that there is no single pronounced difference in the geometry of the force networks.   
In addition, $d_{W^2}$ (picture not shown for brevity), turns out to be quantitatively similar to the 
$d_B$ distance matrices.   Based on the discussion in Sec.~\ref{sec:distance}, this fact indicates that  
the force networks for the bottom and top slices are similar,  but the number of small variations in the birth and 
death of each feature (cluster or loop) from tap to tap are more prominent for the bottom slice.
In particular, the light and dark bands of Fig.~\ref{fig:disks_low_heat}(c) and (d) suggest the existence of further structure in our 
systems that we explore in the next section.

\subsection{Force networks in  the systems exposed to different tapping intensities}
\label{sec:differentTap}

The measures considered in~\cite{paper1} (force PDFs, Betti numbers) do not  identify differences in the properties of the force networks in the systems of disks exposed to different tapping intensities that lead to the same packing fraction. 
However, earlier work~\cite{arevalo_pre13}, based on different type of simulations where the side walls were frictional, which possibly influences the force network structure, suggested that some (rather difficult to observe) 
differences may exist.   
In this section we show that analysis based on persistent homology provide additional insight that allows to understand the origin of the differences, both when considering averaged results, and on the level of individual realizations and their variability.    

Careful inspection of the distance matrices for different distance definitions, normal and tangential
forces, components and loops, uncovers the following facts. First, we do not observe any appreciable differences between
the distances when components/clusters are considered (figures not shown for brevity).   Therefore, we expect that the distribution of 
components between the disk-based systems exposed to high and low tap intensities are similar (however
see below).    This conclusion  does not apply to the distances between $\pd_1$ diagrams.  
Figure~\ref{fig:disks_bottom_heat_hist}(a, b) shows the corresponding results, for $d_{W1}$ distance.   
There is a clear difference between high and low tapping regimes, suggesting that the structure of the loops is very different. 
Figure~\ref{fig:disks_bottom_heat_hist}(c,~d) shows the corresponding
distributions confirming a clear separation between the two cases considered.  In comparison to low tapping states, 
high tapping is characterized by force networks that are significantly 
more similar to each other, both for normal and tangential forces.   We emphasize that this difference is difficult to observe and
quantify by using any other measure we are aware of.    We also observe in Fig.~\ref{fig:disks_bottom_heat_hist}(c,~d) that the distances between
the realizations for low tapping are as large as the distances between low and high tapping regimes; we will discuss this finding in more detail in 
what follows.

\begin{figure}
\centering
\subfigure[$\pd_1$ normal forces.]{\includegraphics[width=1.6in]{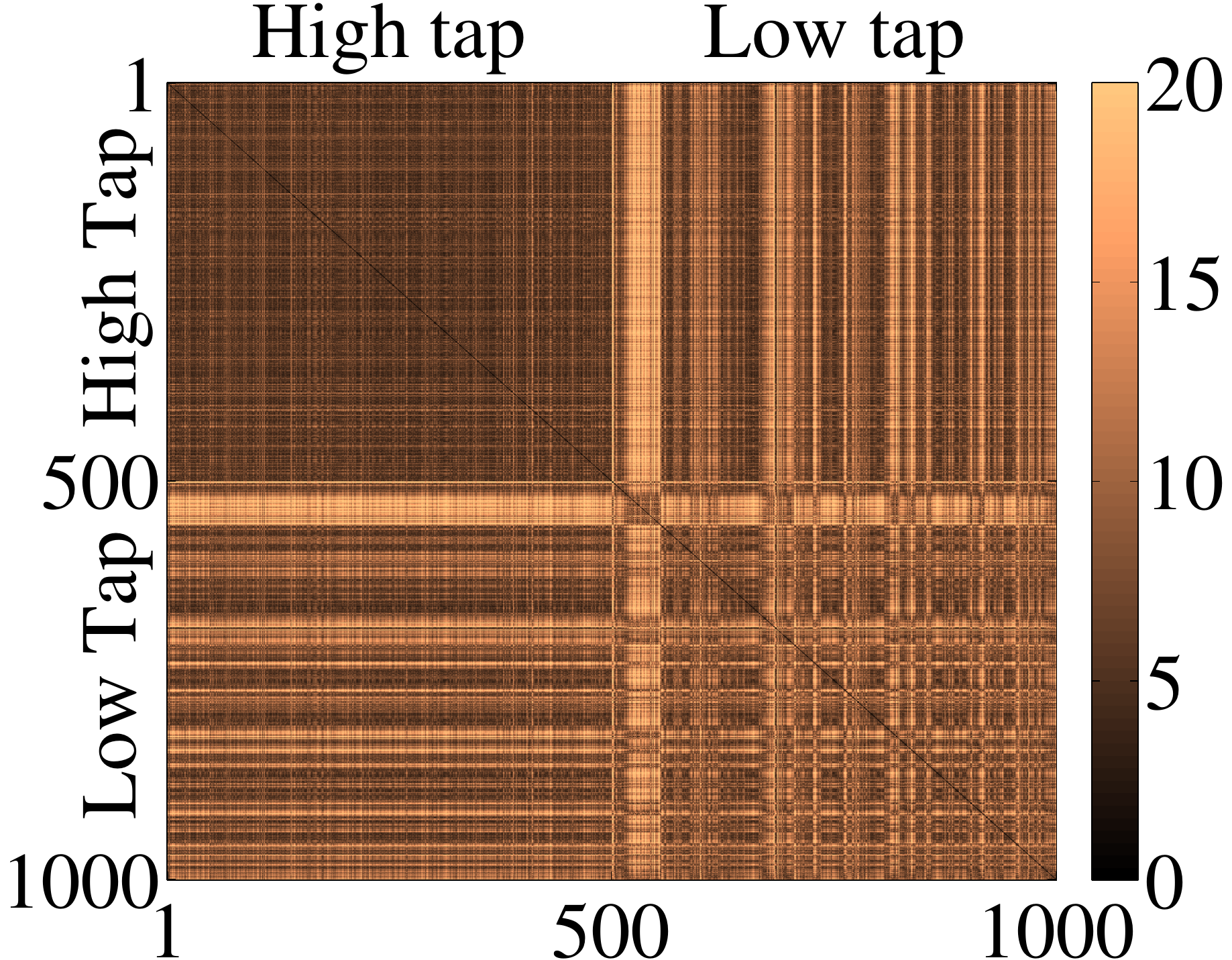}} 
\subfigure[$\pd_1$ tangential forces.]{\includegraphics[width=1.6in]{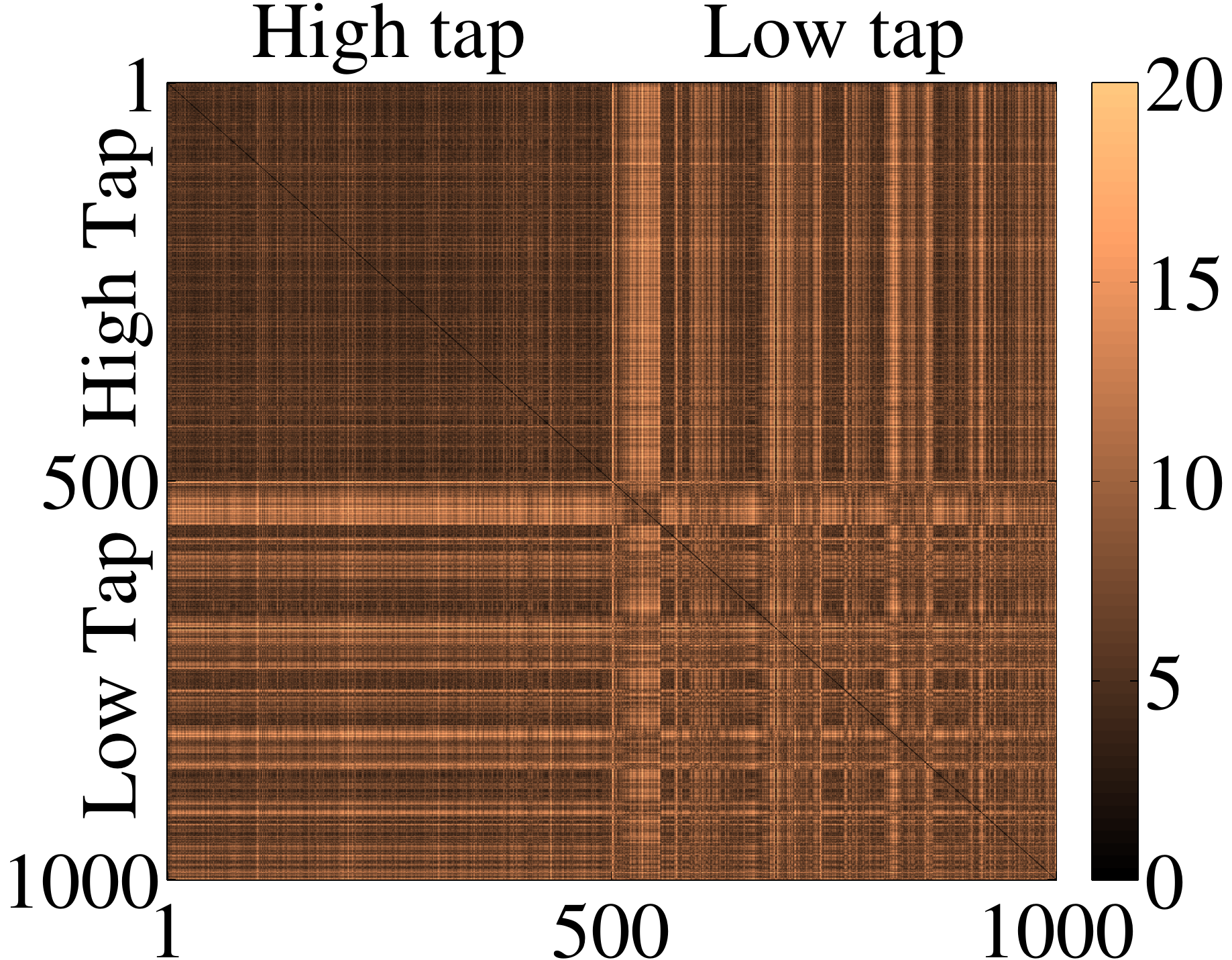}} \\
\subfigure[$\pd_1$ normal forces.]{\includegraphics[width=1.6in]{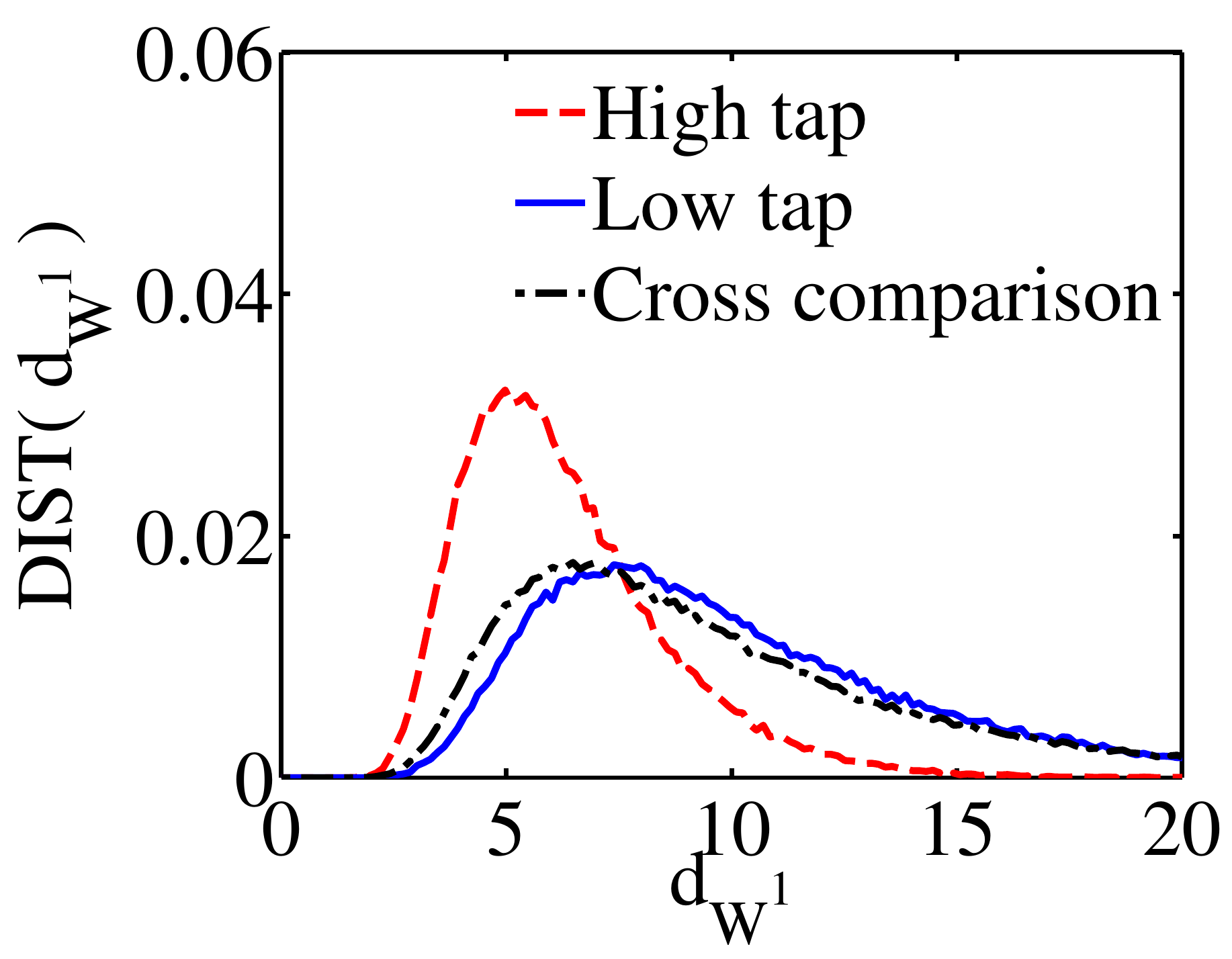}} 
\subfigure[$\pd_1$ tangential forces.]{\includegraphics[width = 1.6in]{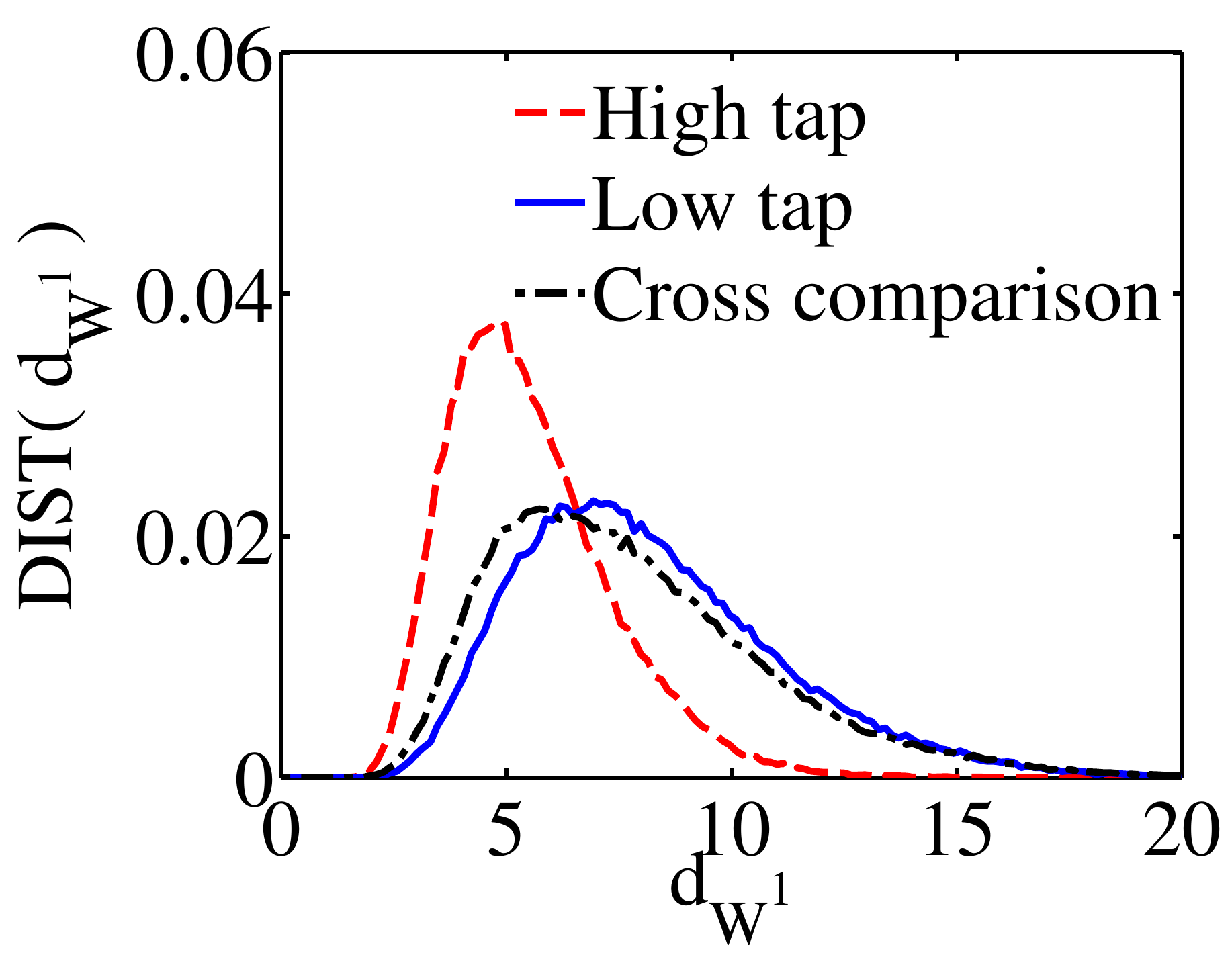}}
\caption{Distance matrices and distributions showing $d_{W1}$ (disks, bottom slice).
}
\label{fig:disks_bottom_heat_hist}
\end{figure}

We now proceed to answer the following two questions: (i) What is  the difference between the structure of the networks for high and low tapping? (ii) Figure~\ref{fig:disks_bottom_heat_hist}(a,~b) shows that there appears to be some structure in the evolution of the distances for $\pd_1$ persistence diagrams.   What is the origin of this structure?

To answer the first question, we now consider other measures that can be computed from ${\pd}$s, starting with birth times.   
Figure~\ref{fig:disks_birth} shows the distribution of birth coordinates for the considered $\pd_0$ and $\pd_1$ persistence
diagrams.   As a reminder, birth time indicates at what force threshold level the considered features (components, loops) appear.   
Perhaps surprisingly, we see that not only the structure of loops is different, but the structure of components differs as well: for low tapping
regime, there is a much more pronounced peak, particularly for the normal forces.    While consistent difference was seen when $\beta_0$
was considered (see Fig. 9 in~\cite{paper1}), the difference is more pronounced when considering birth times.

Figure~\ref{fig:disks_birth} shows that in general there are more features born at every force level for low tapping regime. 
In principle, it could happen that the lifespans of these extra features are relatively small and the systems only differ in minor fluctuations. 
Distributions of lifespans, shown in Fig.~\ref{fig:lifespansHiLow}, demonstrate that this is not the case.  The distributions of lifespans for low tapping regime are typically above the distributions for the high tapping regime. Therefore, even the number of prominent features is typically larger for the low tapping regime. However, there is an important exception. For normal forces the number of  components with lifespan larger than one is larger for the high tapping regime. The crossover of the $\pd_0$ distributions for the normal force around the lifespan equal  one leads to the fact that the corresponding  distributions of  total persistence, shown in Fig.~\ref{fig:totalpers-HiLow}(a), are centered at the same point.  As expected, the remaining distributions for the low tapping regime, shown in  Fig.~\ref{fig:totalpers-HiLow},  are shifted to the  left.

\begin{figure}
\centering
\subfigure[$\pd_0$ normal forces.]{\includegraphics[width=1.6in]{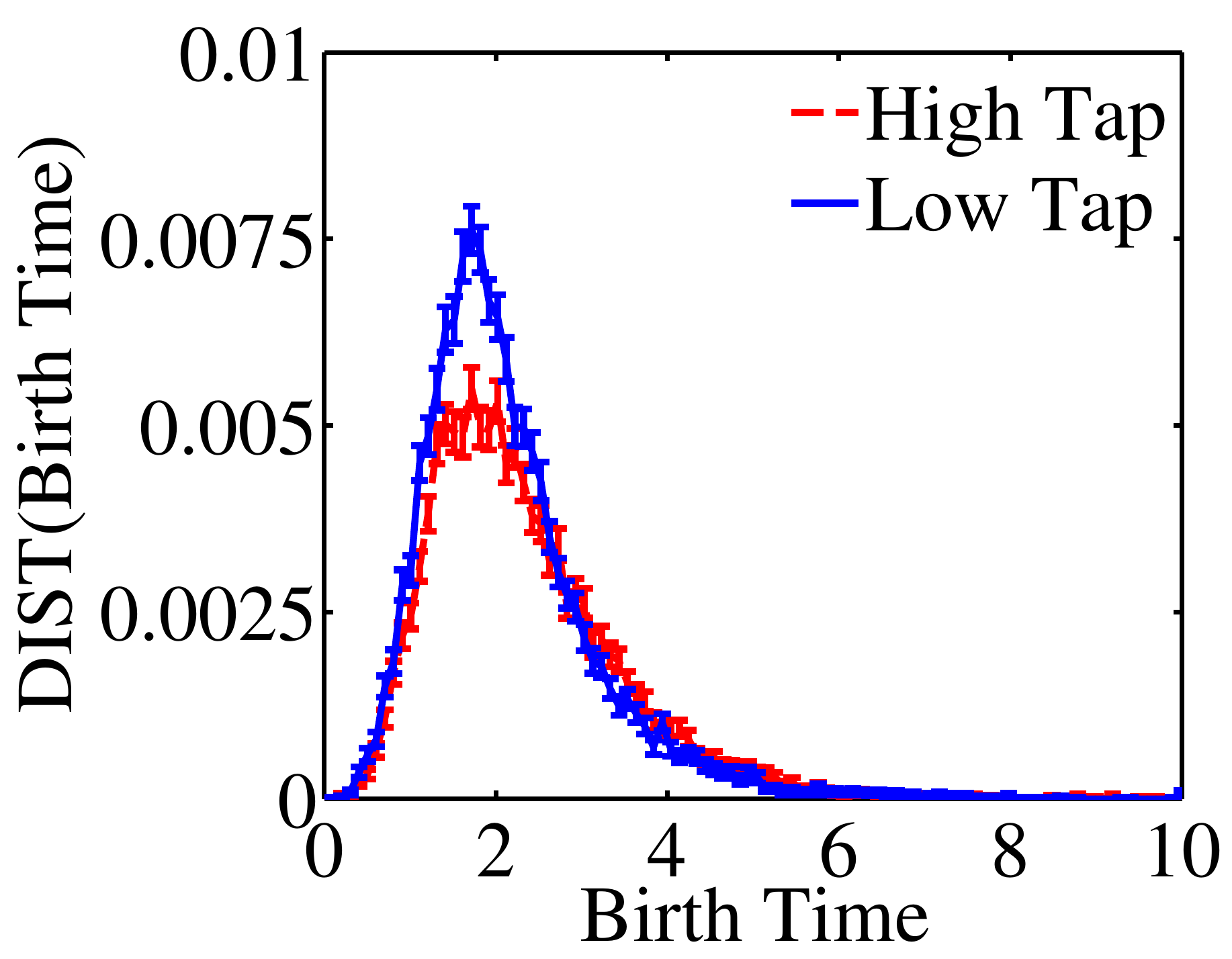}} 
\subfigure[$\pd_0$ tangential forces.]{\includegraphics[width=1.6in]{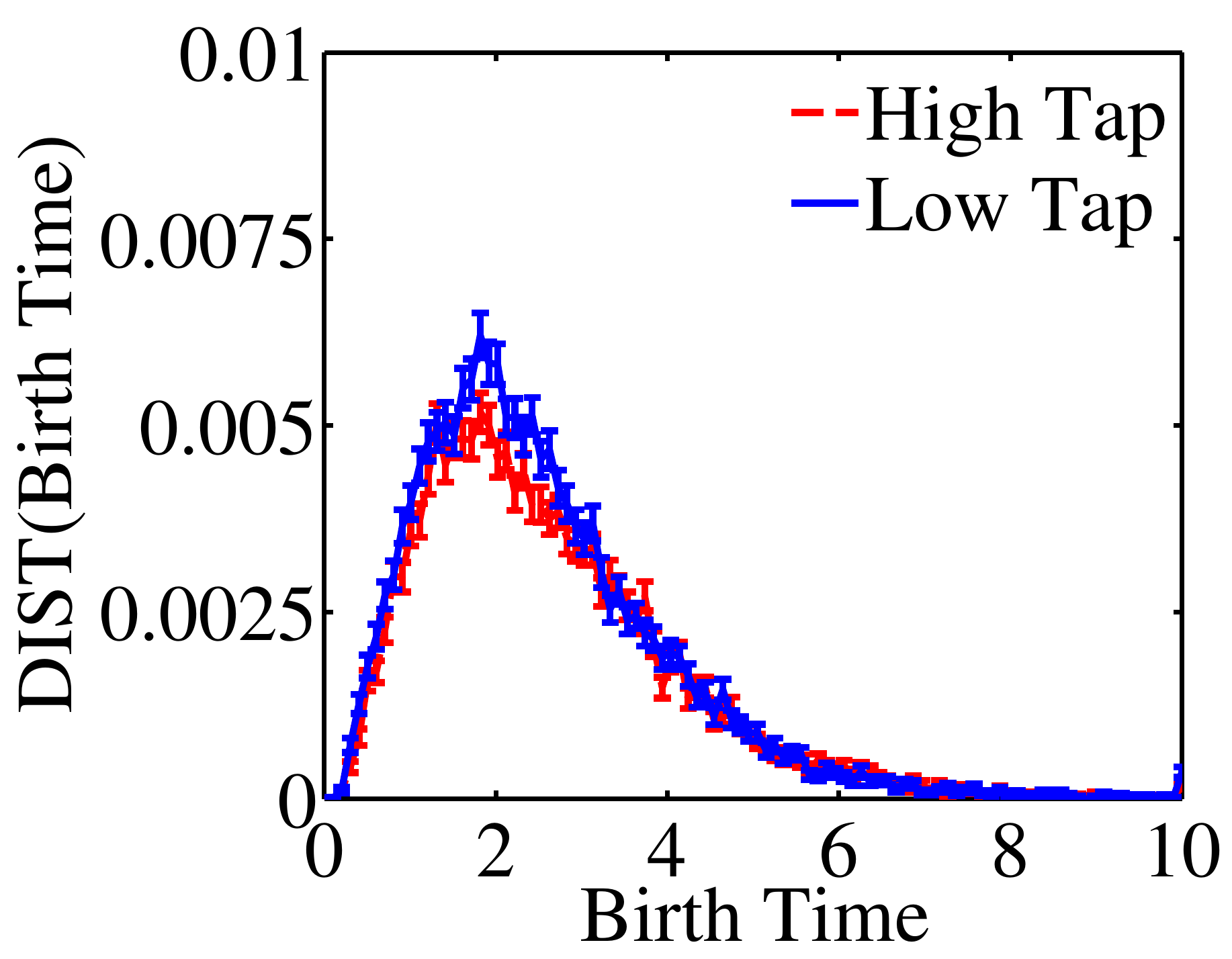}} 
\subfigure[$\pd_1$ normal forces.]{\includegraphics[width=1.6in]{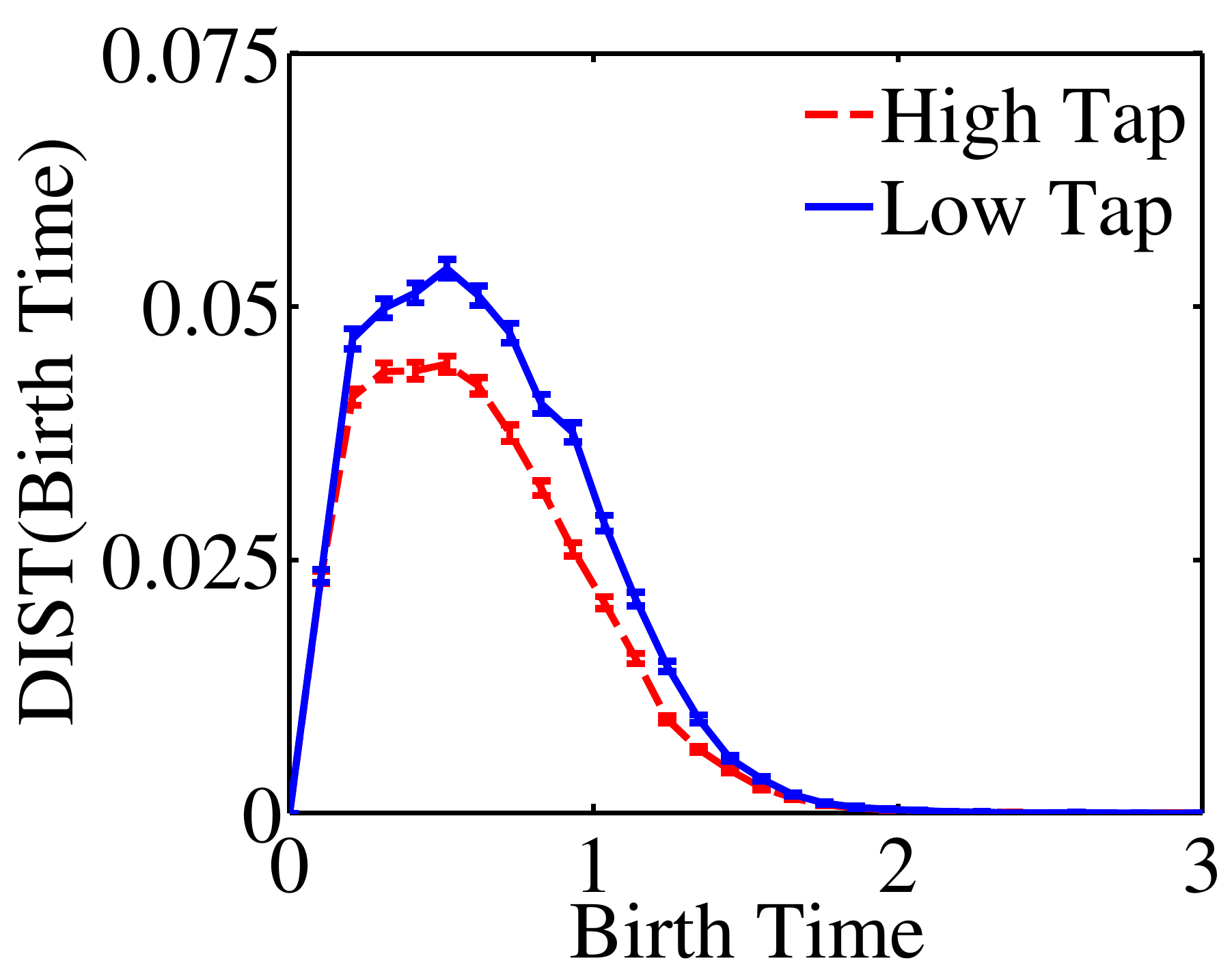}} 
\subfigure[$\pd_1$ tangential forces.]{\includegraphics[width=1.6in]{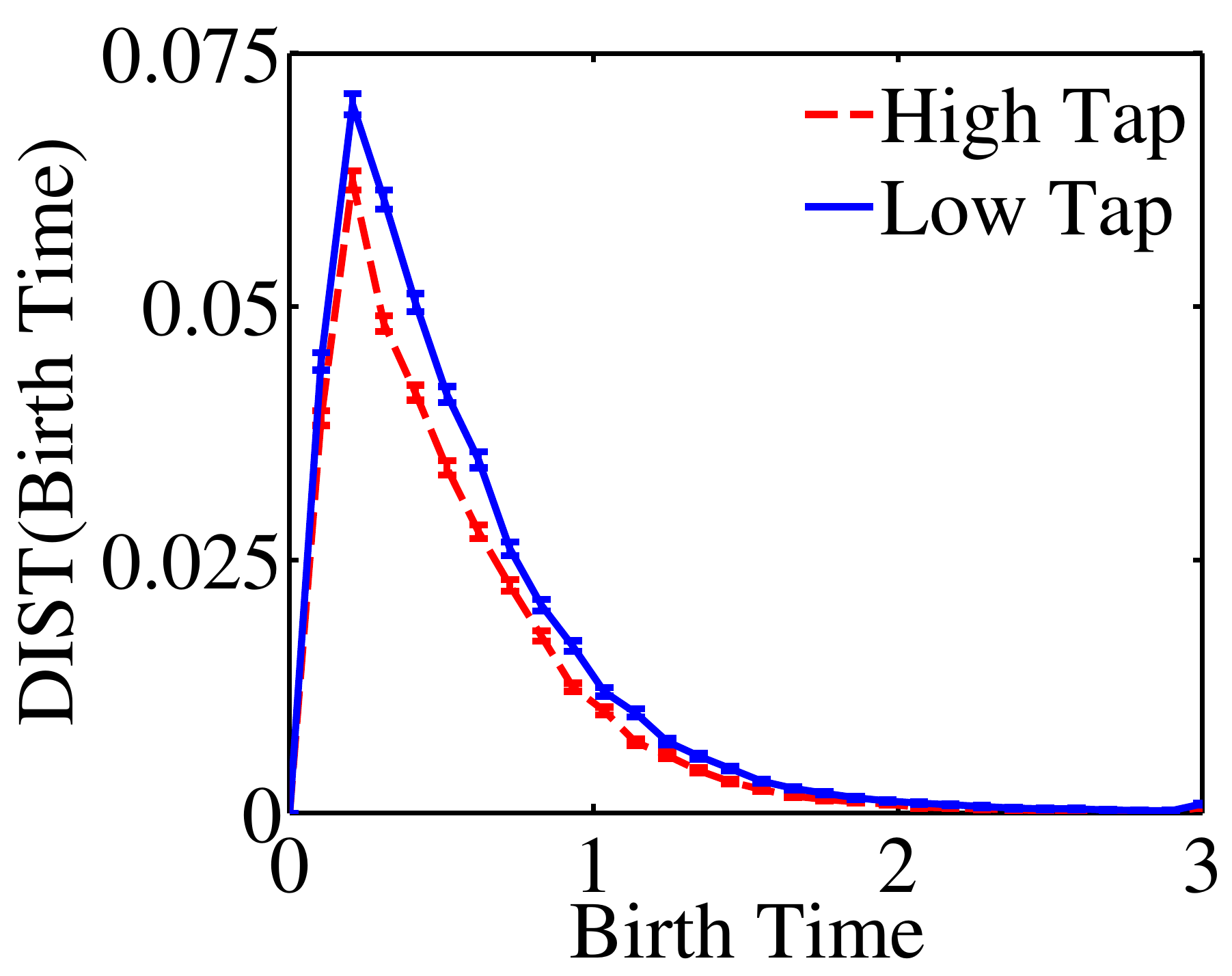}} 
\caption{Distribution of birth times (disks, bottom slice).   Only the features with the lifespan 
larger than $0.1$ are included.}
\label{fig:disks_birth}
\end{figure}

\begin{figure}
\centering
\subfigure[$\pd_0$ normal forces.]{\includegraphics[width=1.6in]{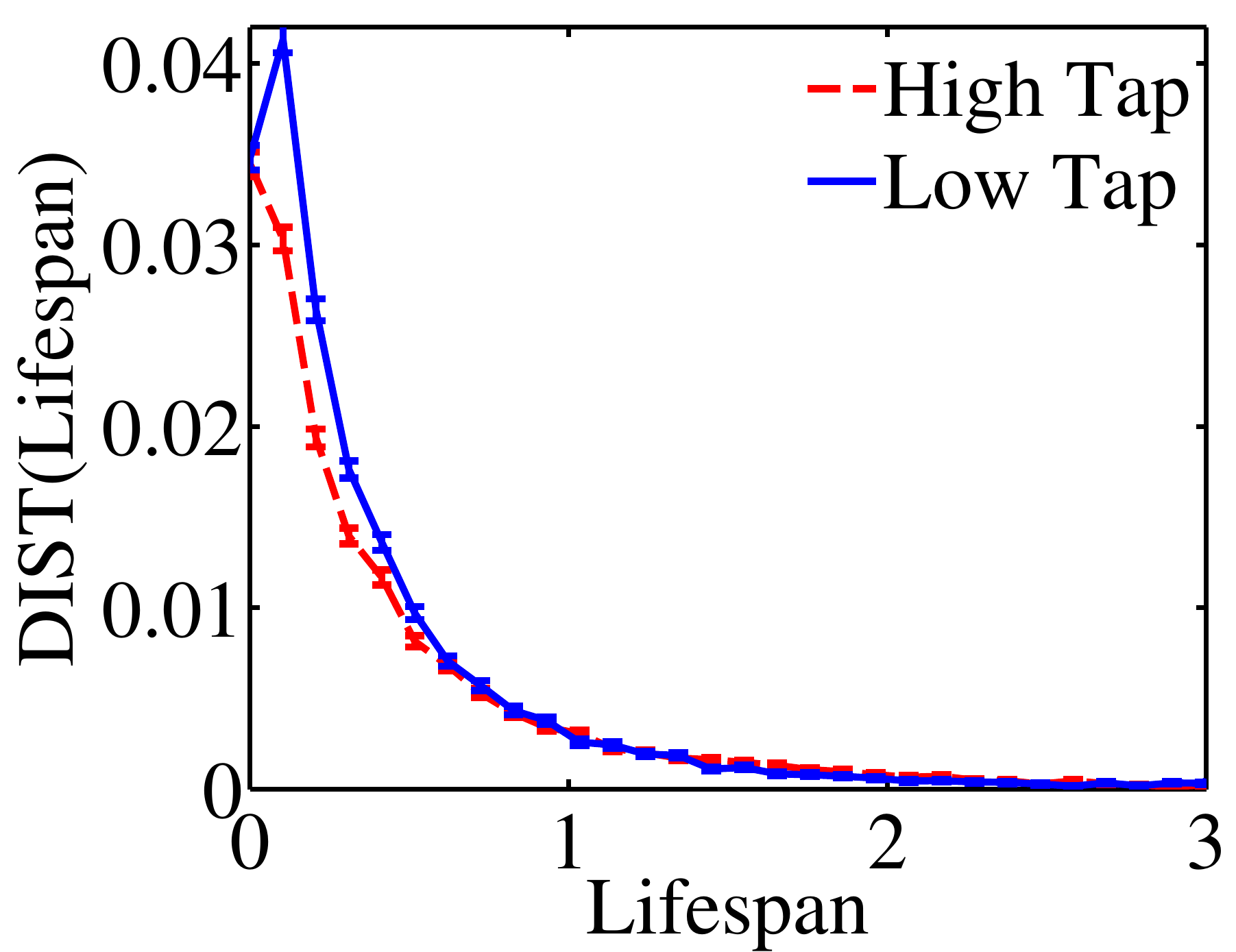}} 
\subfigure[$\pd_0$ tangential forces.]{\includegraphics[width=1.6in]{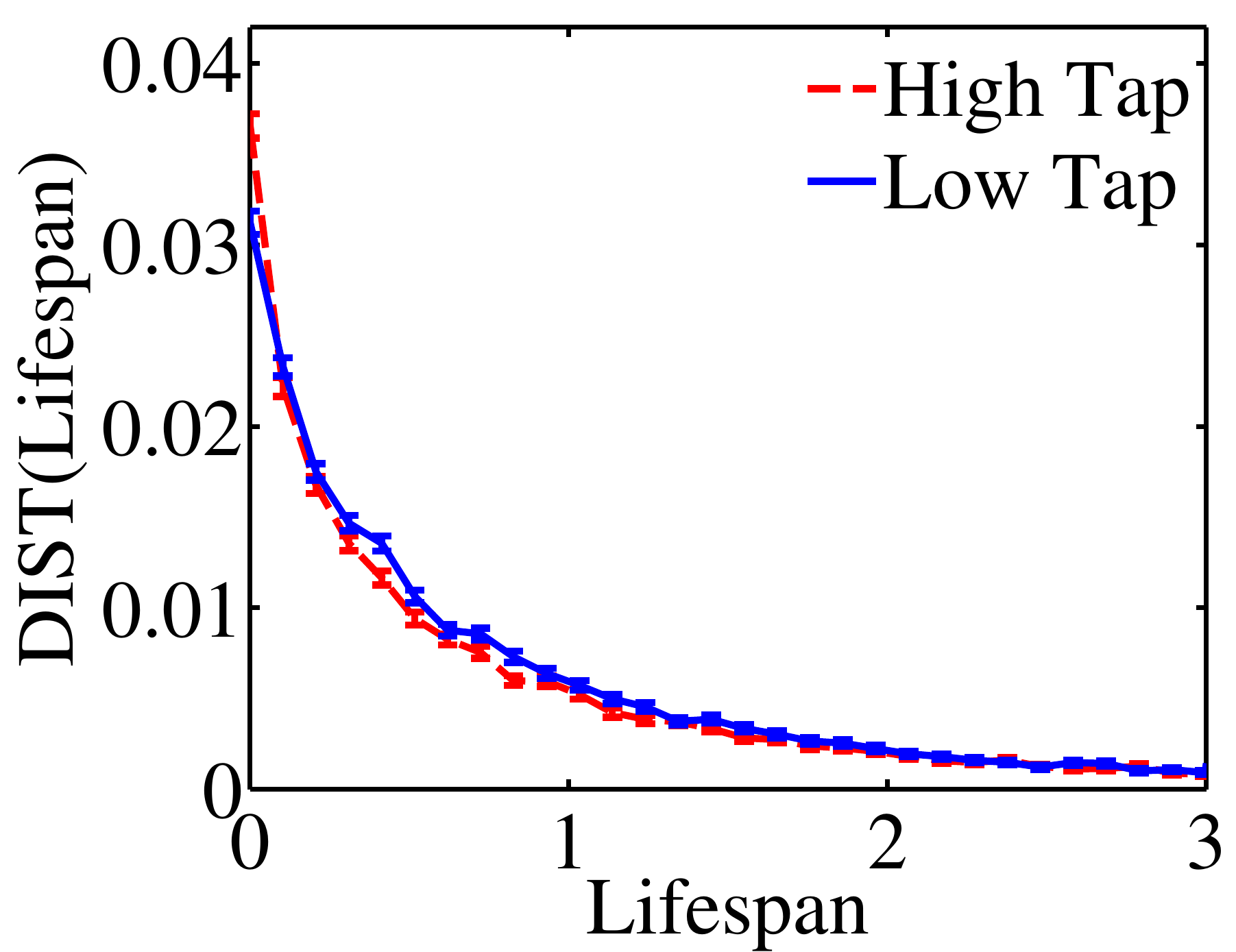}} \\
\subfigure[$\pd_1$ normal forces.]{\includegraphics[width=1.6in]{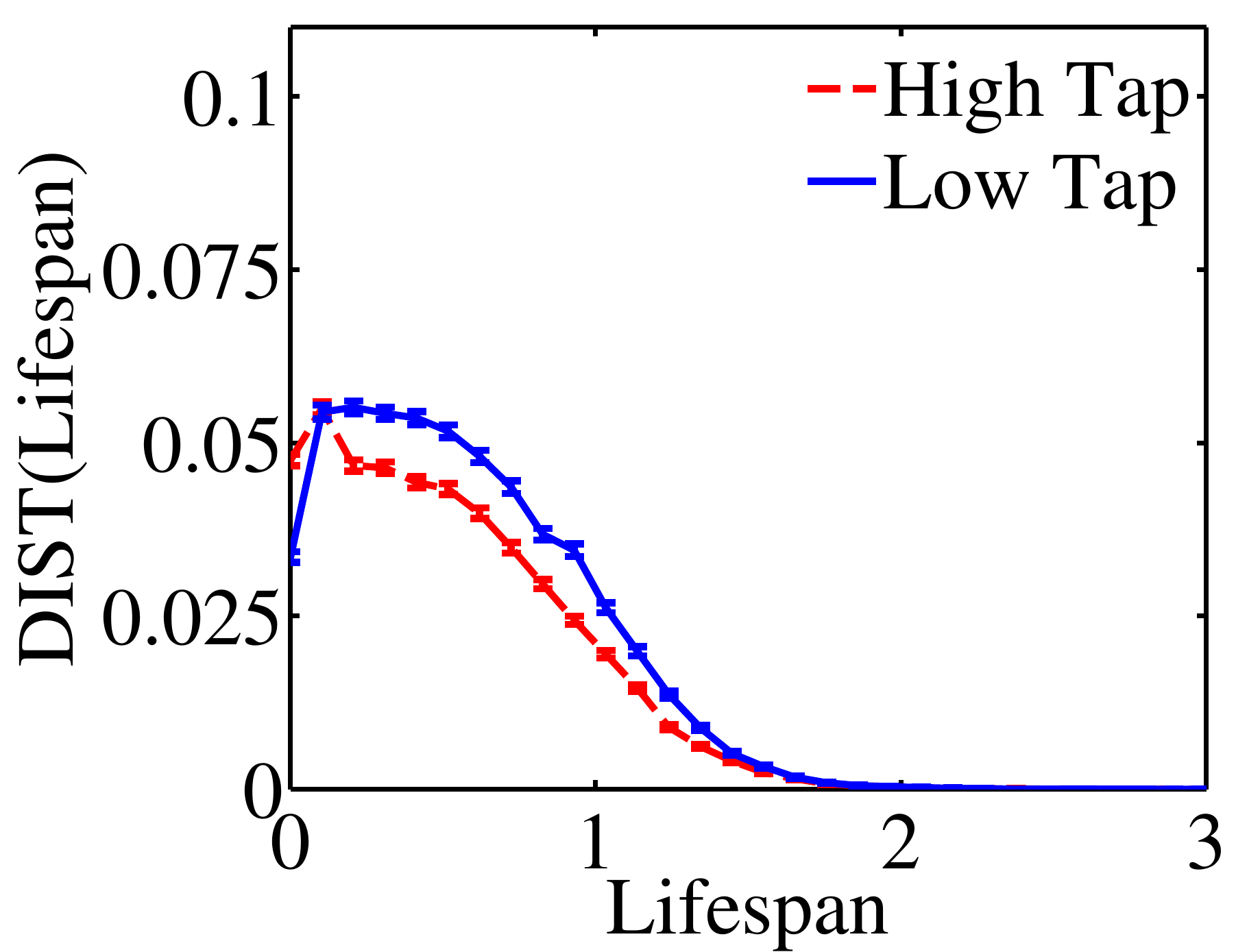}} 
\subfigure[$\pd_1$ tangential forces.]{\includegraphics[width=1.6in]{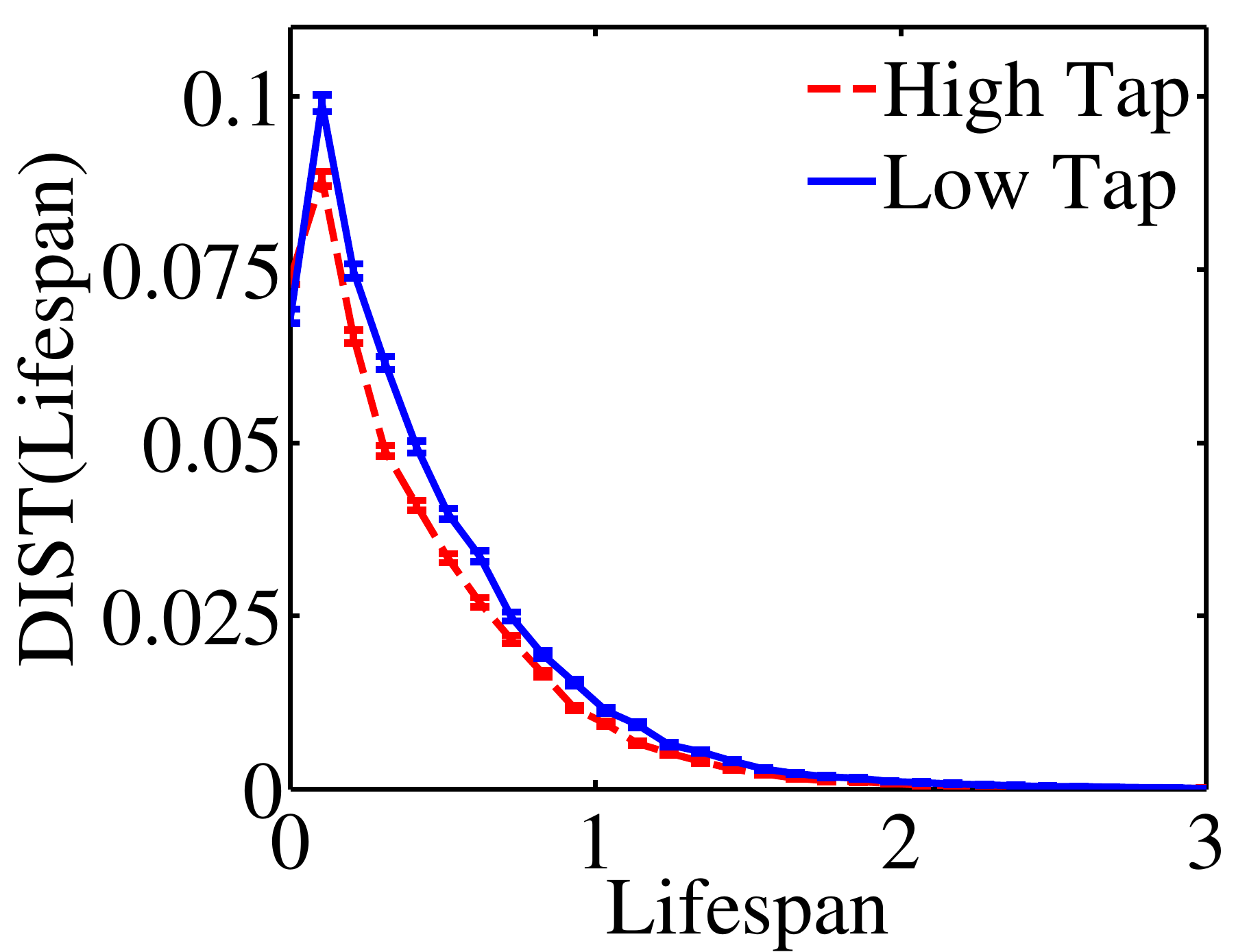}} 
\caption{Distribution of lifespans (disks, bottom slice).}
\label{fig:lifespansHiLow}
\end{figure}

While the interpretation of the results, such as the differences of birth times and lifespans may not be immediately obvious, these measures clearly 
show the ability of the persistence analysis to distinguish between the considered systems.   On a practical side, the fact that the differences
{\it can} be identified suggests that the considered systems  have different macroscopic properties.

\begin{figure}
\centering
\subfigure[$\pd_0$ normal forces.]{\includegraphics[width=1.6in]{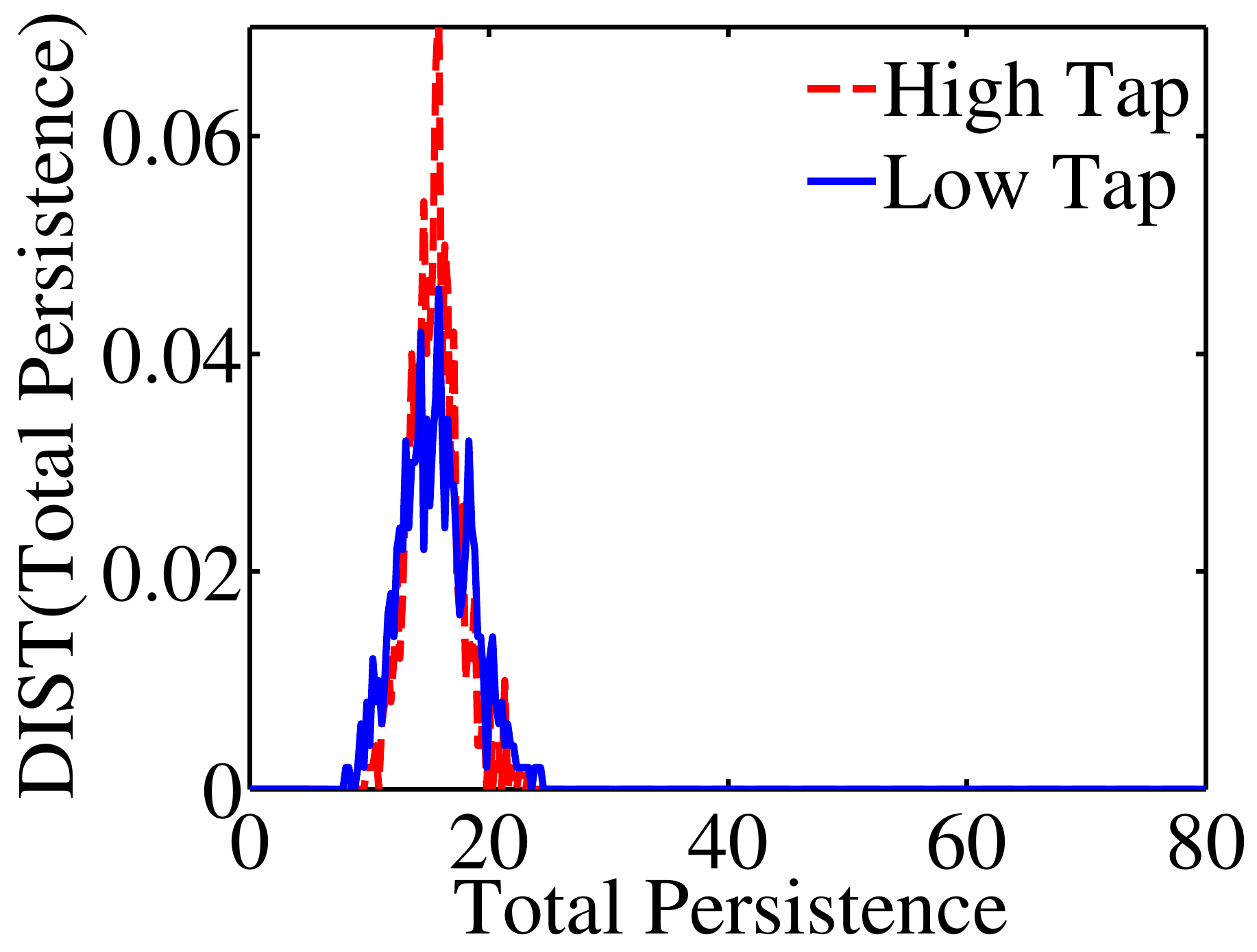}} 
\subfigure[$\pd_0$ tangential forces.]{\includegraphics[width=1.6in]{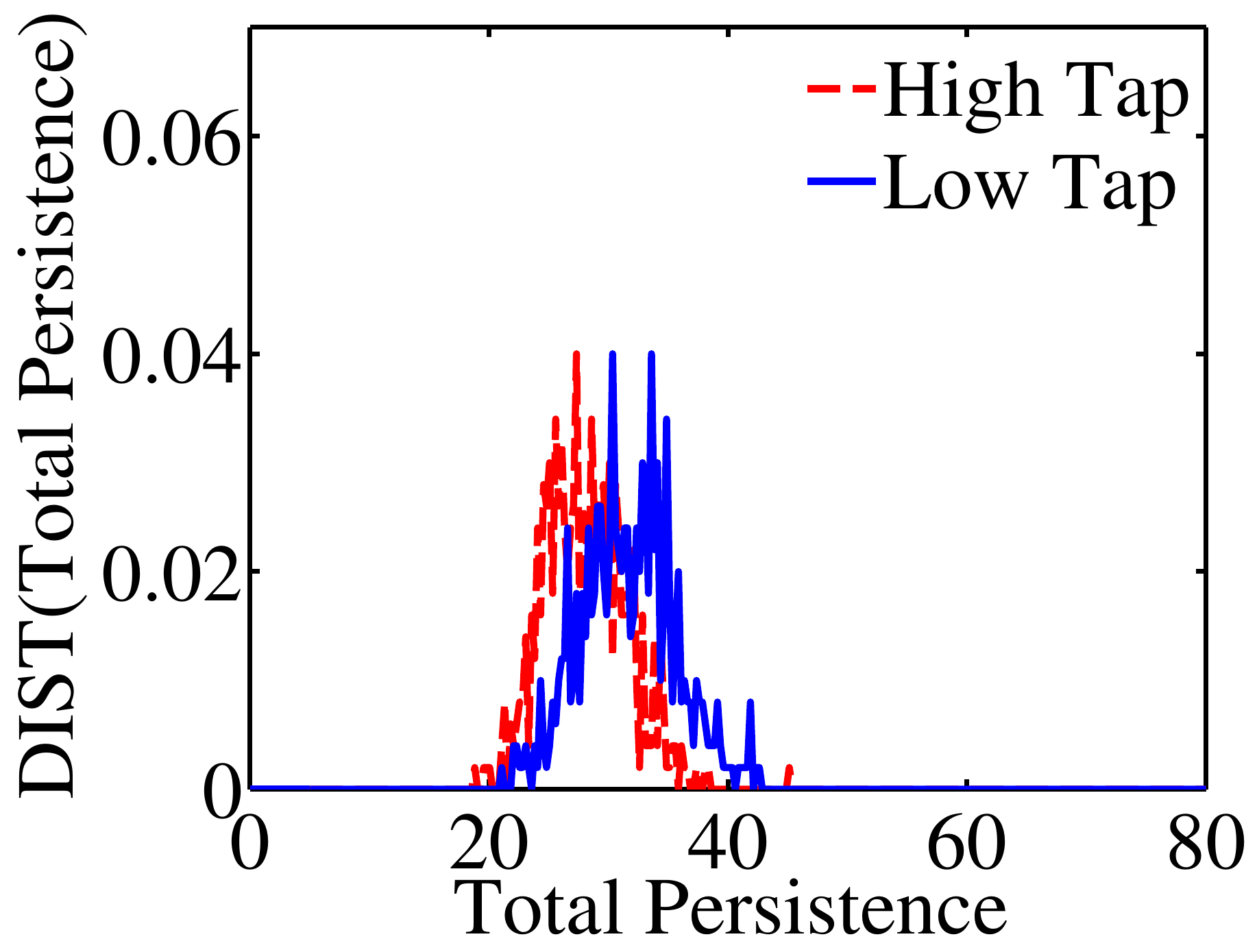}} \\
\subfigure[$\pd_1$ normal forces.]{\includegraphics[width=1.6in]{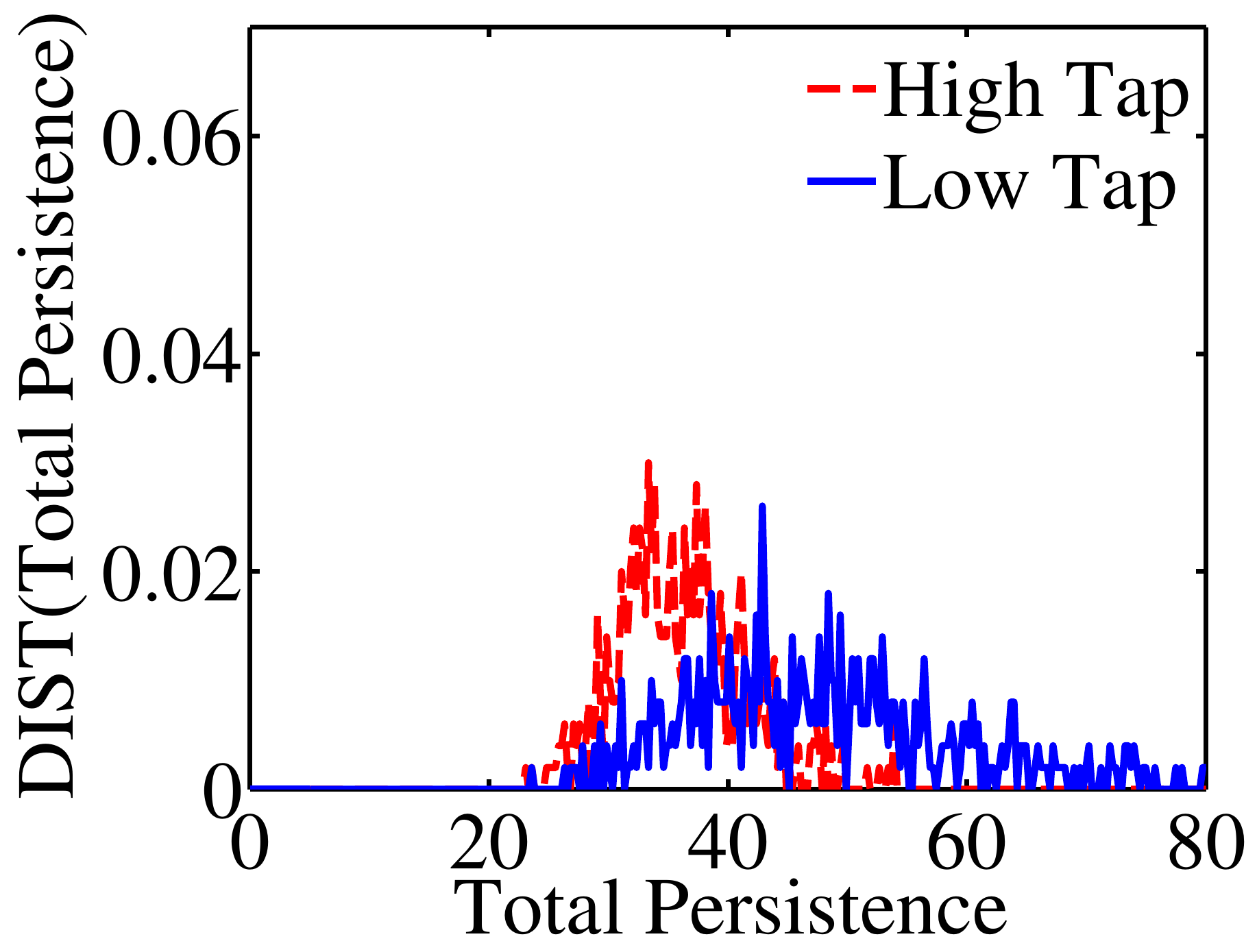}} 
\subfigure[$\pd_1$ tangential forces.]{\includegraphics[width=1.6in]{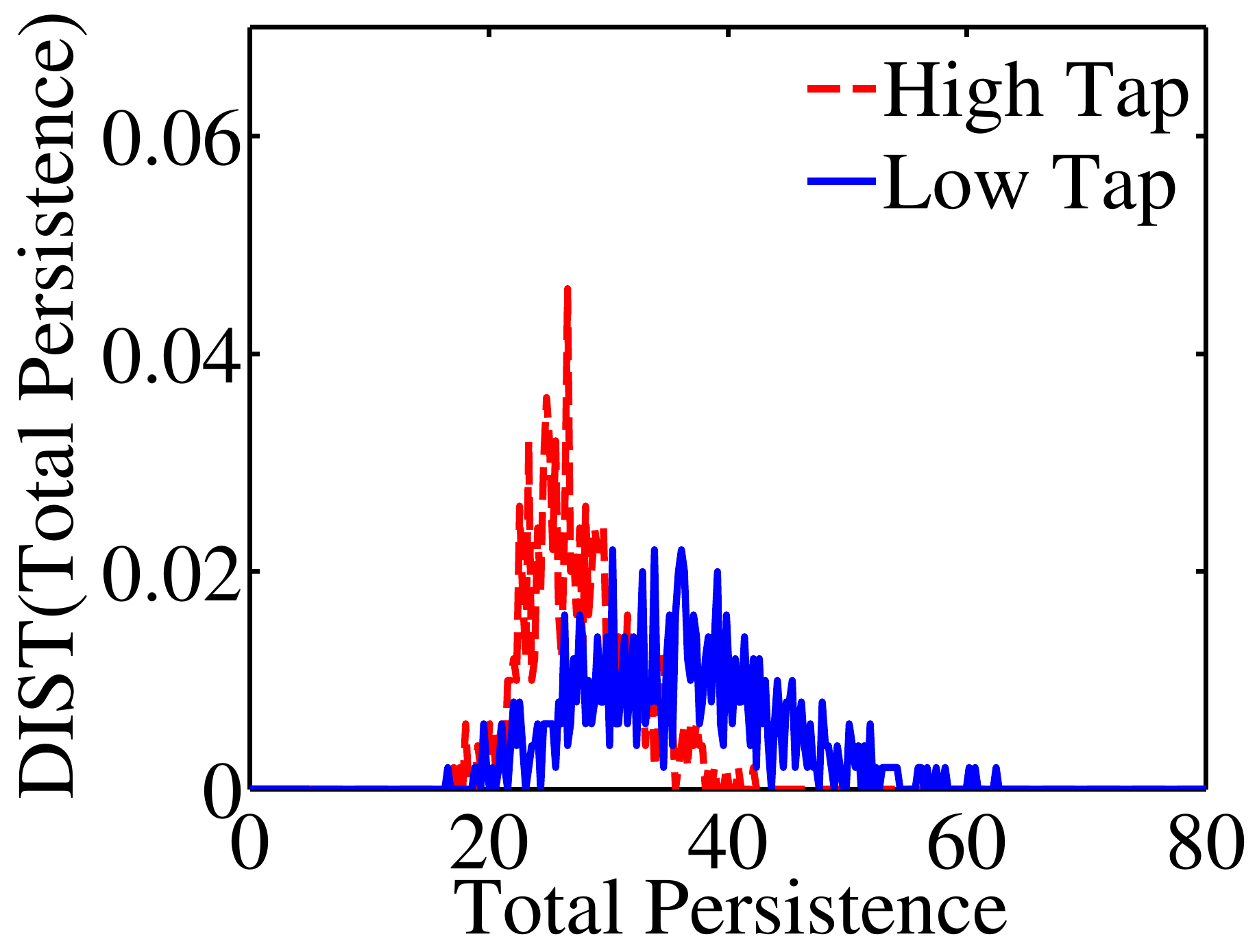}} \\
\caption{Total persistence (disks, low tapping). }
\label{fig:totalpers-HiLow}
\end{figure}

So far, we have discussed one of the questions listed above, regarding the differences between the considered systems.   
Now we proceed to discuss the second one related to the origin of the structure apparent in Fig.~\ref{fig:disks_bottom_heat_hist}  for the low tapping regime.   
Recall (see~\cite[Sec. III B]{paper1}) that low tap intensity often does not lead to significant changes of the packings of disks from one tap to the next.  
Therefore, one may expect that for a certain number of taps, the realizations are correlated, close to each other (with small distances between them), until 
the configuration of the particles, and the corresponding force network, change dramatically. The question is whether these `jumps' from one set of similar packings to another one are captured by the ${\pd}$s.

To show that the answer to this question is positive, we recall that the packing fraction $\phi$ may change from one tap to the next. 
Since $\phi$ values obtained for different taps provide a rather noisy signal, we consider instead its auto-correlation function, as well as instantaneous cross-correlation between $\phi$ and total persistence, $TP$ (for simplicity, we consider here $TP$ instead of the distances). Regarding total persistence, 
we focus on $TP(\pd_1)$. Figure~\ref{fig:correlation} gives the results for the auto-correlation functions defined for any descriptor $f$ as $c(t) = c(0)^{-1} [\langle f(0)f(t) \rangle - \langle f\rangle^2]$, where $\langle\rangle$ indicates an average over all time origins, and $c(0)$ is used to normalize so that $c(t=0)=1$.   Auto-correlation curves  for $\phi$ show that while 
for high tapping there is no observable correlation between taps, for low tapping there is a clear correlation for up to  $10-15$ taps.  This correlation is  consistent with the structure of the distance matrices for low tapping, see Fig.~\ref{fig:disks_bottom_heat_hist}(a,~b).  The results for $TP(\pd_1)$ auto-correlation functions are very similar.  We
can go further and compute the instantaneous cross-correlation $c$ between $TP(\pd_1)$ and $\phi$, defined as 
\[c= \frac{\langle [\phi(t)-\langle\phi\rangle] [TP(\pd_1)(t)-\langle TP(\pd_1) \rangle] \rangle}{\sigma_{\phi}\sigma_{TP(\pd_1)}}, \]
where $\sigma$ indicates the variance of the variable. We find significant correlation between these two 
quantities: $c_{\phi TP(\pd_1)}$ reaches the values of $\approx 0.6 - 0.7$  for normal and tangential forces (here $1$ means perfect correlation and $0$ lack of 
correlation).  

To rationalize the correlation of $\phi$ and $TP(\pd_1)$, we recall the results obtained by considering the systems of 
disks exposed to compression~\cite{pre13}.  In that system it was found that for monodisperse disks, which are more likely to crystallize (therefore having
larger $\phi$), there is a larger number of points in $\pd_1$}, consistently with the results presented here.  We note that in~\cite{pre13} it was found that 
larger number of points in $\pd_1$ occurred for strong forces (with the idea that strong loops form at the boundaries of the fault zones 
separating crystalline regions); we expect a similar reason for the larger values of $TP(\pd_1)$ in the present setting. 

To conclude this section, we note that persistence analysis allows to clearly identify  differences between the systems of disks exposed to different 
tapping intensities leading to the same (average) packing fraction:  these differences are particularly clear when considering the structure of 
loops.   The differences are apparent for the averaged persistence diagrams but they are even more prominent when considering individual taps and their variability.    This variability is much stronger for the systems exposed
to low tapping.   As already noted in the context of the results shown in Fig.~\ref{fig:disks_bottom_heat_hist}(c,~d) 
the differences between different realizations for low tapping may be as large as the differences between low and 
high tapping ones. 

\begin{figure}
\centering
\subfigure[Low tapping.]{\includegraphics[width=1.6in]{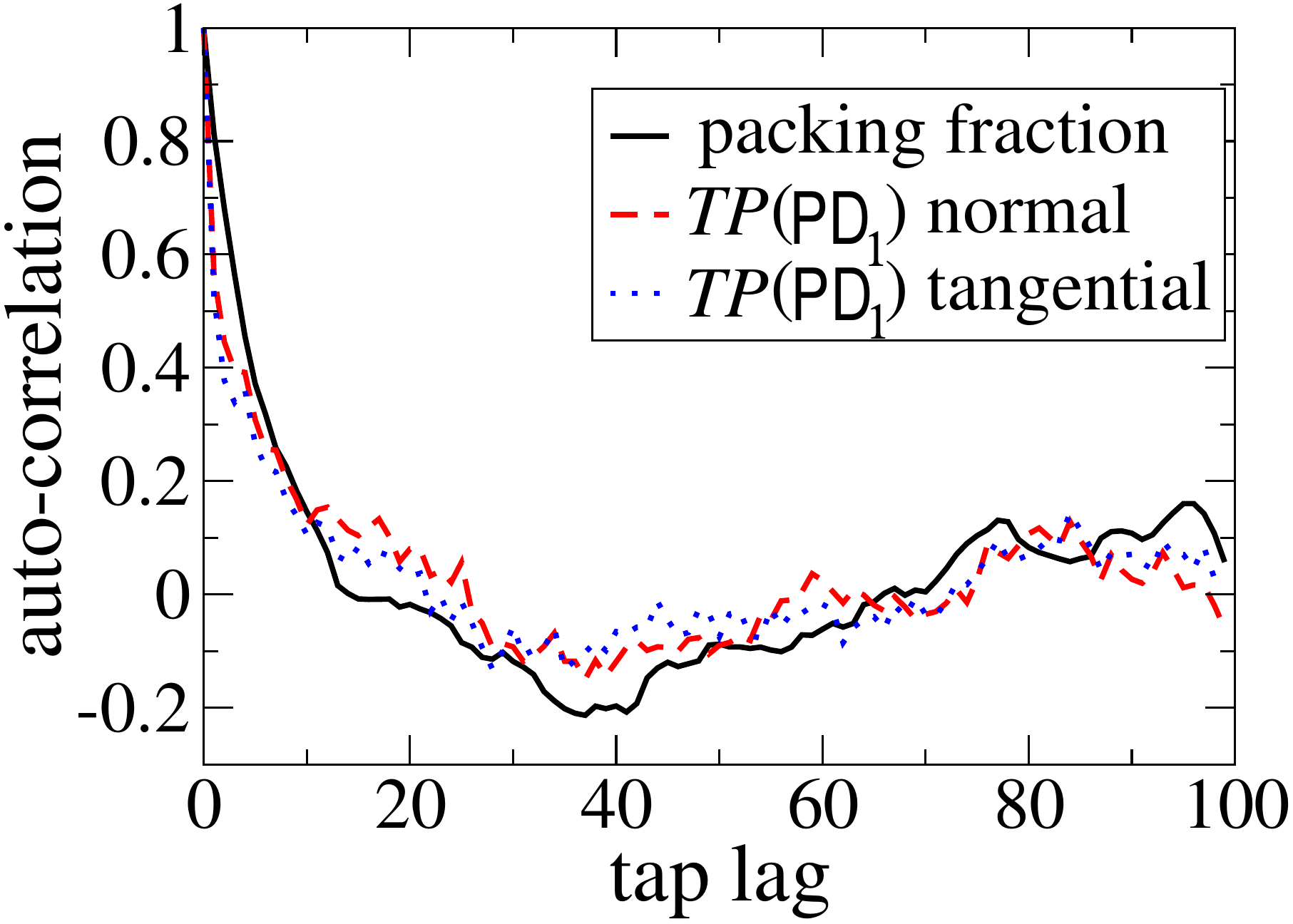}} 
\subfigure[High tapping.]{\includegraphics[width=1.6in]{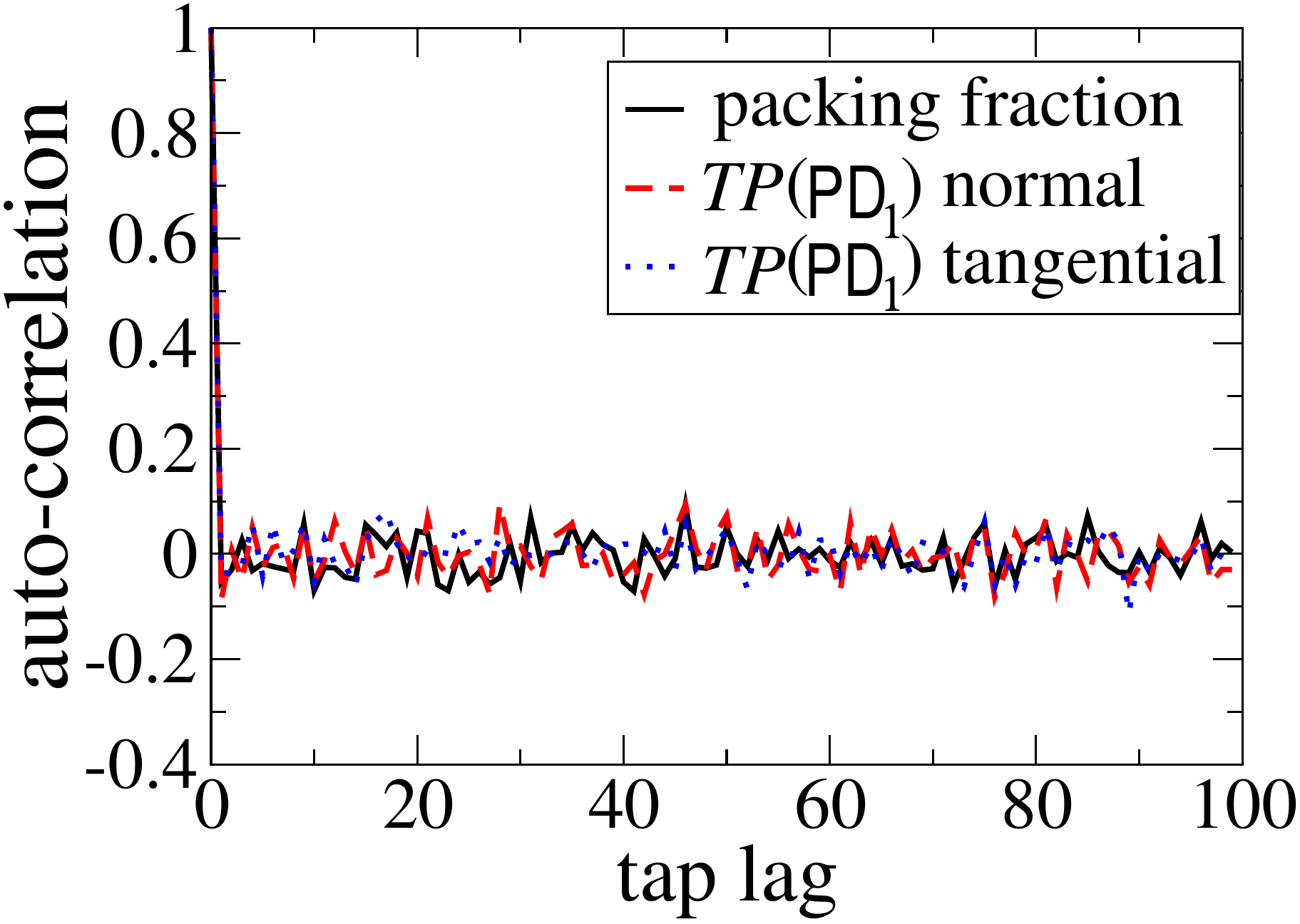}} 
\caption{Auto-correlation functions for (a) low and (b) high tapping.  
Each figure shows auto-correlation functions for packing fraction, $\phi$, and total persistence, $TP(\pd_1)$, for 
normal and tangential forces (disks, bottom slice). 
}
\label{fig:correlation}
\end{figure}

\subsection{Force networks in the systems of  disks and pentagons}
\label{sec:forcenet}

In~\cite{paper1} we discussed some of the differences in the structure of the force networks between disks and pentagons.  The main findings
reported in that paper are that the differences between these systems manifest themselves particularly in the structure of tangential 
force networks measured by $\beta_0$ (although PDFs of the forces are almost indistinguishable), and by the number of loops, measured by 
$\beta_1$, for both normal and tangential forces.  The number of loops in the disk-based system is consistently larger.  This finding supports
the idea that the clusters are larger for disks, and therefore can support larger number of loops.   In the present work, we will discuss additional
insight that can be reached by persistence analysis.  

\begin{figure}
\centering
\subfigure[$\pd_0$ normal forces.]{\includegraphics[width=1.6in]{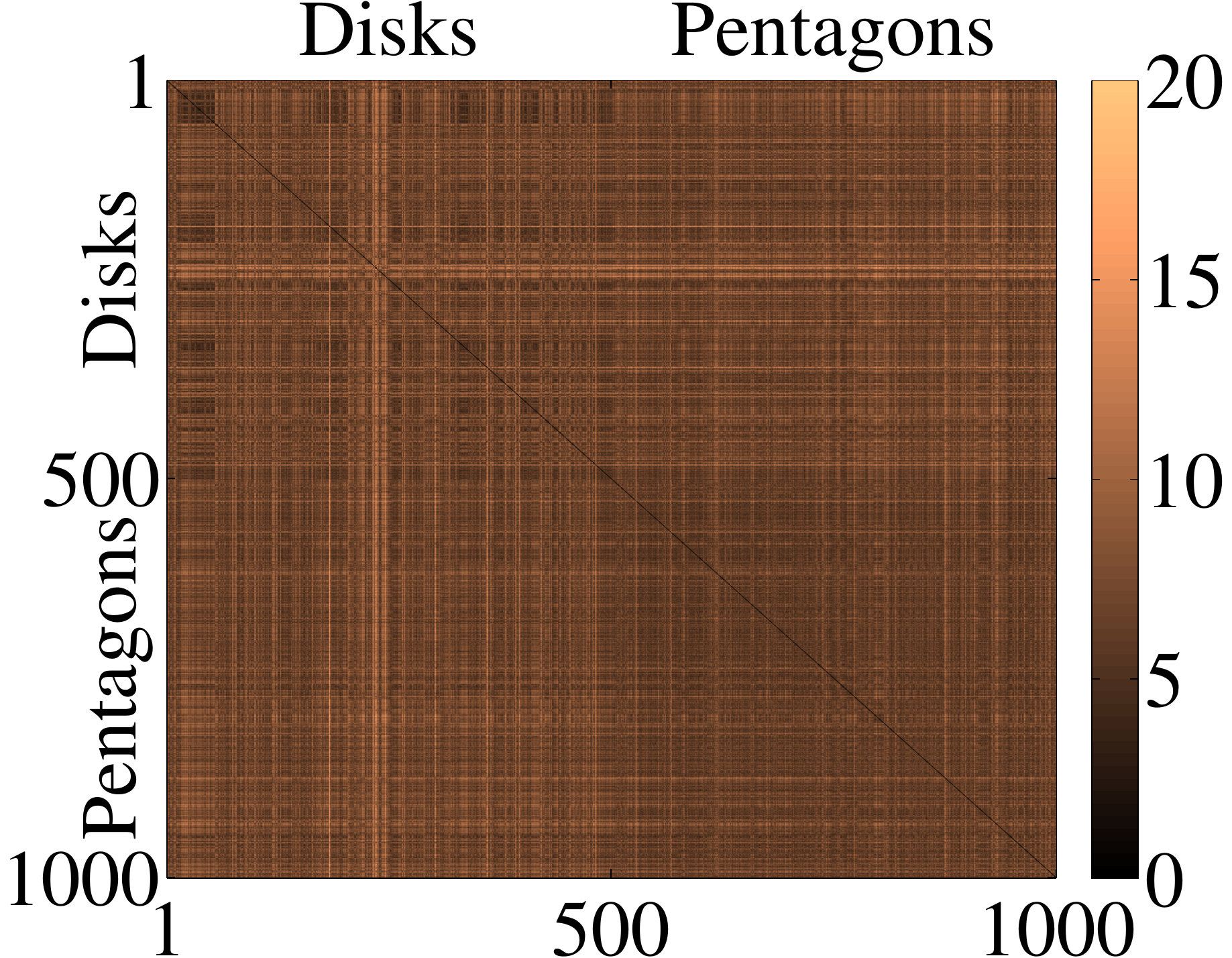}} 
\subfigure[$\pd_0$ tangential forces.]{\includegraphics[width = 1.6in]{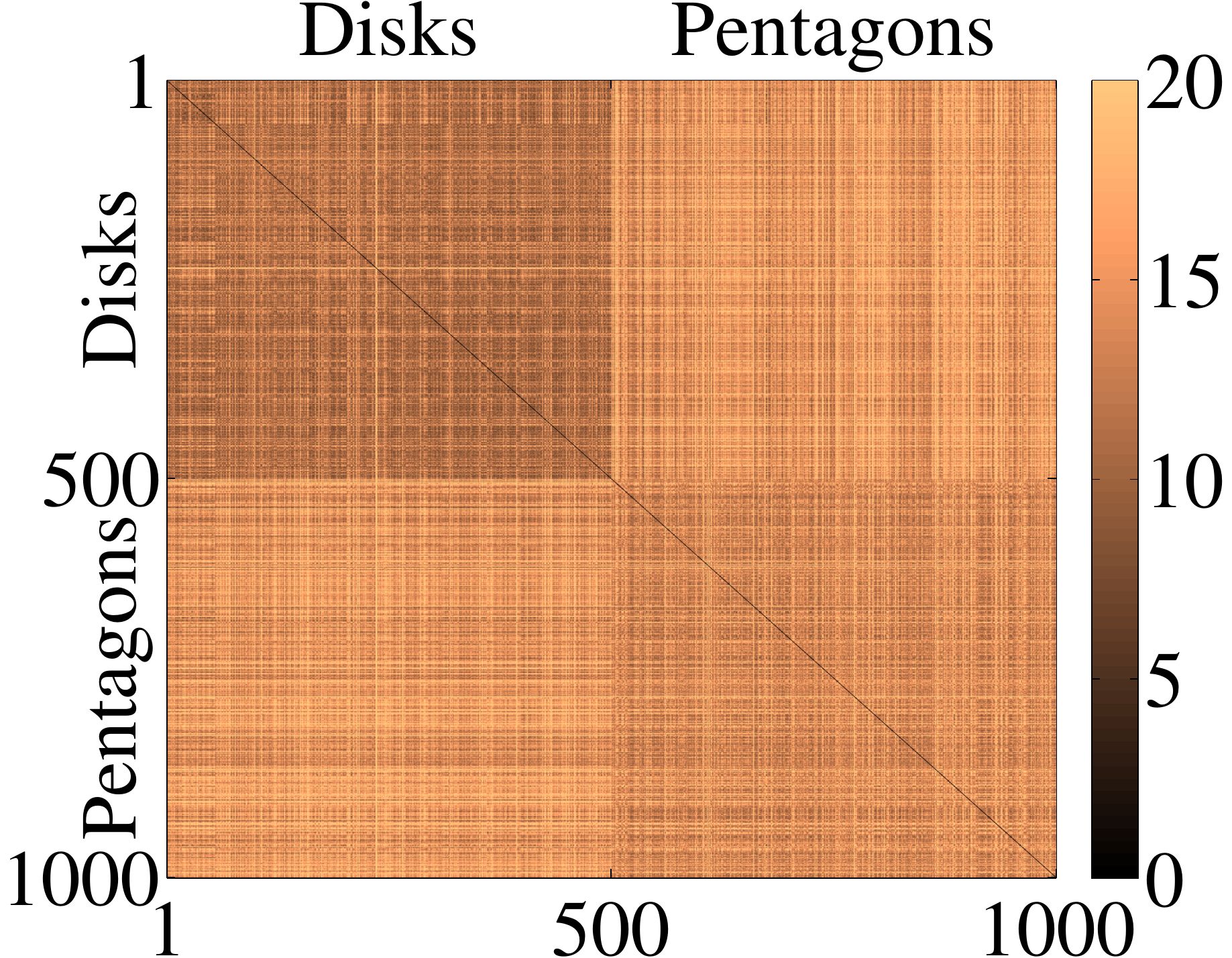}}\\
\subfigure[$\pd_1$ normal forces.]{\includegraphics[width=1.6in]{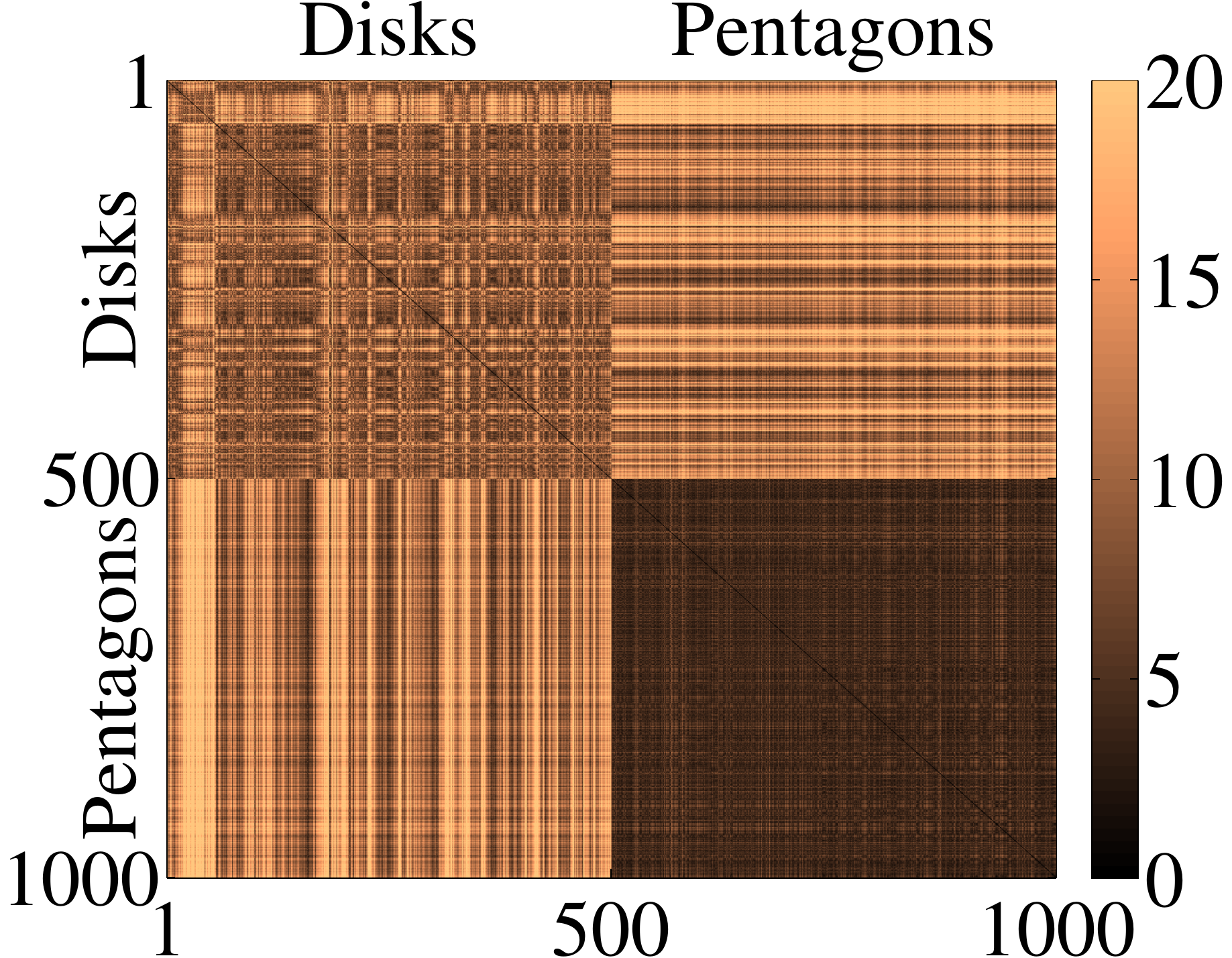}} 
\subfigure[$\pd_1$ tangential forces.]{\includegraphics[width = 1.6in]{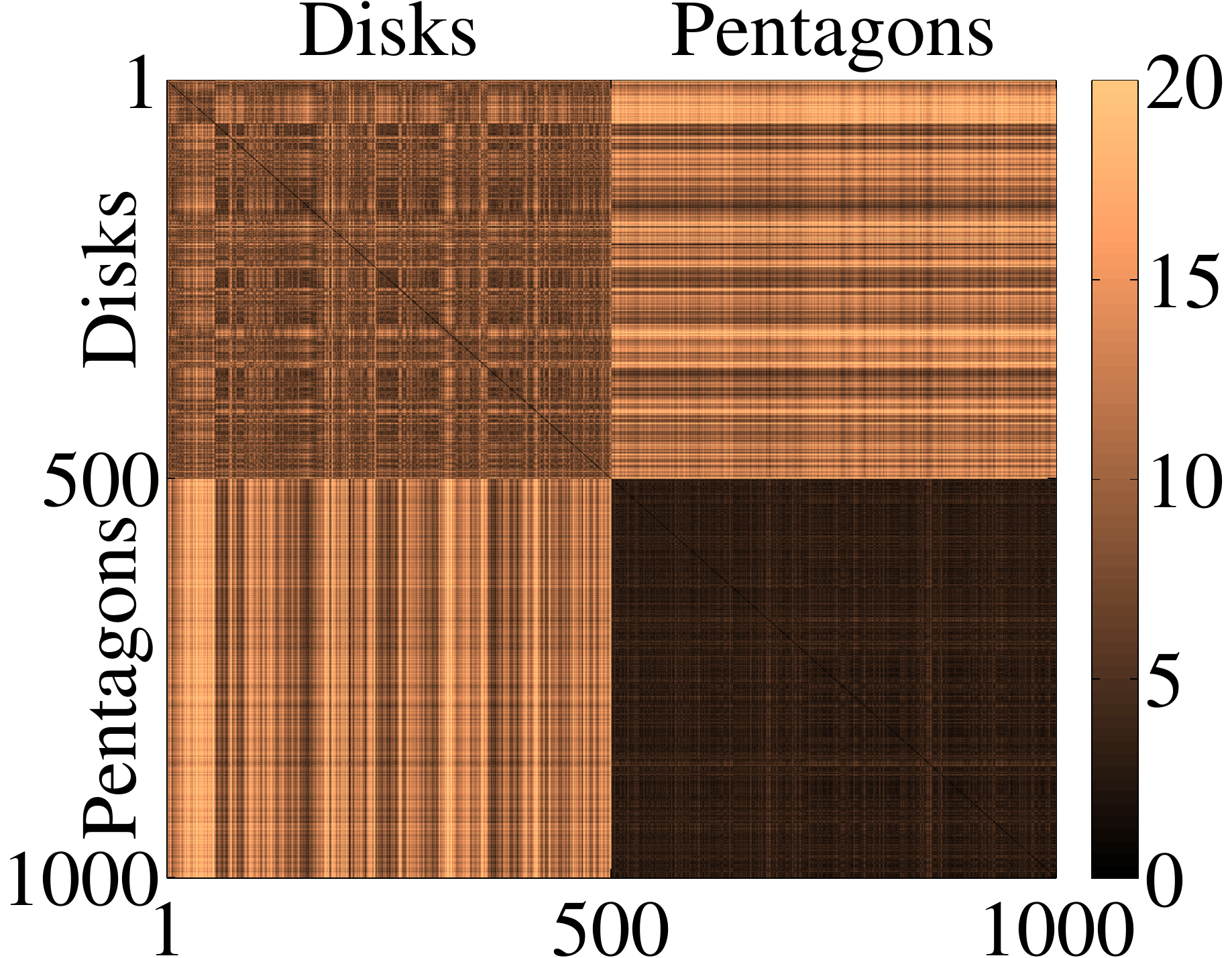}}
\caption{Distance matrices showing $d_{W1}$ (bottom slice, low tapping).
}
\label{fig:disks_pents_heat_W1}
\end{figure}

\begin{figure}
\centering
\subfigure[$\pd_0$ normal forces.]{\includegraphics[width=1.6in]{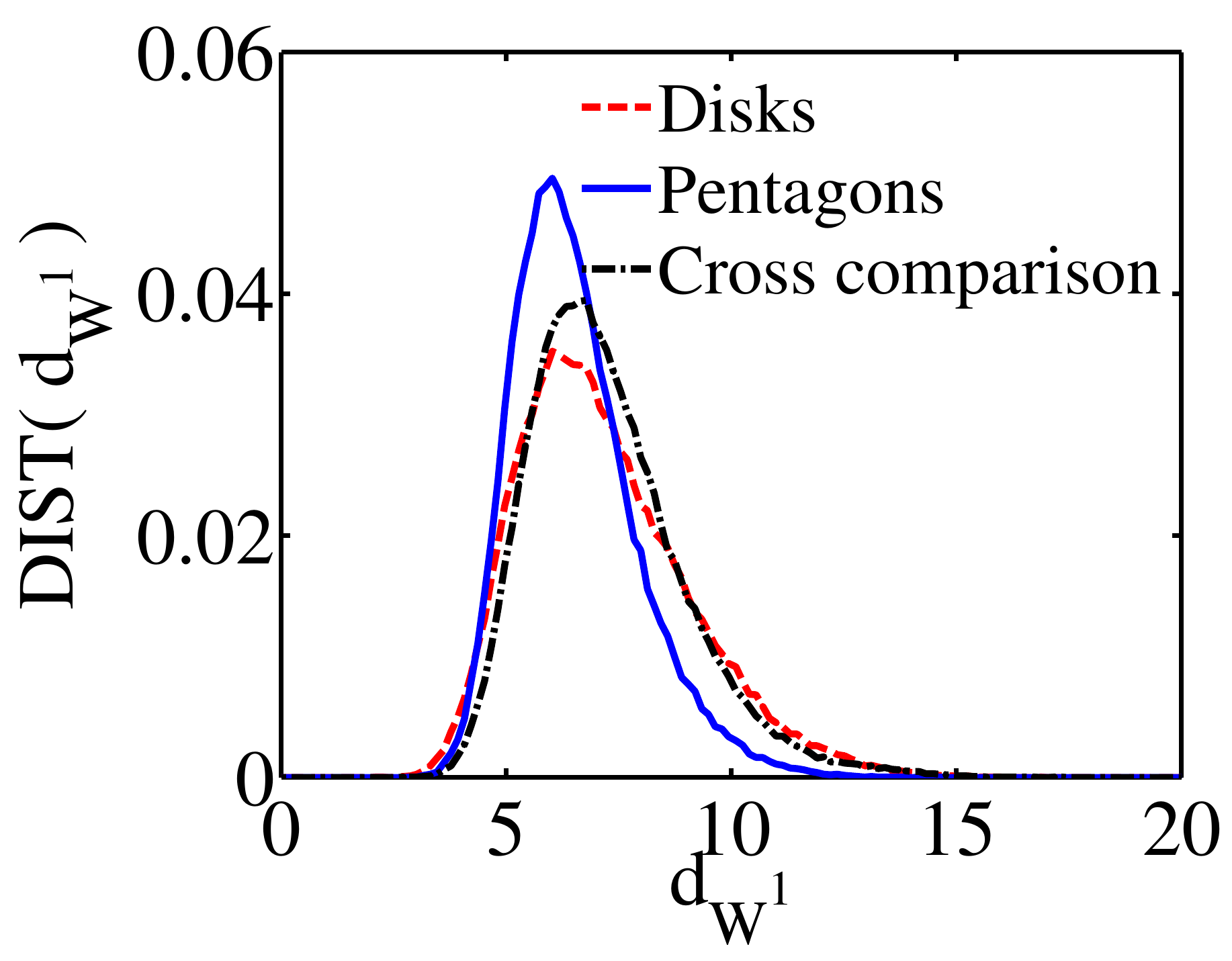}} 
\subfigure[$\pd_0$ tangential forces.]{\includegraphics[width=1.6in]{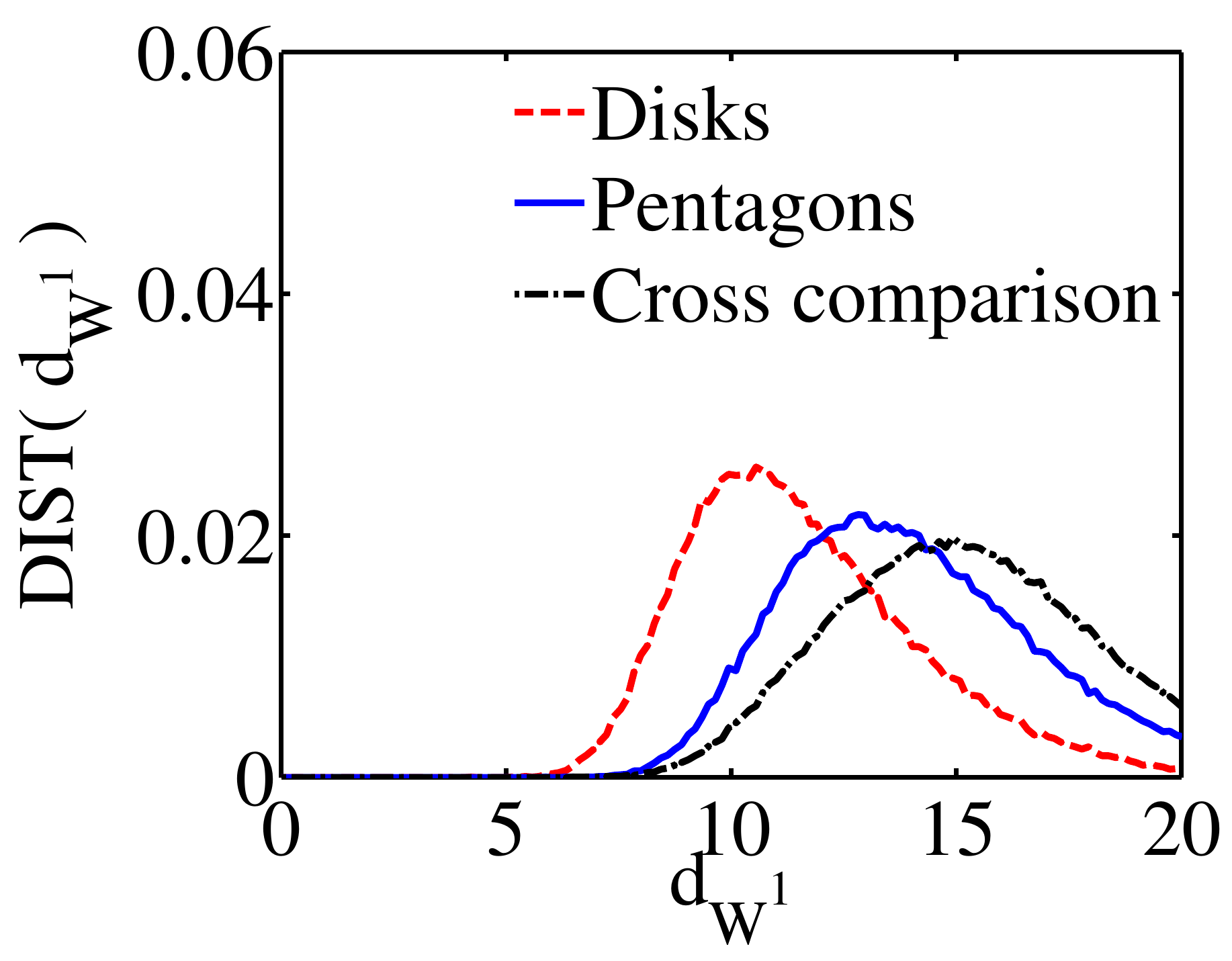}}\\
\subfigure[$\pd_1$ normal forces.]{\includegraphics[width=1.6in]{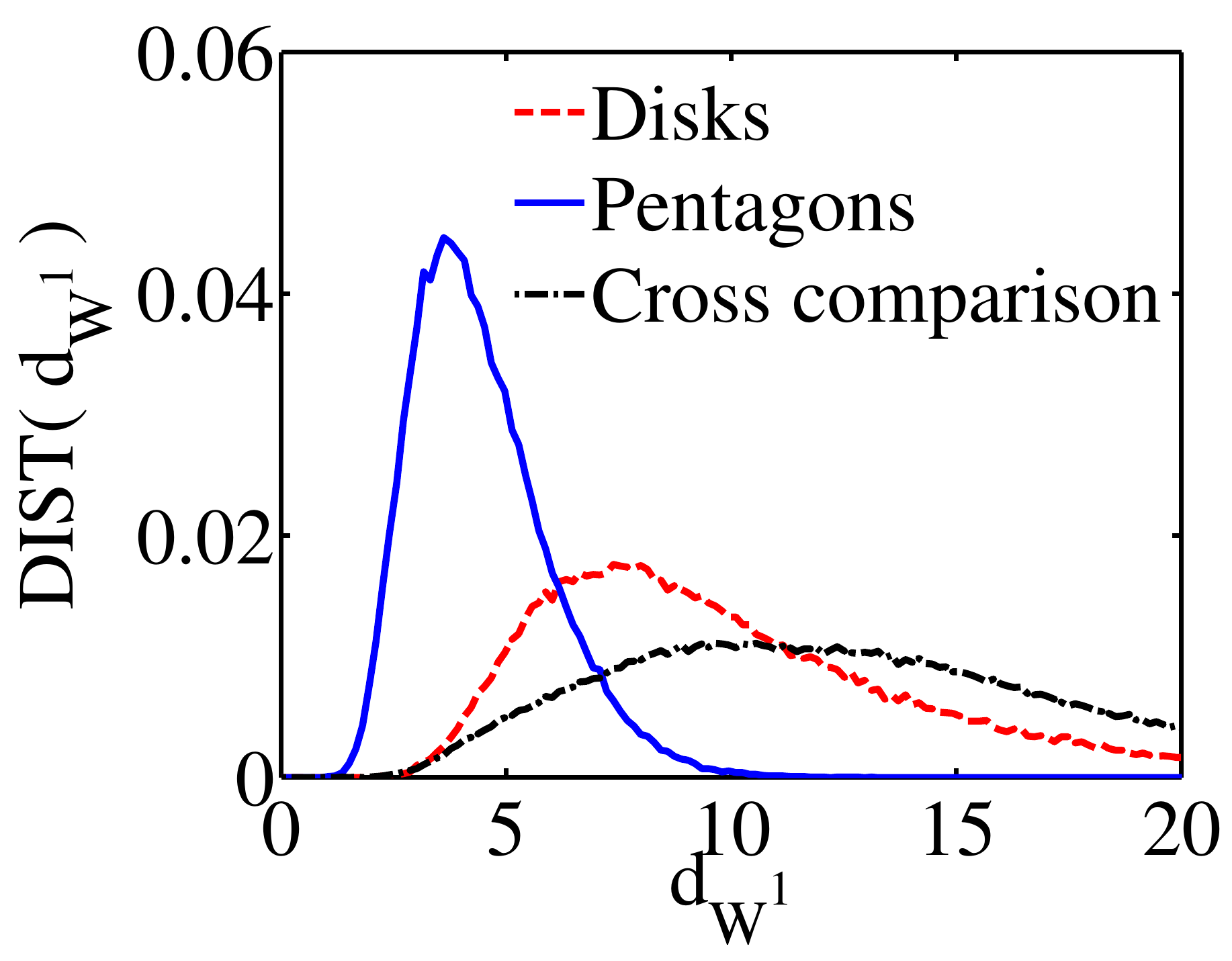}} 
\subfigure[$\pd_1$ tangential forces.]{\includegraphics[width=1.6in]{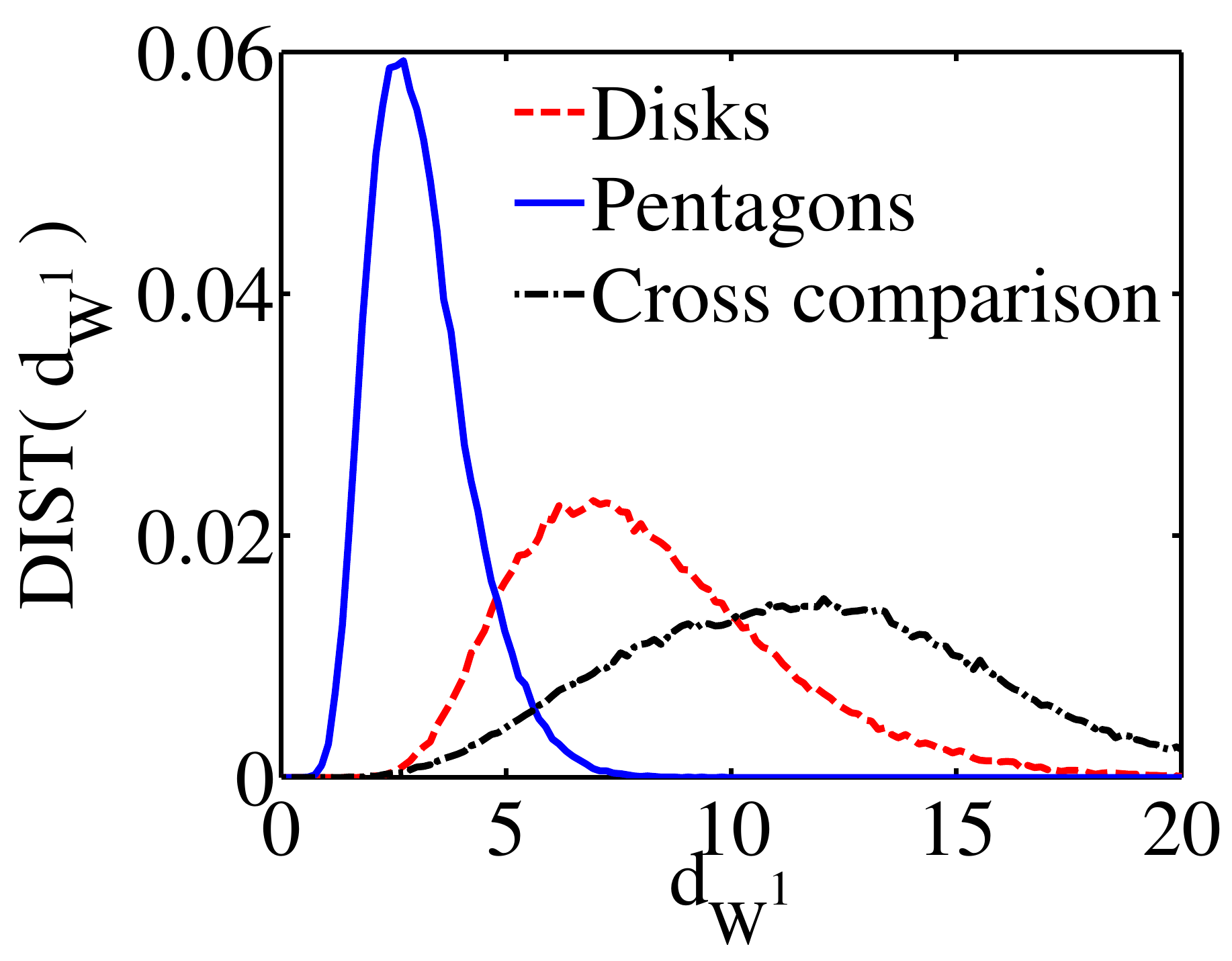}} 
\caption{Distributions of $d_{W1}$ distance (bottom slice, low tapping). }
\label{fig:disks_pents_hist_W1}
\end{figure}

Figures~\ref{fig:disks_pents_heat_W1} and~\ref{fig:disks_pents_hist_W1}  show the distance matrices and corresponding distributions comparing 
disks and pentagons exposed to the same (low) tapping  intensity.   In agreement with the results from~\cite{paper1}, the differences between
the components (the parts (a) and (b) of these figures) are relatively minor.   
Considering loops, these figures show that the distances between pentagon-based systems are
much smaller than for the disk-based ones.    In particular,  Fig.~\ref{fig:disks_pents_hist_W1} shows that 
the distances between pentagon-based systems are centered at much smaller values, and their distribution is much narrower than for disks.   We
also note that the distances between disks and pentagons are much larger than between different disk realizations, 
showing that persistence  analysis clearly distinguishes these systems.

We note  that consideration of other distances, such as bottleneck, $d_B$, that measures only the largest difference, are consistent with the ones presented for 
$d_{W^1}$ distance (figures not shown for brevity).  
In particular, the distributions of $d_B$ for loops are similar to the ones shown in Fig.~\ref{fig:disks_pents_hist_W1}, with the maximum
and the spread for pentagons smaller than for disks.   This is as expected, since the loops form at lower force level in pentagon-based system, compare 
Fig.~\ref{fig:diagrams_all}(c) and (d). 

\begin{figure}
\centering
\subfigure[$\pd_0$ normal forces.]{\includegraphics[width=1.6in]{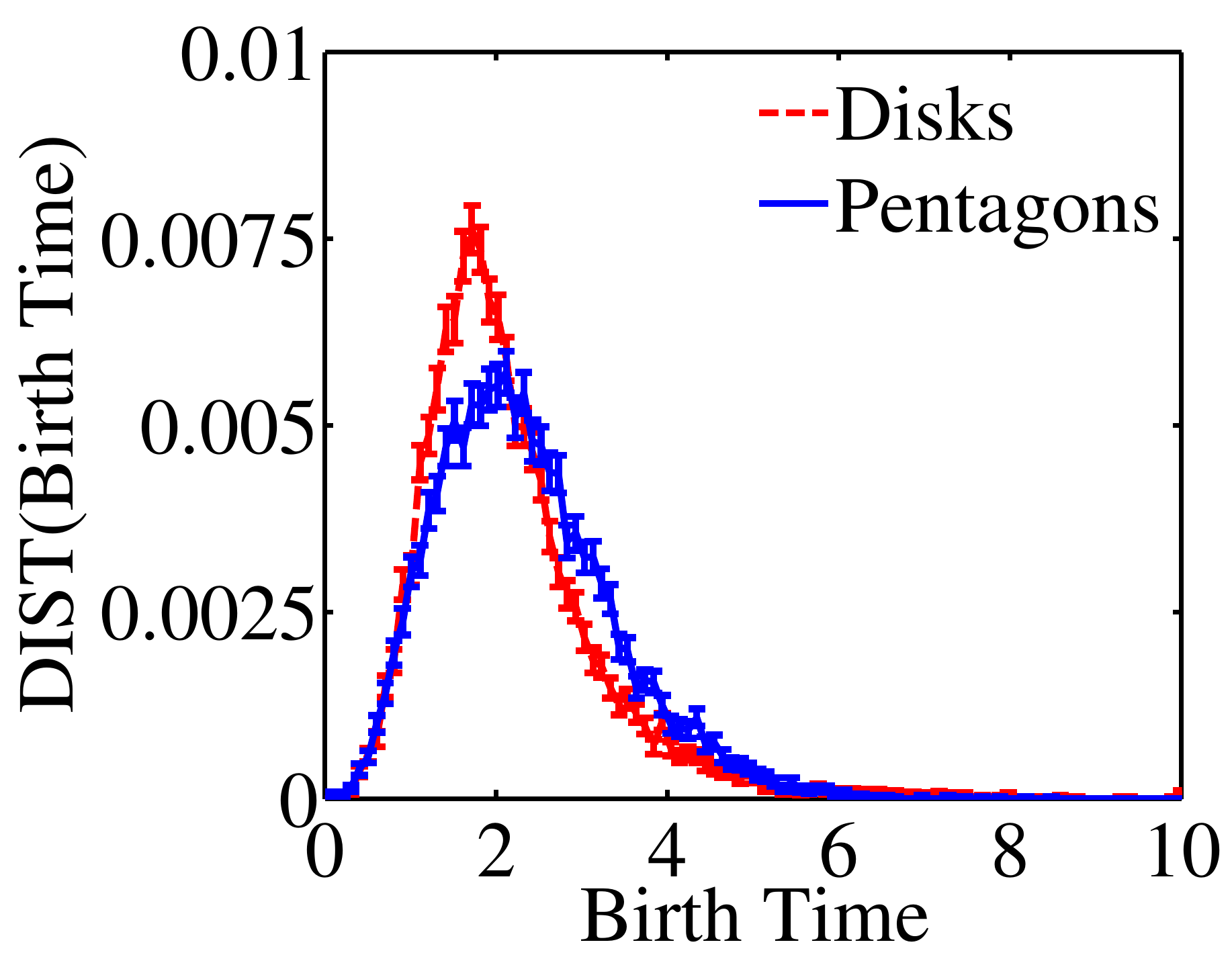}} 
\subfigure[$\pd_0$ tangential forces.]{\includegraphics[width=1.6in]{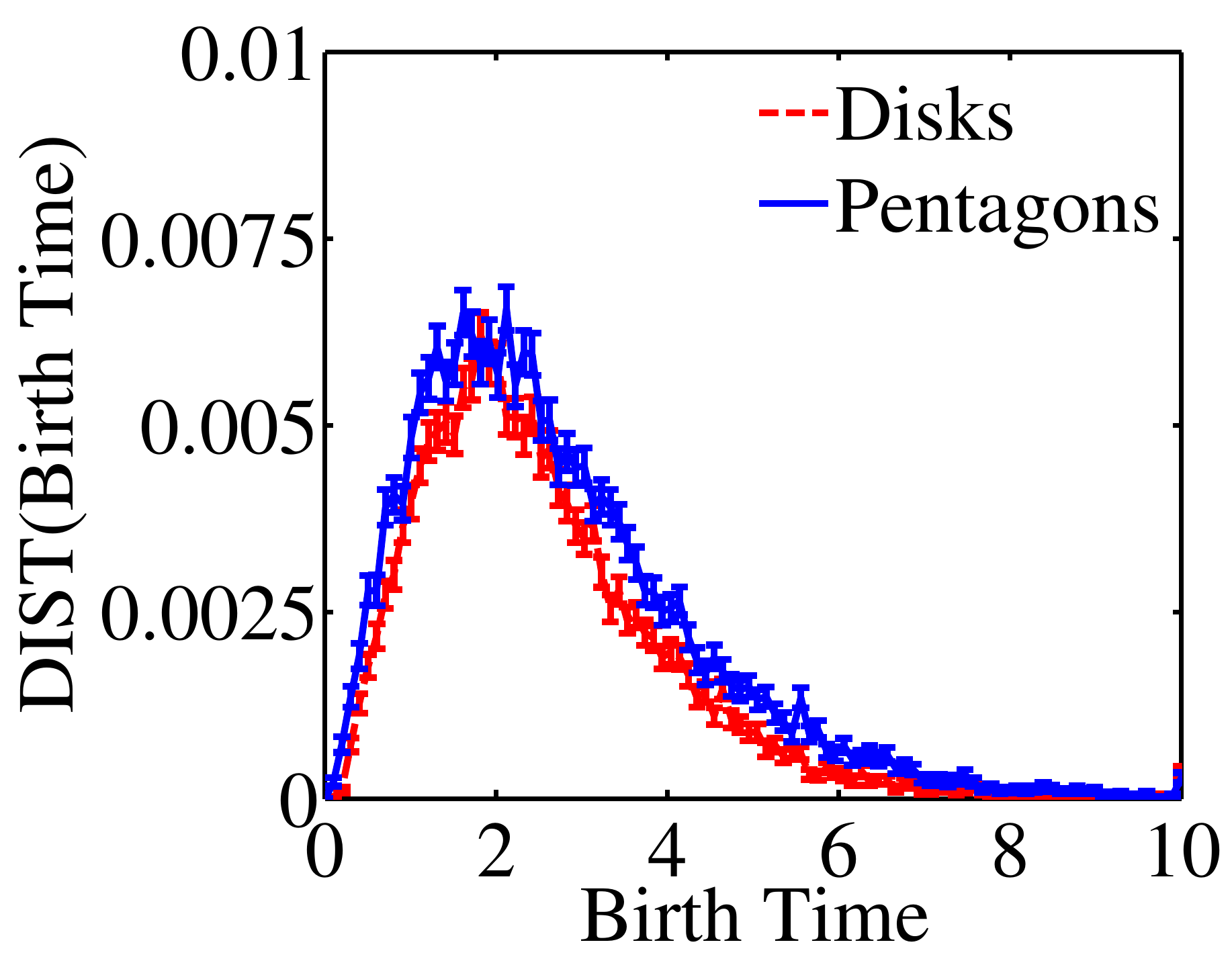}} 
\subfigure[$\pd_1$ normal forces.]{\includegraphics[width=1.6in]{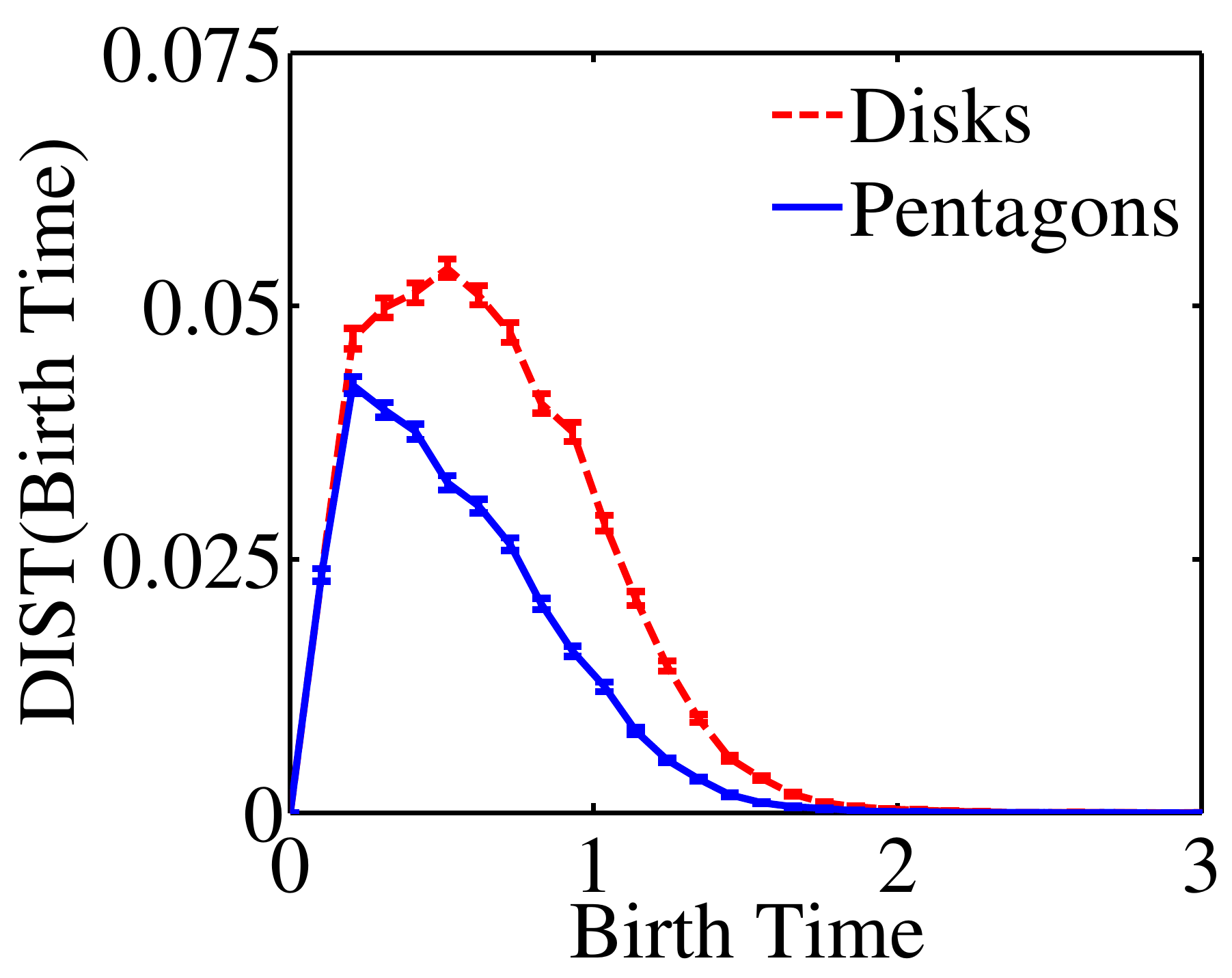}} 
\subfigure[$\pd_1$ tangential forces.]{\includegraphics[width=1.6in]{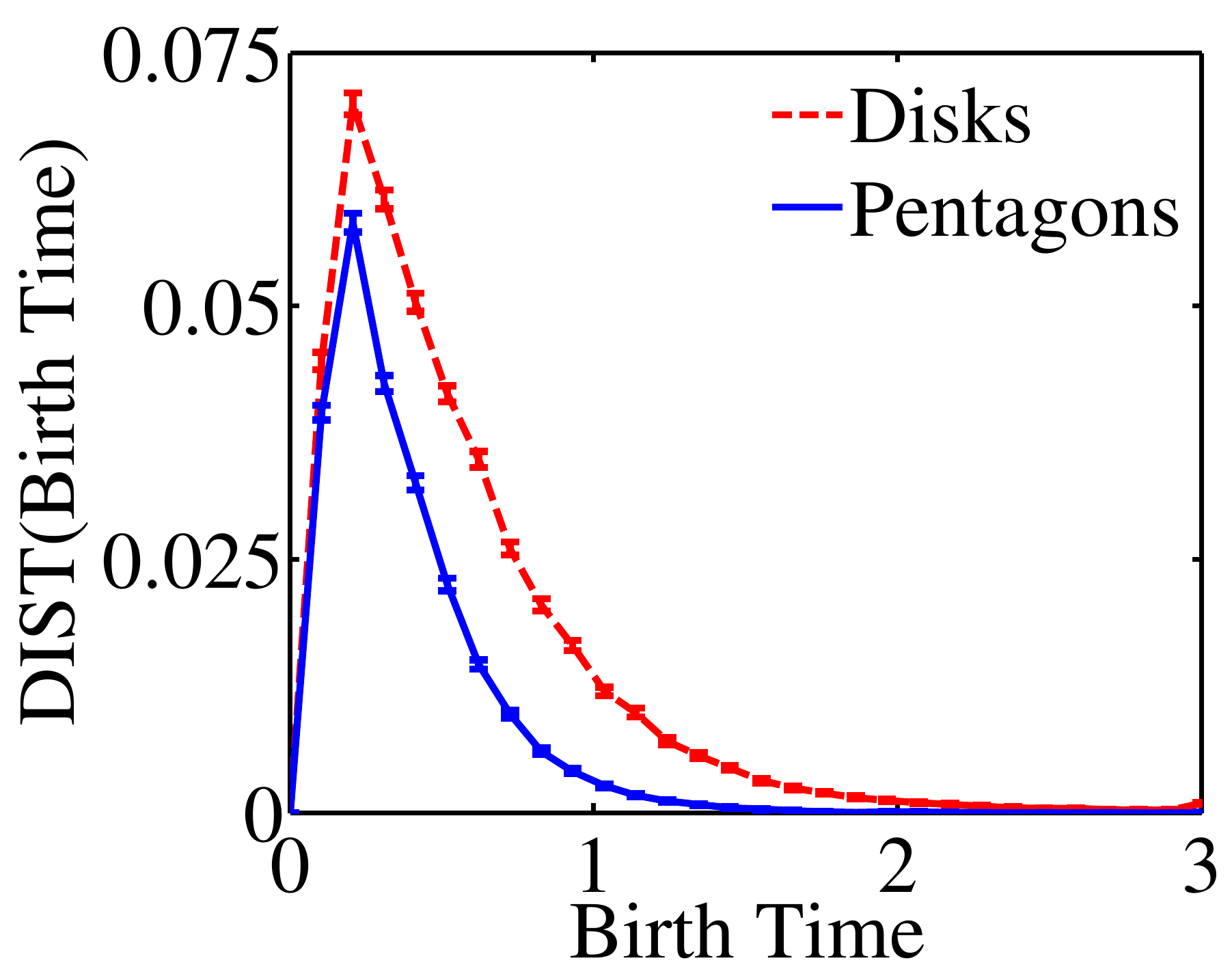}} 
\caption{
Distribution of birth times (bottom slice, low tapping).   Only the features with the lifespan larger than $0.1$ are included.   Compare to Fig.~\ref{fig:disks_pents_birth0}.
}
\label{fig:disks_pents_birth0.1}
\end{figure}

\begin{figure}
\centering
\subfigure[$\pd_0$ normal forces.]{\includegraphics[width=1.6in]{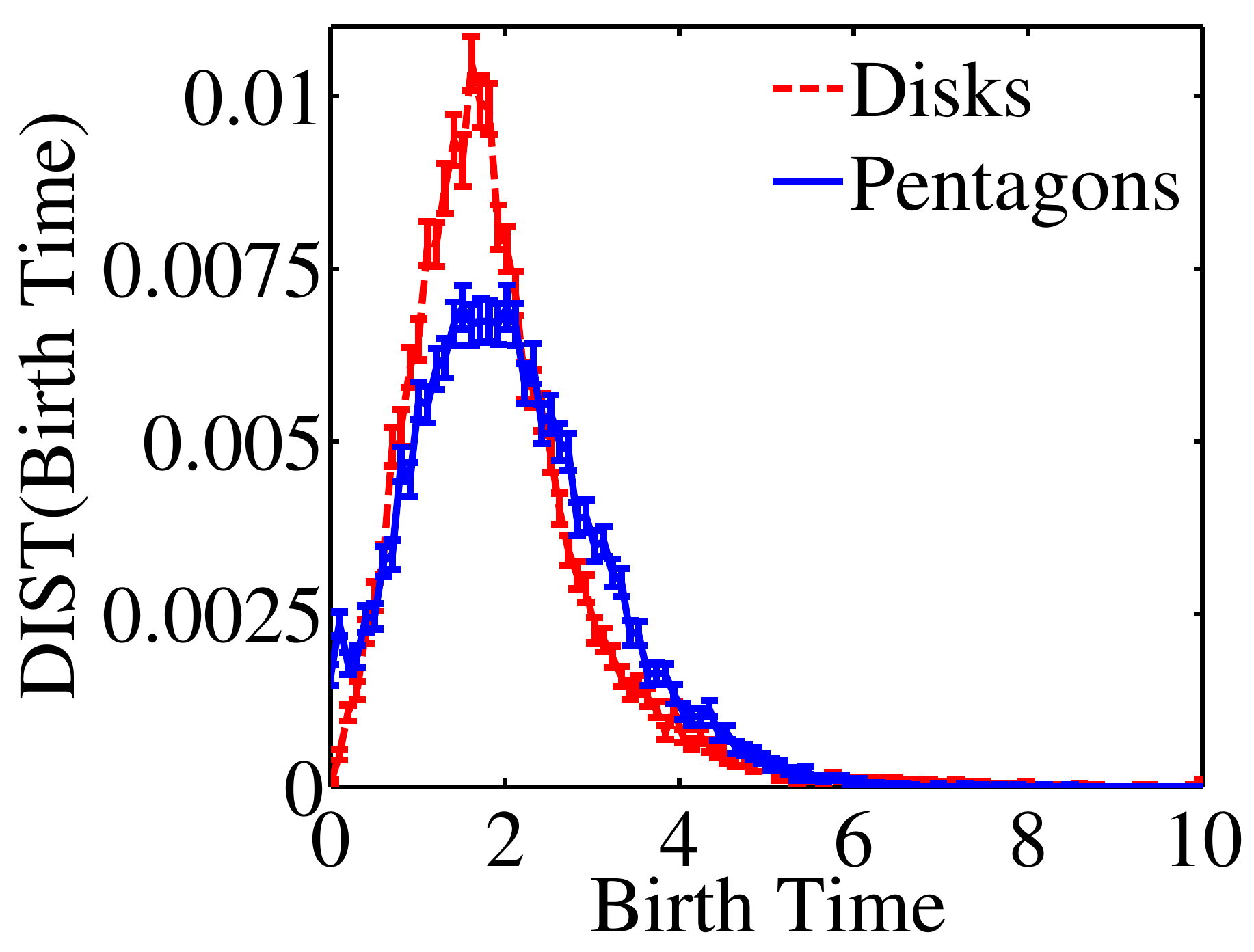}} 
\subfigure[$\pd_0$ tangential forces.]{\includegraphics[width=1.6in]{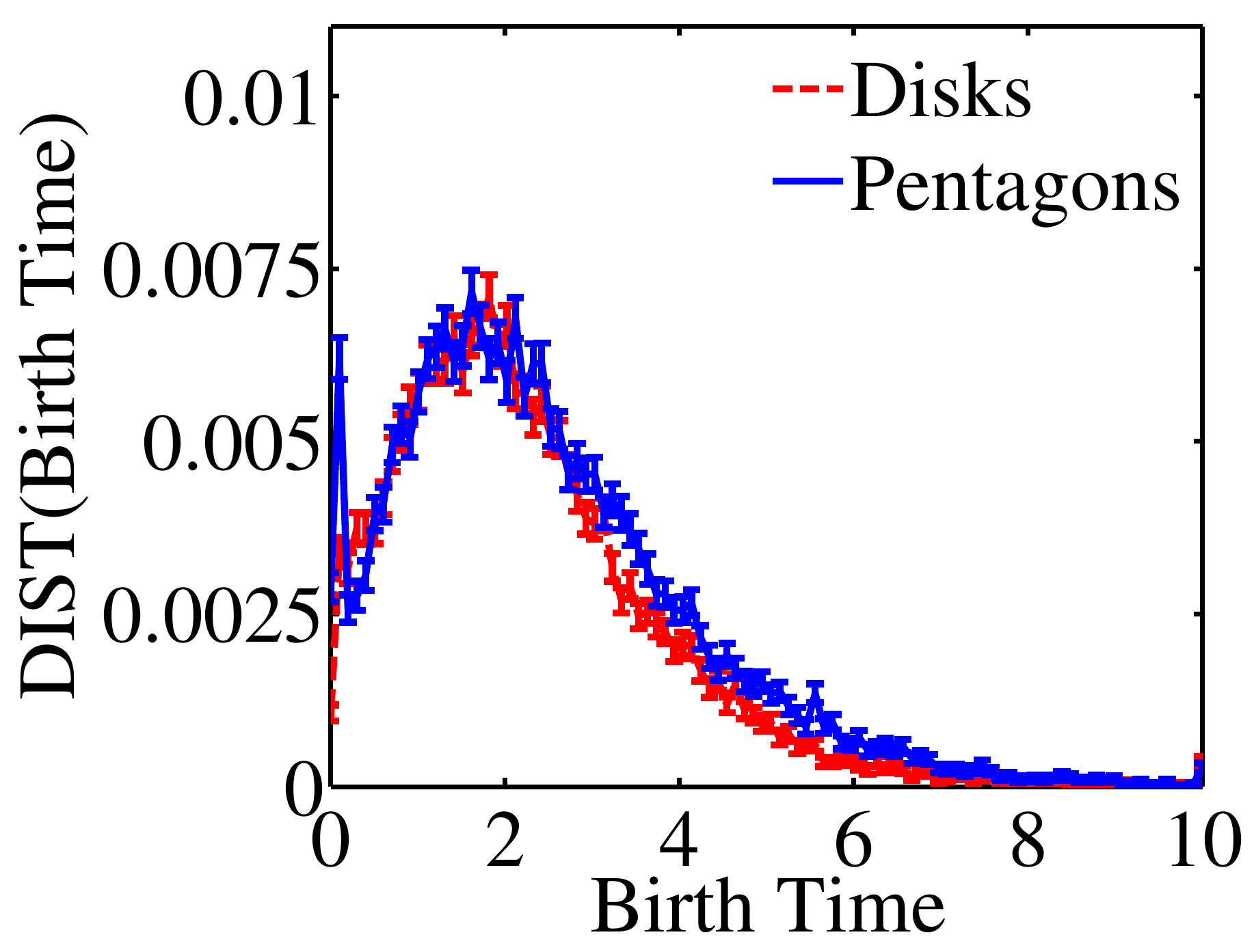}} 
\subfigure[$\pd_1$ normal forces.]{\includegraphics[width=1.6in]{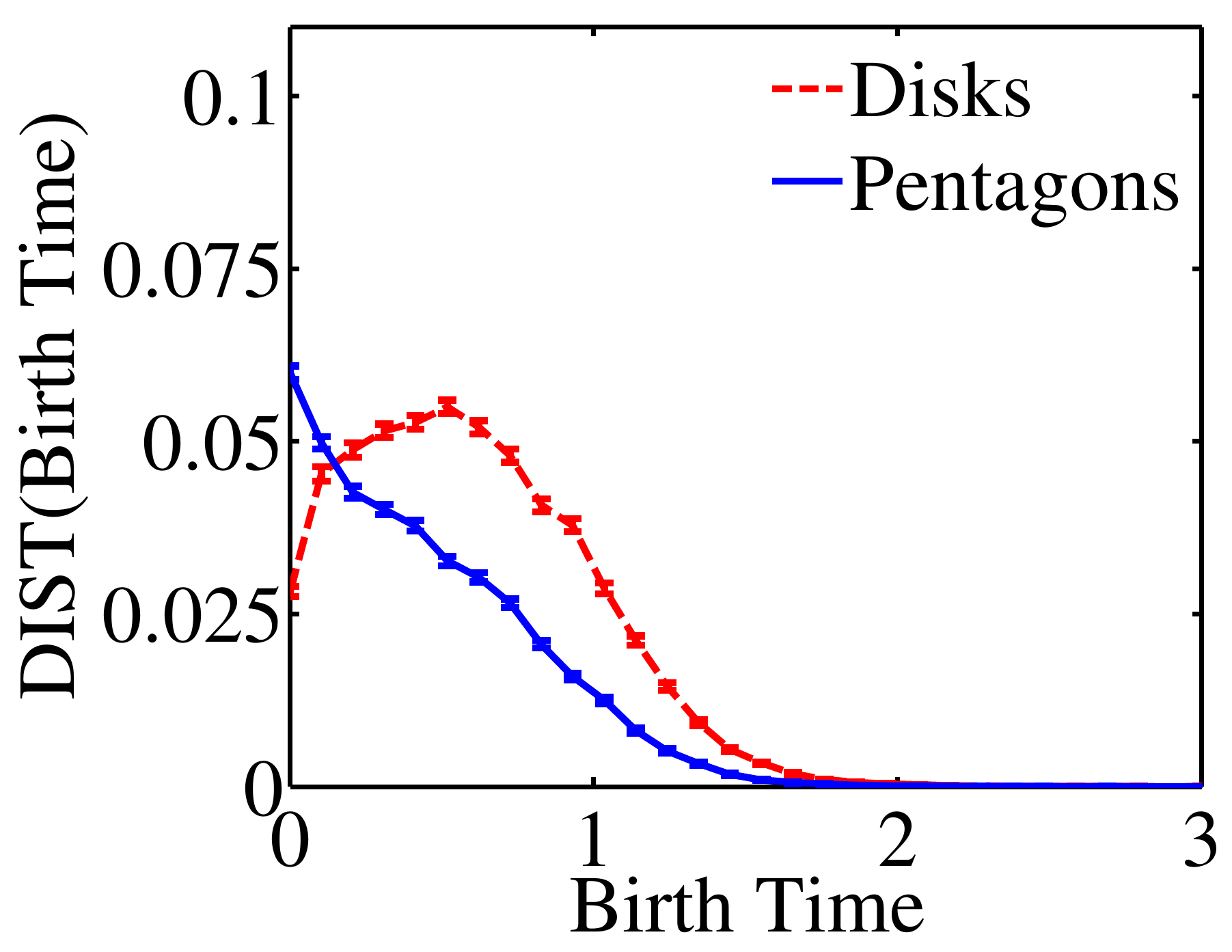}} 
\subfigure[$\pd_1$ tangential forces.]{\includegraphics[width=1.6in]{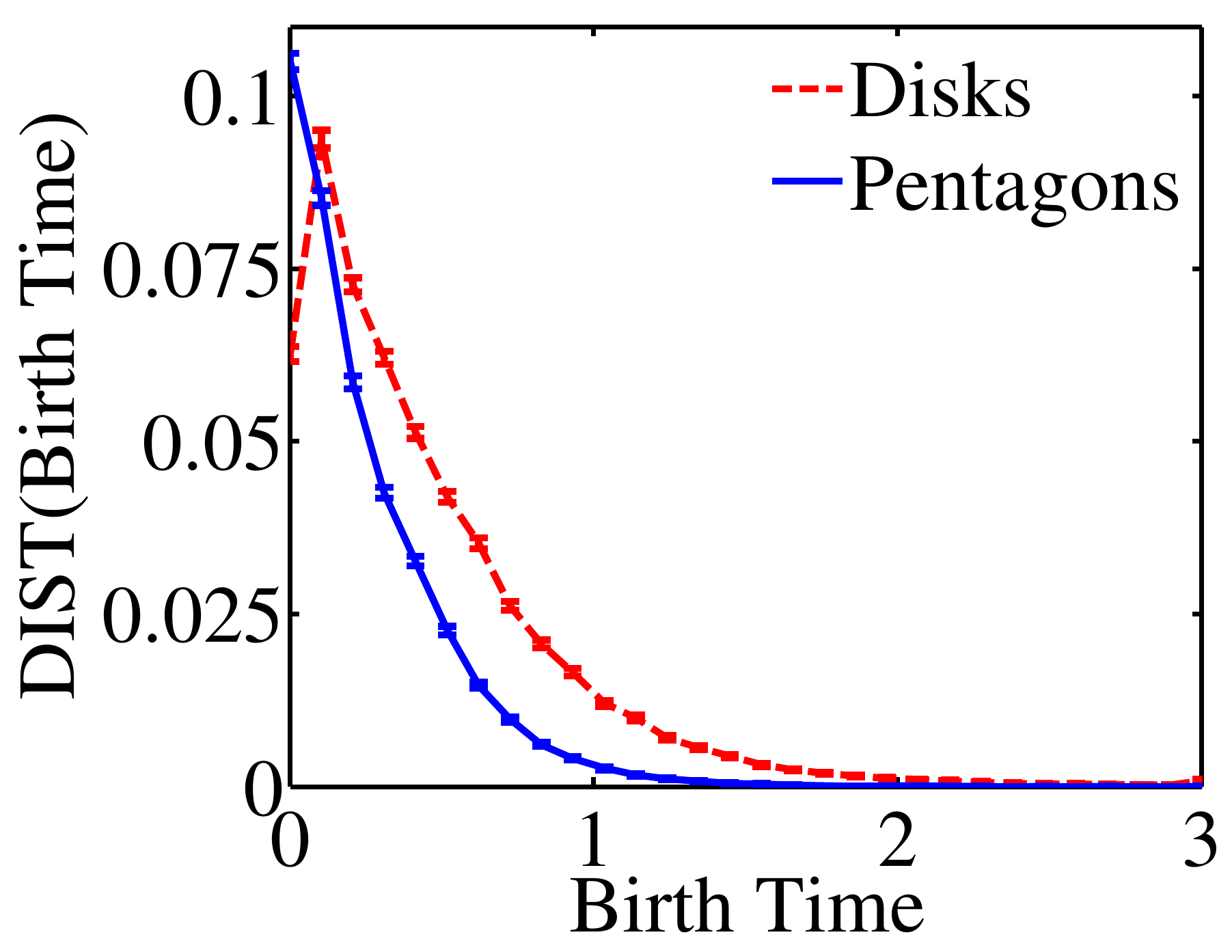}} 
\caption{
Distribution of birth times (bottom slice, low tapping).   
All the features, independent of lifespan, are shown.  
Compare to Fig.~\ref{fig:disks_pents_birth0.1} (note different range on the vertical axes). 
}
\label{fig:disks_pents_birth0}
\end{figure}

We proceed by discussing the source of the differences between the disk- and pentagon-based systems considered so far.   First, we focus on the distributions
of birth times.  Figsures~\ref{fig:disks_pents_birth0.1} and~\ref{fig:disks_pents_birth0} show the corresponding results. The only difference 
between these figures is that in Fig.~\ref{fig:disks_pents_birth0.1} we consider only the points with the lifespan larger than $0.1$, while in 
Fig.~\ref{fig:disks_pents_birth0} we consider all the points.  The reason for showing both figures is that the differences between the two provide additional information about
the points with short lifespan. Considering components for normal forces, parts (a) in these two figures, we observe that birth times 
capture some differences between the
two systems that were not obvious when considering distances.  There are more points in $\pd$s for disks that are 
born around $F \sim 2$, and more points in $\pd$s for pentagons born at
larger forces. This is consistent with the PDFs for disks and pentagons shown in Fig.~10 of~\cite{paper1}.   
For the tangential forces, parts (b), we do not see much if 
any difference in the birth times.   Regarding loops, the parts (c) and (d) of Figs.~\ref{fig:disks_pents_birth0.1} and~\ref{fig:disks_pents_birth0}, one consistent 
observation is that there are more points in $\pd$s for disks than for pentagons for the whole range of forces considered.
Moreover, for disks loops start appearing at higher  force level than for pentagons.    The differences between these two 
figures show how many of the points have a short lifespan; these differences
are particularly interesting for loops, parts (c) and (d): we note a significantly larger number of points for pentagons at small birth times, suggesting that loops
for pentagon-based systems form at very small or vanishing force, consistently with the discussion in~\cite{paper1}.   
This finding holds both for loops formed by normal and tangential forces.

\begin{figure}
\centering
\subfigure[$\pd_0$ normal forces.]{\includegraphics[width=1.6in]{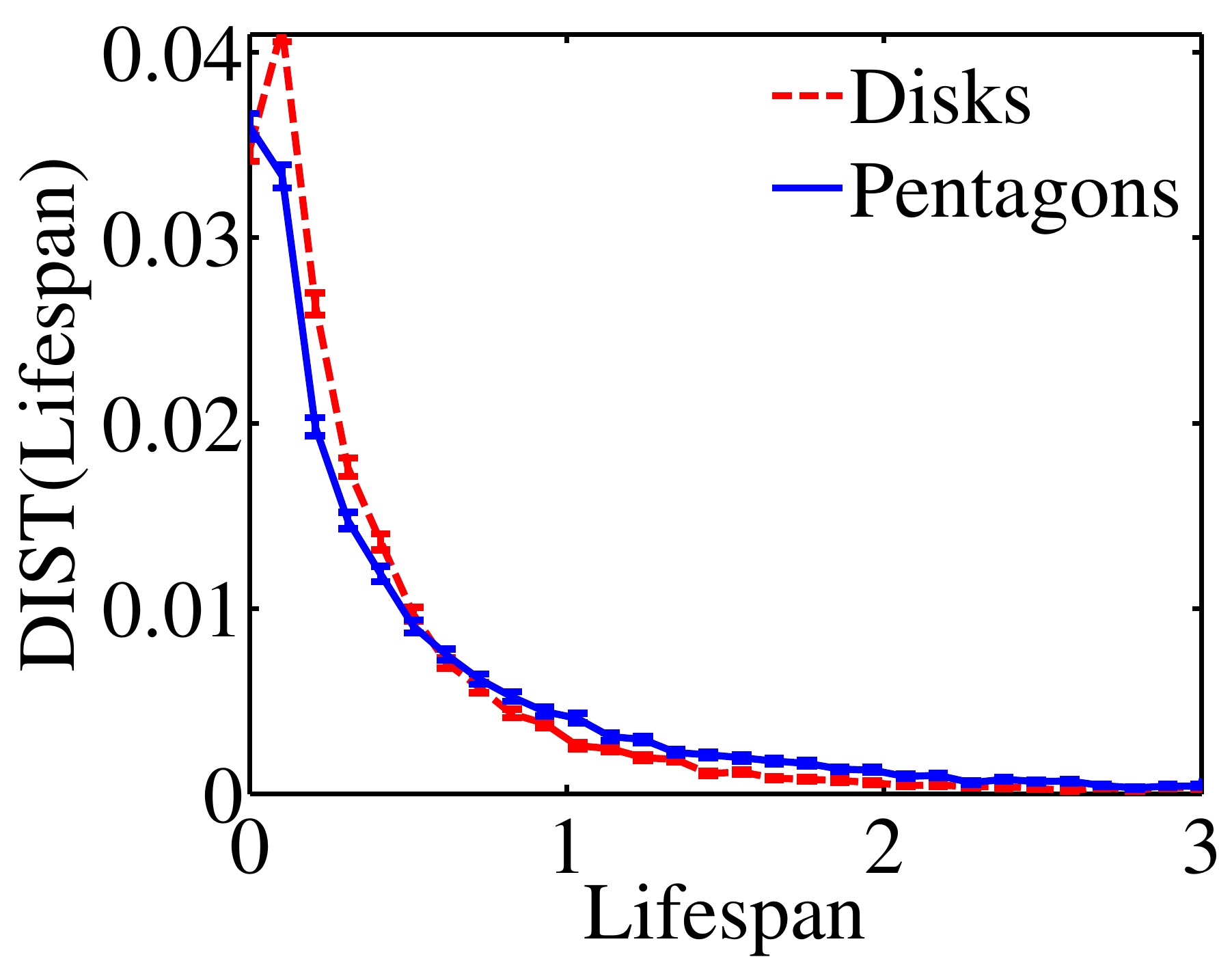}} 
\subfigure[$\pd_0$ tangential forces.]{\includegraphics[width=1.6in]{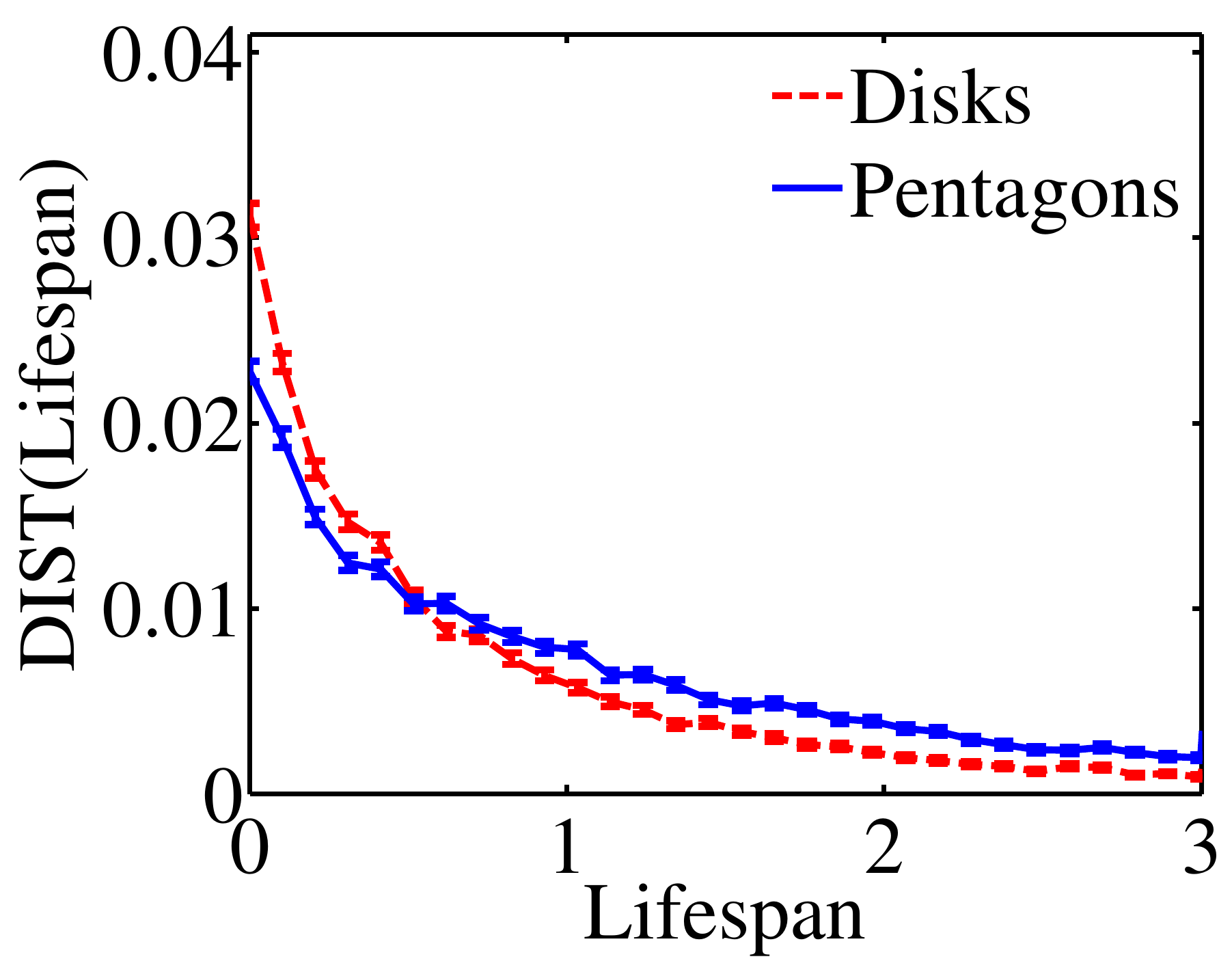}} \\
\subfigure[$\pd_1$ normal forces.]{\includegraphics[width=1.6in]{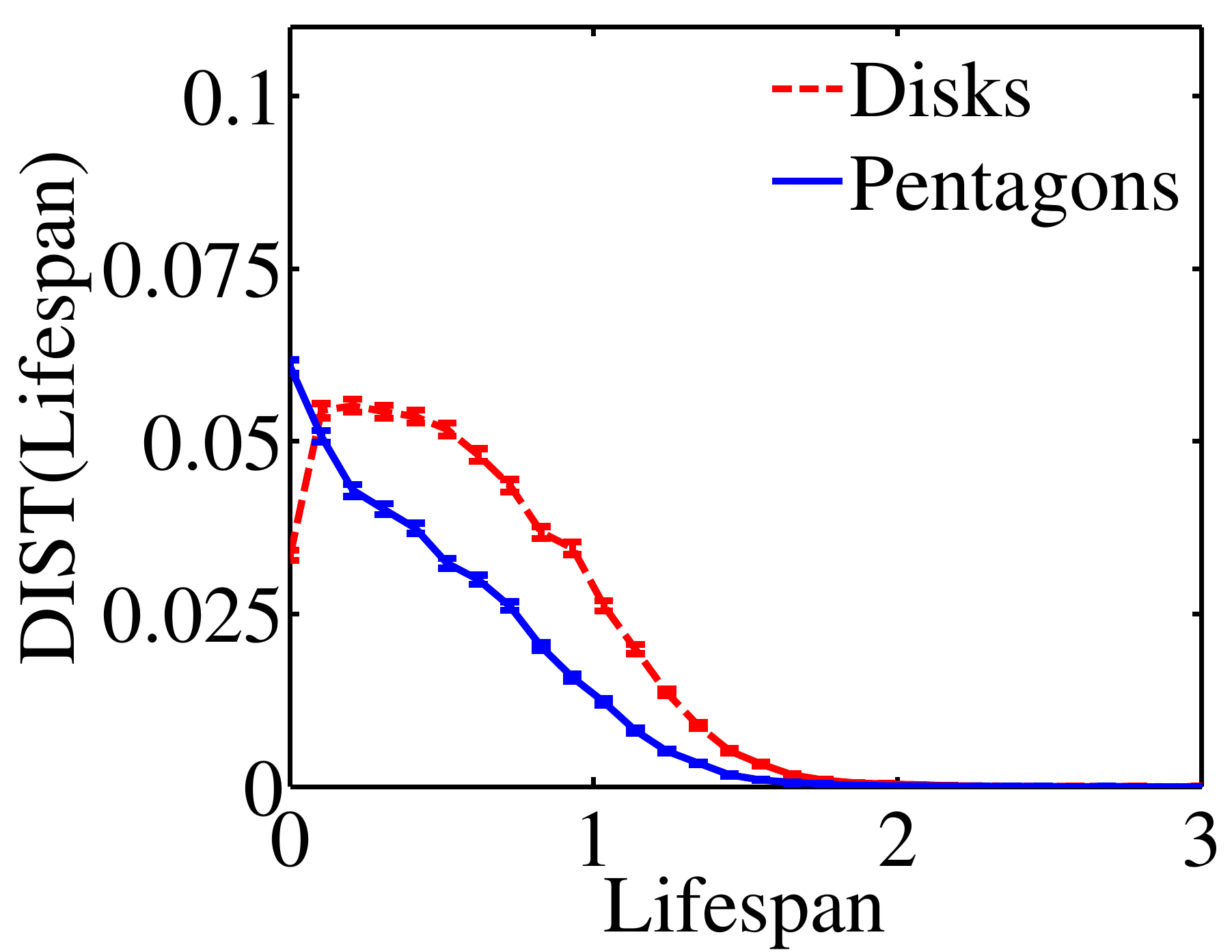}} 
\subfigure[$\pd_1$ tangential forces.]{\includegraphics[width=1.6in]{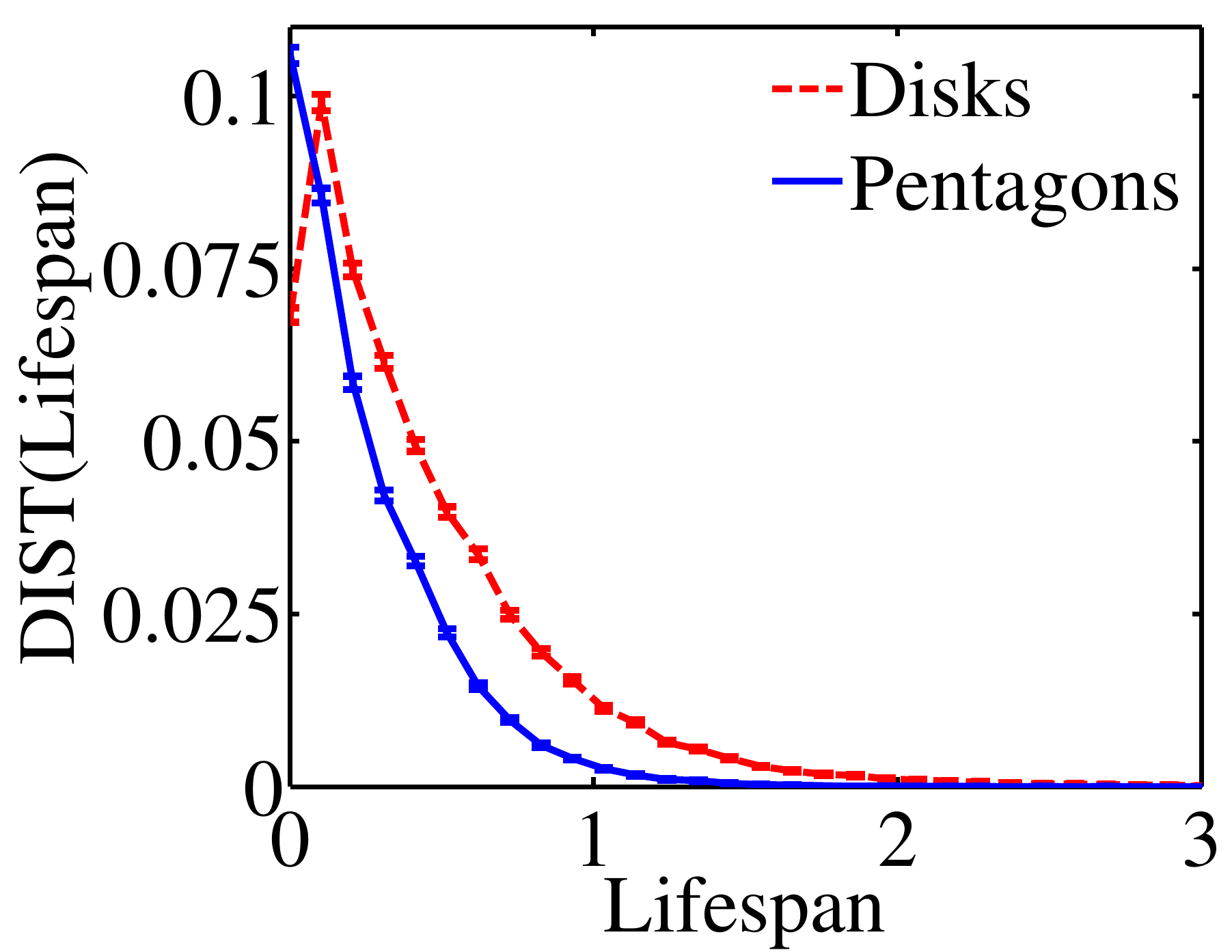}} 
\caption{Distribution of lifespans (bottom slice, low tapping).    
}
\label{fig:lifespans}
\end{figure}

Figure~\ref{fig:lifespans} presents distributions of the lifespans for disks and pentagons.  
From $\pd_0$ diagrams we conclude that for both disks and pentagons, the dominant number of components is characterized by rather short
lifespans.   We also observe a cross-over (more pronounced for tangential forces) between disk and pentagon distributions, although the difference is not large. We note that the lifespans larger than $\approx 0.75$ are more probable for pentagons than for disks.   Therefore, the components live longer for pentagon-based system in particular when tangential forces are considered. 
To use the landscape analogy, this result says that mountain peaks in the tangential force network are more pronounced for pentagon-based systems.   
Observe from Fig.~\ref{fig:lifespans}(c,~d) that the lifespan curves are similar to the birth time curves shown in Fig.~\ref{fig:disks_pents_birth0}.
This is because for both disks and pentagons, most of the loops disappear very close to the zero force level, and thus the death time 
provides no additional information.

\begin{figure}
\centering
\subfigure[$\pd_0$ normal forces.]{\includegraphics[width=1.6in]{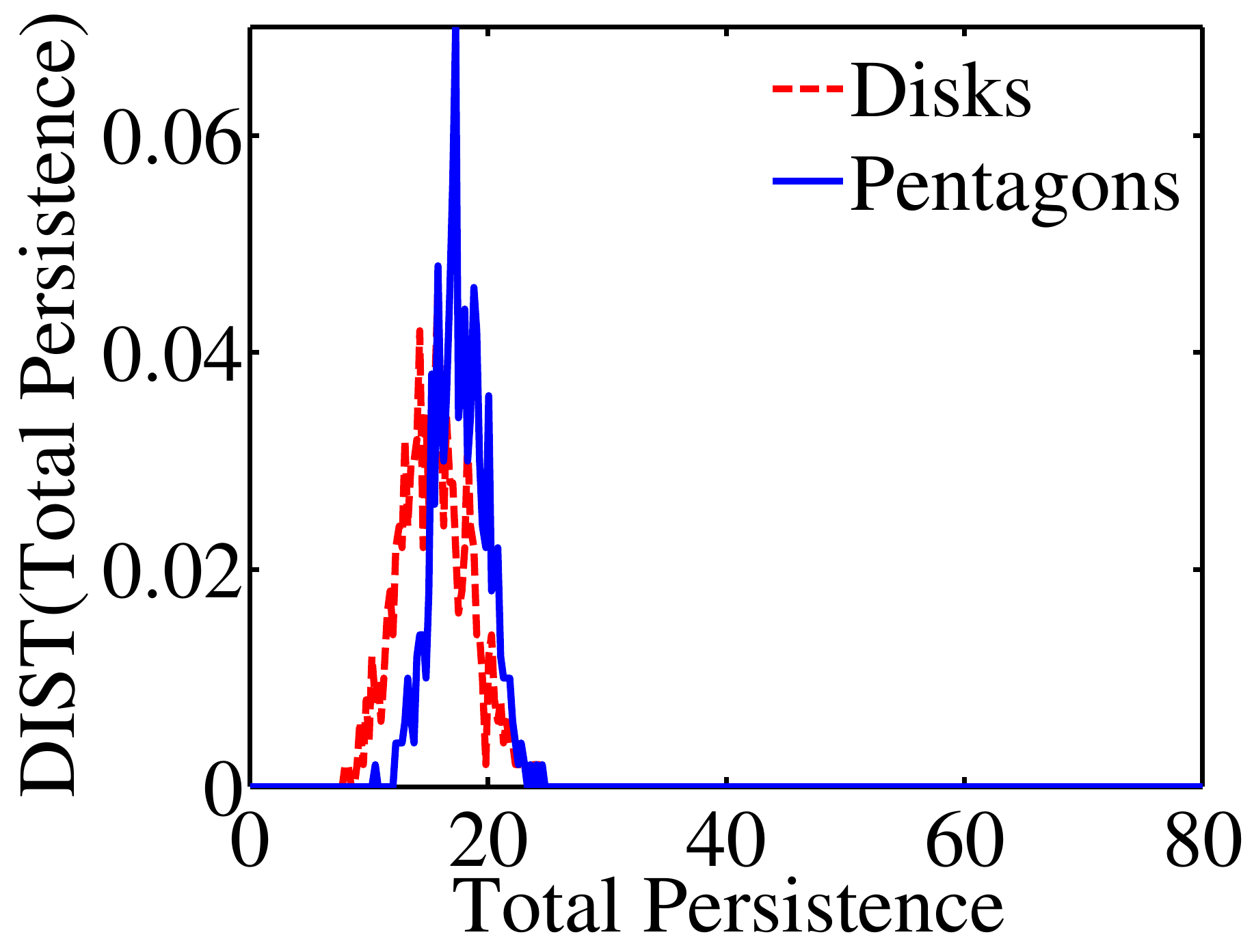}} 
\subfigure[$\pd_0$ tangential forces.]{\includegraphics[width=1.6in]{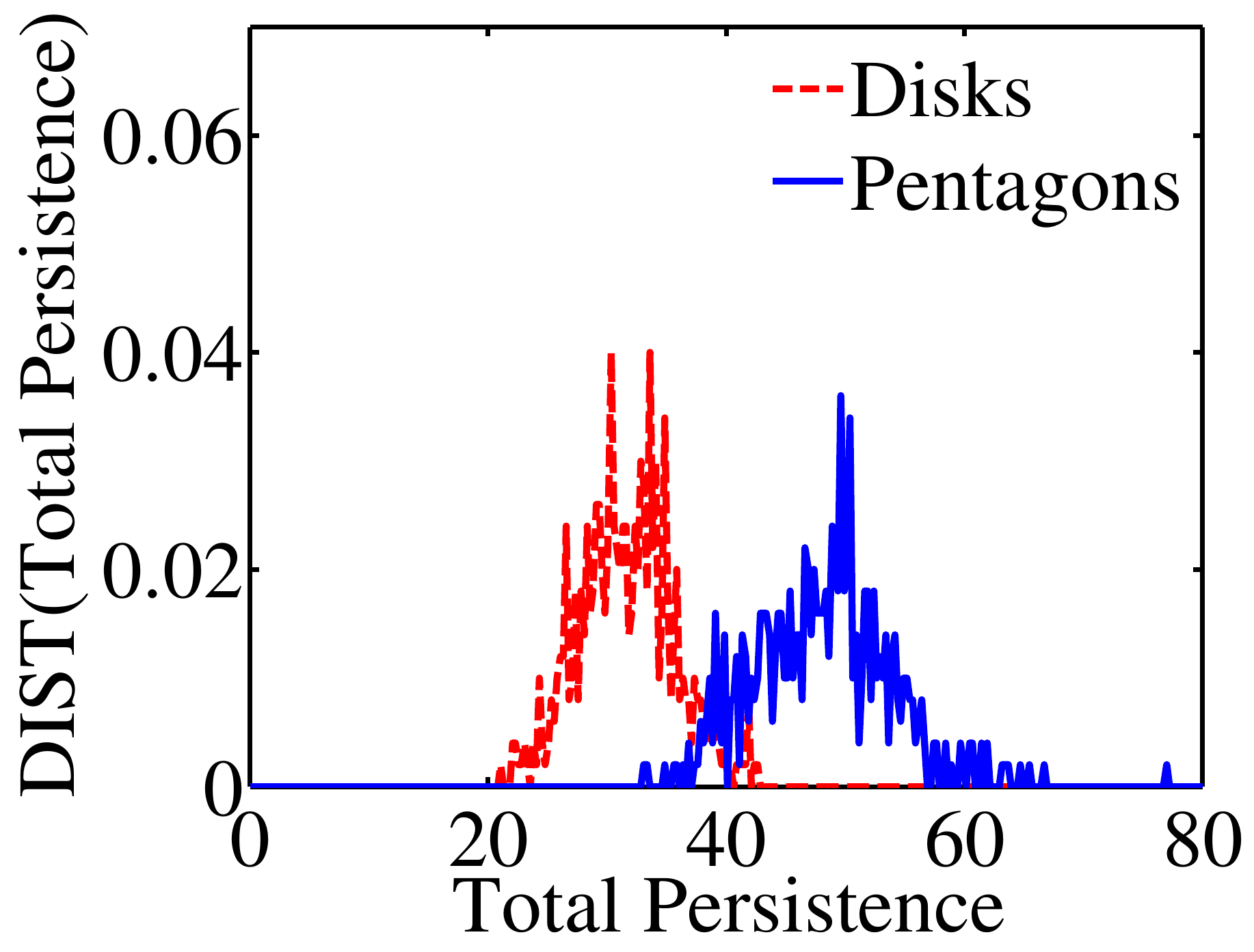}} \\
\subfigure[$\pd_1$ normal forces.]{\includegraphics[width=1.6in]{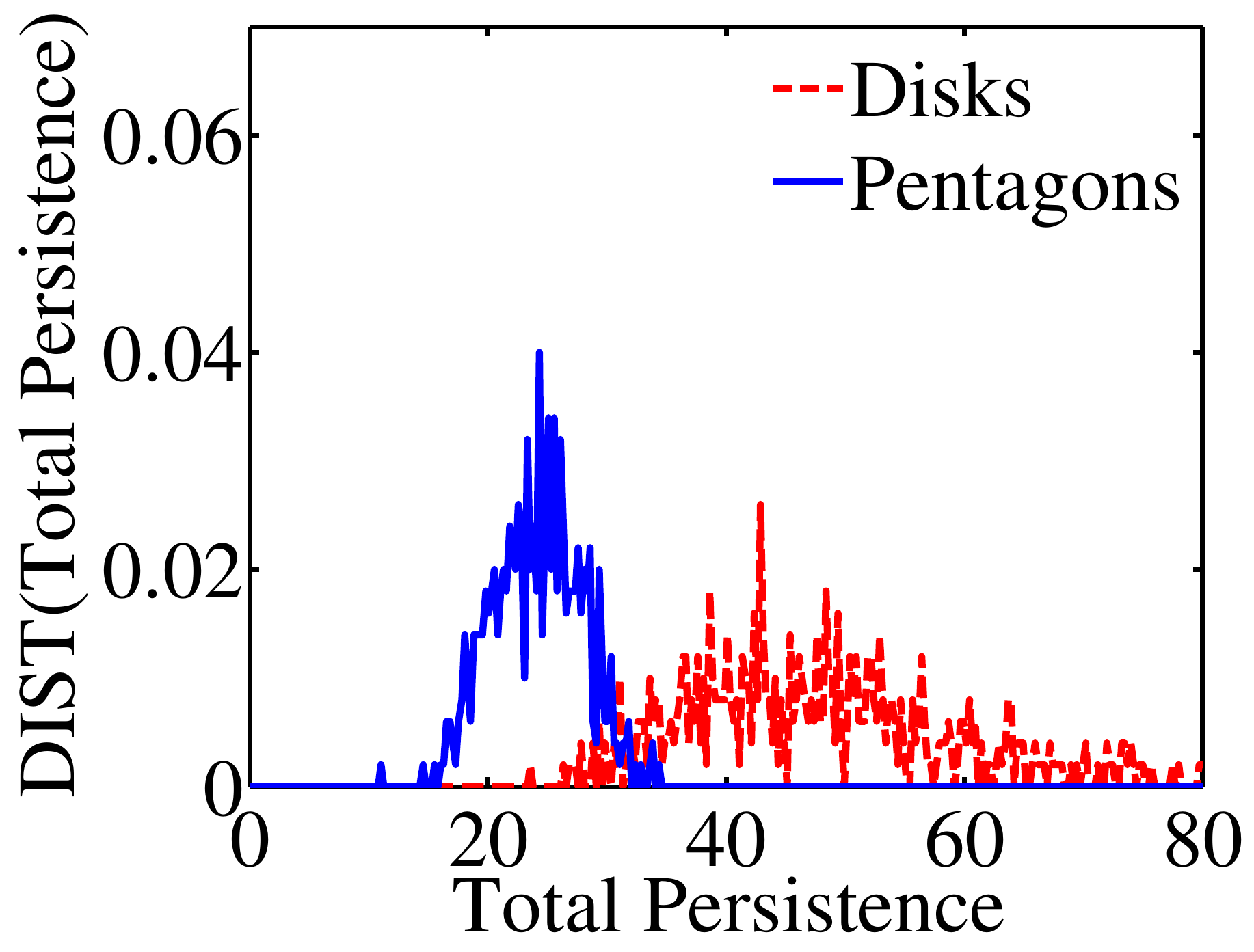}} 
\subfigure[$\pd_1$ tangential forces.]{\includegraphics[width=1.6in]{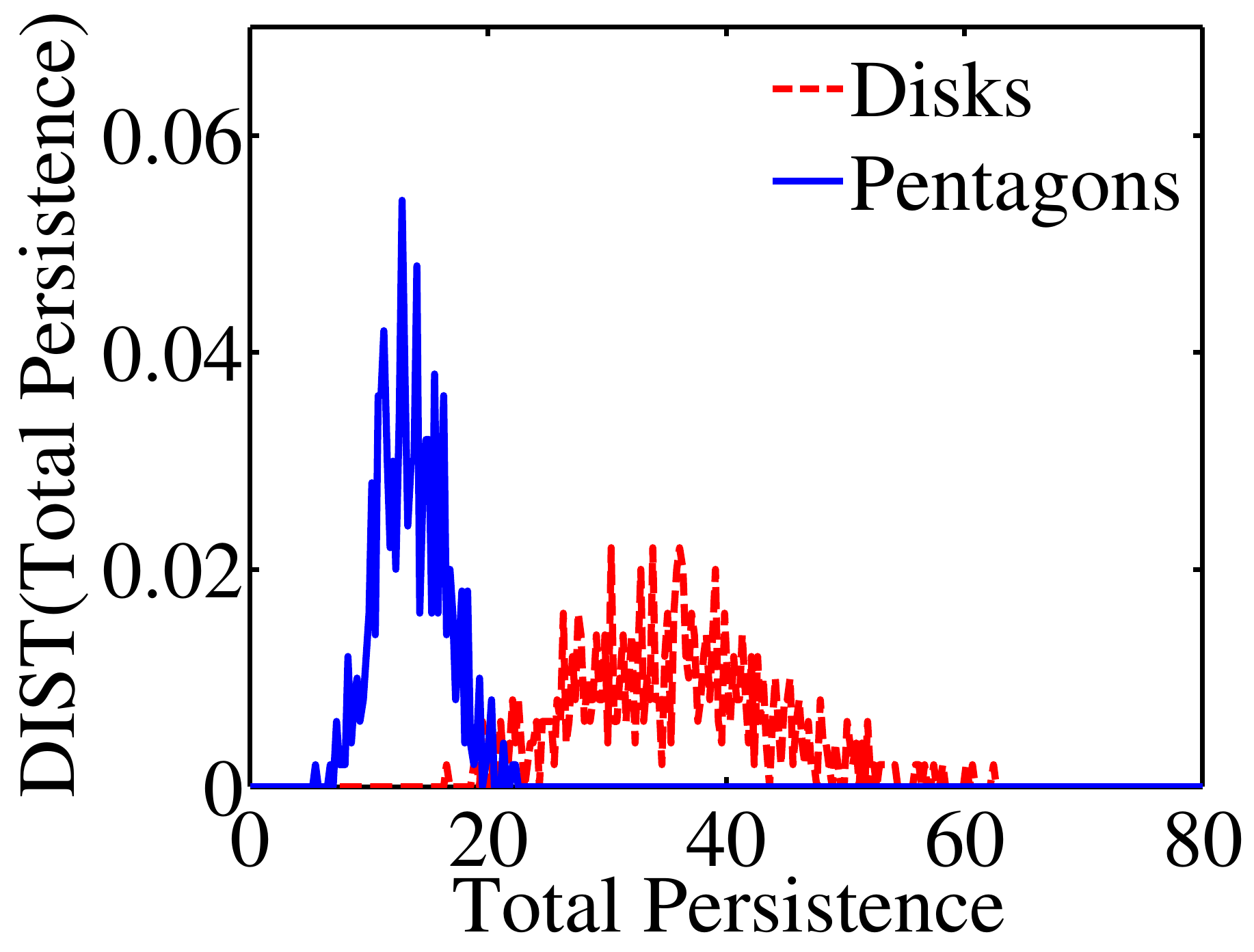}} \\
\caption{Total persistence (bottom slice, low tapping).   Note large differences between disks and pentagons. }
\label{fig:totalpers}
\end{figure}

Figure~\ref{fig:totalpers}  shows the total persistence, $TP$, that to a large degree summarizes many of the findings discussed so far.   
We recall that $TP$  corresponds to the sum of the lifespans, see Sec~\ref{sec:methods}, 
so considering the results shown in this figure together with 
the ones shown in Fig.~\ref{fig:lifespans} is useful.   For  $TP(\pd_0)$ diagrams, there is only a minor difference 
between disks and pentagons in the normal force network;  however, for tangential forces, there are significant differences. 
This reflects the larger lifespans of the components for pentagon-based system.
For $TP(\pd_1)$, the differences are very obvious for both normal and tangential forces, and in contrast to $TP(\pd_0)$ results, 
here we find that the distribution of $TP(\pd_1)$ is shifted to larger values and is much broader for disk-based systems.   

Figure~\ref{fig:totalpers} shows clearly significant differences in the structure of force networks in the systems of tapped disks and pentagons.  
Pentagon systems tend to form new components (clusters) at higher force levels and these endure longer before they merge, in comparison to 
disk-based ones. This is particularly evident for the tangential force network.   In contrast, loops are formed at relatively low force levels in 
pentagon-based systems. Hence, one could expect that the clusters that form at higher force levels are more stretched  (because they do not 
contain loops) for pentagons.  Since most loops persist down to zero force levels, the $TP(\pd_1)$  for pentagons is significantly lower than for disks.

\section{Conclusions}
\label{sec:conclusions}

In the present paper, we discuss and describe properties of force networks in tapped particulate 
systems of disks and pentagons.   
Our analysis is based on persistent homology that allows to precisely measure and quantify a number of properties of these networks.  
The persistence diagrams record the distribution and connectivity of the 
features (components, loops) that develop in the force landscape as the  force  threshold is decreased. These
diagrams can then be analyzed and compared by a number of different means, some of them 
described and used in the present work. 

One of the considered concepts is the distance between the persistence diagrams that allows for their 
direct comparison.  The comparison can be carried out on the level of individual diagrams, 
allowing to compare between different configurations of nominally
the same system, between different parts of a given system, or between completely different systems.   
In addition, one can compare the distributions of the distances. These comparisons allow us
to identify, in a precise manner, the differences between persistence diagrams, and therefore
force networks.   

In addition to distances, we have defined and used other measures, such as birth times, 
showing at which force level features appear; lifespans, showing how long the features
persist as force threshold is modified; and finally total persistence to describe essentially 
how `mountainous' the force landscape considered is.  The listed measures were computed
both for components/clusters that could be in a loose sense related to force chains, and for 
loops that could be related to `holes' in between the force chains.

The use of the outlined measures has allowed us to identify a number of features of force
networks.   We use these
measures, for example, to identify and explain the differences between the systems of 
disks exposed to different tapping intensities that lead to (on average) the same packing
fraction.  In addition to identifying the differences between these systems, the implemented
measures have also shown that the systems of disks, when exposed to low tapping intensity, 
evolve in a nontrivial manner, with  the subsequent taps possibly correlated to the preceding 
ones.   We have shown that the oscillations in the measures built upon persistence diagrams 
are correlated with small oscillations in the packing fraction.     More generally, the finding
is that if the system is tapped strongly and therefore
the force network is rebuilt from scratch at each tap, the resulting force networks are similar; however,
under low tapping regime, the system (and the resulting force network) appears to be stuck 
in a certain state, and jumps out of it only infrequently.   This nontrivial finding and its consequences
will be explored in more detail in our future works. 

Another comparison that we carried out involves tapped systems of 
disks and pentagons.  One important finding here is that the differences between disks and 
pentagons are significant when the structure of loops is considered: presence of loops is much more common
for the systems of disks than for pentagons, independently of whether normal or 
tangential forces are considered.    On the other hand, the differences between the persistence diagrams
based on components/clusters are minor and relatively difficult to identify.   
Therefore, the force networks that form in tapped systems of disks and pentagons 
are similar when only components are considered, but significantly different when loops 
are included.  

On a more general note, it should be emphasized that the measures described here allow to 
directly compare force networks, and quantify the differences.   For example, we can
now quantify in precise terms variability of force networks between one system and another. 
This ability  opens the door for developing more elaborate comparisons, measures, and 
also connections between the force network properties and mechanical response on 
the macroscale.   Furthermore, the analysis that we presented here can be easily applied
to the three dimensional systems, where any direct visualization may be difficult.  Our 
future research will proceed in this direction.

\begin{acknowledgments}
KM and MK were partially supported by NSF grants No. DMS-0915019, 1125174, 1248071, and contracts from AFOSR and DARPA.  LK acknowledges support by the NSF grant No. DMS-0835611 and DMS-1521717. 
\end{acknowledgments}

\bibliographystyle{apsrev}

\end{document}